\documentclass[3p,11pt,sort&compress,authoryear]{elsarticle}

\usepackage{natbib}
\usepackage{enumerate}
\usepackage{amsmath, amsthm, amssymb, bm}
\usepackage{appendix}
\usepackage{graphicx}
\usepackage{wrapfig}
\usepackage{multirow}
\usepackage{caption}
\usepackage{subcaption}

\usepackage{afterpage}
\usepackage{float}
\usepackage{url}
\usepackage{hyperref}

\makeatletter
\newcommand\footnoteref[1]{\protected@xdef\@thefnmark{\ref{#1}}\@footnotemark}
\makeatother

\begin{document}

\begin{frontmatter}

\title{Examining Travel Patterns and Characteristics in a Bikesharing Network\\ and Implications for Data-Driven Decision Supports:\\ Case Study in the Washington DC Area}

\vspace{10 mm}

\journal{Journal of Transport Geography, 71: 84-102, 2018 {\em [DOI: \href{http://dx.doi.org/10.1016/j.jtrangeo.2018.07.010}{10.1016/j.jtrangeo.2018.07.010}]}}

\author[label1]{Xiao-Feng Xie}
\ead{xie@wiomax.com}
\author[label1]{Zunjing Jenipher Wang}
\ead{wang@wiomax.com}
 
\address[label1]{WIOMAX LLC, Rockville, MD 20848}

\begin{abstract}
\label{pap:abstract}
Bikesharing has gradually become an adopted form of mobility in urban area recent years as one sustainable transportation mode to bring us many social, environmental, economic, and health-related benefits and rewards. There is increased research toward better understanding of bikesharing systems (BSS) in urban environments. However, our comprehension remains incomplete on the patterns and characteristics of BSS. In this paper, aiming to help improving sustainability in multimodal transportation through BSS, we perform a systematic data analysis to examine underlying patterns and characteristics of the system dynamics in a bikeshare network and to acquire implications of the patterns and characteristics for decision making. As a case study, we use trip history data from the Capital Bikeshare system in the Washington DC area and some additional data sources. The study covers seven important aspects of bikeshare transportation systems, which are respectively trip demand and flow, operating activities, use and idle times, trip purpose, origin-destination flows, mobility, and safety. For these aspects, by using appropriate statistical methods and geographic techniques, we investigate travel patterns and characteristics of BSS from data to evaluate the qualitative and quantitative impacts of the inputs from key stakeholders on main measures of effectiveness such as trip costs, mobility, safety, quality of service, and operational efficiency, where key stakeholders include road users, system operators, and city. We also disclose some new patterns and characteristics of BSS to advance the knowledge on travel behaviors. Finally, we briefly summarize our findings and discuss the implications of the patterns and characteristics for data-driven decision supports from the relations between BSS and key stakeholders for promoting bikeshare utilization and transforming urban transportation to be more sustainable.

\end{abstract}

\begin{keyword}
Bikeshare Systems \sep Urban Mobility \sep Data Analytics \sep Origin-destination (O-D) Flows
\end{keyword}

\end{frontmatter}

\section{Introduction}

Nowadays, transportation is predominantly dependent on motor vehicles, which has resulted in a practical problem in urban areas, traffic congestion. In 2014, congestion caused urban Americans to spend a cost of \$160 billion on substantial delays and extra fuel consumption \citep{schrank20152015}, besides the detrimental impact on environment from the increased vehicle emissions. To tackle the challenges in maintaining urban sustainability, bicycle use as an emission-free substitute for motor vehicles was encouraged and has become an increasing trend in cities around the world~\citep{fishman2014bike}. Bicycling can either replace driving for the short-to-medium-distance trips, or provide first- and last-mile connections to other transportation modes to facilitate an intermodal transportation system~\citep{demaio2009bike,shaheen2013public,ma2015bicycle}. Use of bicycles is hugely rewarding from social, environmental, economic, and health-related aspects for cities, communities and bike users. Benefits of alleviating congestion and mitigating associated environmental damages accrue from the vehicle miles traveled (VMT) in transportation reduced by bicycling~\citep{hamilton2017bicycle,wang2017bike}. Accessibility to neighborhoods is enhanced by bicycling to boost economic opportunities to local businesses~\citep{buehler2015business}. Moreover, as an active transportation mode, bicycling not only plays a unique role in supporting recreational trips, but also provides substantial public health advantages \citep{shaheen2013public,mueller2015health}. In terms of the increased physical activity, bicycling is shown quite effective for reducing potential health risks \citep{mueller2015health,fishman2016bikeshare}. 

In recent years, many cities have improved their cycling infrastructure. The adoption of the bicycling mode in transportation has experienced significant growth. According to American Community Survey (ACS) 2008--2012 \citep{mckenzie2014modes}, commuting by bike had a percentage increase about 61\% from 2000, which increased larger than any other commuting mode. More cities around the world have invested substantially in public bicycle programs or bikesharing systems (BSS) \citep{demaio2009bike,shaheen2013public,o2014mining,fishman2016bikeshare}. Incorporated with information technology, BSS allows users to immediately reserve, pickup, and drop-off public bikes in the network of docking stations at an affordable cost of paying some user or membership fee for bike riding services. Compared to private bikes, BSS not only makes users freed from ownership and regular maintenance of bike, but also allows users to bike one way to connect with other transportation modes with more flexibility on intermodal trips. In 2017, over 1000 cities have offered bikeshare programs and over 4.5 million public bicycles have been in use \citep{Meddin2017bike}. 

As a new form of mobility that gradually emerged, BSS has attracted much interest and attention in research. Analysis of surveys and data \citep{romanillos2016big} have been performed by the operators and analysts who aim to achieve a better understand on the system states and key factors influencing the user experiences and the effectiveness of BSS, and on the role of BSS in transforming future urban mobility. Various aspects of bikesharing have been studied. \cite{o2014mining} classified 38 BSSs based on an analysis of variations in occupancy rate. A few studies provided basic insights concerning the impacts of seasonal weather and temporal trends on bicycling in urban environments \citep{gebhart2014impact,el2017effects}. Corresponding to bikeshare users, several studies summarized data and surveys on some significant differences of the user behaviors in terms of their demographic characteristics~\citep{zhao2015exploring,bikeshare2016capital,bhat2017spatial}. 
Some other research analyzed the difference of the trip attributes to gain some understandings on trip purposes, especially between round and one-way trips \citep{zhao2015exploring,Noland2017} and between casual and member users \citep{buck2013bikeshare,Wergin2017,Noland2017} among other factors \citep{fishman2016bikeshare}. 

Mobility and safety are two main factors for road users to make their mode choice of transportation and for urban planners and policy makers to improve transportation systems. Concerning the bikeshare mobility, some initial studies have been conducted in literature. As shown in the survey by \cite{moritz1997survey}, the average speed of bicycle commuting was 14.6 mph. \cite{Wergin2017} estimated the average speed using a small sample of 3,596 trips with GPS tracking data. \citet{Perez2017} studied the impact of mobility from the viewpoint of accessibility in the bike lane network. As sharing road with vehicles, cyclists are vulnerable users that are more likely to be injured when involved in traffic collisions. According to \cite{NHTSA2015TSF}, 818 bicyclists were killed and an additional estimated 45,000 were injured in traffic crashes in USA in 2015. \cite{lowry2016prioritizing} classified bike roads in a network in terms of stress levels \citep{Rixey2017}. Other studies were performed to understand the crash risk of bikeshare users~\citep{martin2016bikesharing,fishman2016global}. 

It has been a common interest for researchers and operators to push BBS into demand-responsive operations. A few studies examined BSS usage and traffic patterns at different levels of spatio-temporal aggregation to recognize the impacts of contributing indicators on BSS demand \citep{fournier2017sinusoidal,jestico2016mapping,faghih2016incorporating}. \citet{vogel2011understanding} applied clustering-based data mining to explore activity patterns, which revealed the imbalances in the spatial distribution of bikes in BSS. \citet{o2014mining} documented the redistribution problem of bikes in BSS from the variations in load factor. Some studies \citep{de2016bike,fishman2016bikeshare,faghih2017empirical} proposed bike rebalancing methods (e.g., trucks and corral services of BSS) to help solving the imbalance between demand and supply at bike stations so that to improve the operational efficiency for BSS and to meet the service level agreements (SLA) and guarantee the quality of service (QoS) for users. Finding a more cost-effective way for sustainable operations requires us to harness the spatio-temporal flow patterns of bikesharing. 

However, our understanding remains incomplete on the patterns and characteristics of BSS. For example, the impacts from some operational activities of BSS on the patterns and characteristics have not been investigated. Some fundamental questions remain open even on the known patterns and characteristics of BSS, particularly on their implications for the potential decision supports toward sustainable transportation in complex urban environments.
As a function of moving people in the spatio-temporal dimensions using shared bikes, BSS outputs the patterns and characteristics according to the integrated inputs combining many critical factors provided by the stakeholders, including infrastructures, policies, operating activities, management agreements, trip information (such as purpose, route, origin, destination) and etc. Incorporating these inputs from key stakeholders into BSS modeling and analysis is essential to understand the patterns and characteristics for the decision making of improving sustainability in multimodal transportation through BSS.
Some corresponding unsolved questions include (but are not limited to), for example, how to link these inputs with the patterns and characteristics to better understand their impacts on BSS for decision making? Referring to the inputs from different stakeholders, what do the patterns and characteristics imply on key measures of effectiveness (MoE) and supports to BSS? What roles would the patterns and characteristics from data play for decision making of BSS? 
Notice that urban transportation is a complex system, and in contrast, our available data and computational resources are rather limited. It is challenging to promptly provide a fully automated data-driven decision making that is realistic and efficient for BSS, although some successful efforts on data-driven decision supports (DDDS) have been put in the recent non-BSS transportation research \citep{cesme2017data,yi2018data,zhou2017data} to prove the value of DDDS in offering intelligence and performance monitoring for decision making \citep{power2008understanding}. To unleash the potentials of BSS in fostering sustainable multimodal urban transportation, we need take initial steps to bridge the gap between the current comprehension on the patterns and characteristics of BSS and the needs from the BSS modeling and applications for practical and effective data-driven decision supports.

In this paper, we perform a comprehensive data analysis to examine the underlying patterns and characteristics of BSS embedded in a complex urban environment. Aiming to help improving sustainability in multimodal transportation through BSS, we also investigate the implications of the patterns and characteristics for data-driven decision supports (DDDS). We choose the trip history data from \citet{CaBi2017Data} as our main data source of case study. The Capital Bikeshare (CaBi) system is a public-private venture operating more than 3,500 bicycles to casual and member users at over 400 stations in the Washington metropolitan area \citep{bikeshare2016capital}. The data contains 14 million anonymous individual bike trips between 2012-2016. Beyond CaBi, we also extract related information from auxiliary data sources, including the Google Maps application program interfaces (APIs) from \citet{GMap2017API}, LEHD Origin-Destination Employment Statistics (LODES) from \cite{LEHD2017LODES}, and the crash data from Open Data DC in \citet{DC2017Data}. We use data visualization, data fusion, data analysis, and statistical analysis to systematically investigate BSS scheme and examine travel patterns and characteristics on seven important aspects, which are respectively trip demand and flow, operating activities, use and idle times, trip purpose, origin-destination (O-D) flows, mobility, and safety. For each aspect, we explore the results to discuss qualitative and quantitative impacts of the inputs from various stakeholders of BSS on key measures of effectiveness (MoE) such as trip costs, mobility, safety, quality of service, and operational efficiency. We are also interested in revealing new patterns and characteristics of BSS to expand our knowledge on travel behaviors. Finally, we briefly discuss the implications of the patterns and characteristics and some critical roles for data-driven decision supports from the relations between BSS and key stakeholders to show and summarize the values of our findings for transforming urban transportation to be more sustainable, where key stakeholders include road users, system operators, and city planners and policymakers.

\section{Data Description}

The major data source for this analysis is from the CaBi system. CaBi offers bicycle sharing service in the Washington metropolitan area. Like other BSSs, CaBi consists a network of docking stations and a fleet of bikes. The bikes can be rented from and returned to any station of the system, which gives their users flexibility to use BSS to both round and one-way trips and on different purposes. The program offers single trip or multiple options on use time for casual and member users. In the choice of membership, trips under 30 minutes are free of charge, and incremental charges are added for the extra use time afterwards. Obviously, this pricing strategy encourages short trips rather than extended long trips. 

We consider the CaBi trip history data in the recent 5-year fully operational period between 2012-2016. In this period, the data contains 14 million anonymous individual bike trips between docking stations. Any trips lasting less than 60 seconds or taken for system testing or maintenance are excluded by the data provider. Let the set of stations be $S$, the set of bikes be $B$, and the set of trips be $L$. Each station $s\in S$ has an associated geolocation. Each trip $l \in L$ is described by a tuple $l=<t_o, t_d, s_o, s_d, b, u>$, where $t_o$ and $t_d$ are respectively the start and end times, $s_o, s_d \in S$ are respectively the origin and destination stations, $b \in B$ is the bike used for the trip, and $u \in \{Casual, Member\}$ indicates whether the user was a {\em casual user} (i.e., Single Trip, 24-Hour Pass or 3-Day Pass) or a {\em member user} (i.e., Day Key, 30-Day or Annual Member). The system has kept expanding over the years \citep{bikeshare2016capital}. The numbers of stations are from 186 in 2012 to 435 in 2016, the  number of bikes in service are from 1746 in 2012 to 4449 in 2016, and the number of trips grows from 2.05 millions in 2012 to 3.33 millions in 2016. 

We also consider the following data sources: (1) Google Maps APIs are used to extract additional trip information between locations; 
(2) The 2014 LEHD Origin-Destination Employment Statistics (LODES) dataset is used to extract commuting information (where LODES is produced by \cite{LEHD2017LODES} using an extract of the Longitudinal
Employer Household Dynamics (LEHD) data); and (3) The 2016 Crash Data in Open Data DC is used for bike safety information. 

\section{Data Analysis and Results} \label{sec:anaysis_results}

\begin{figure} [htb]
\centering

\begin{subfigure}{.535\textwidth} 
\centering \includegraphics[width=0.95\textwidth]{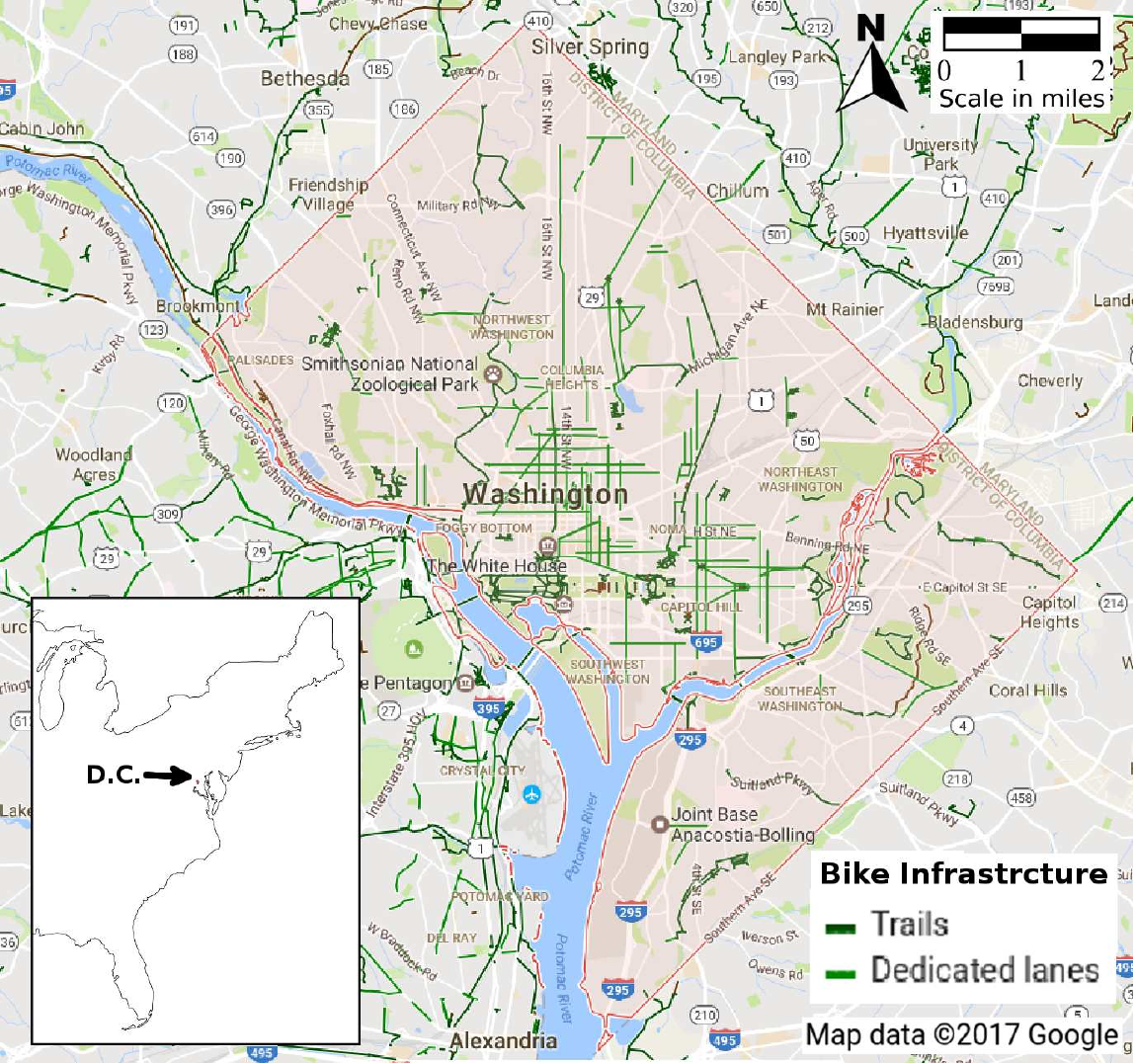} \caption{D.C. Area and Its Cycling Infrastructure.}
\label{fig:dc_bikelanes} 
\end{subfigure}
\begin{subfigure}{.455\textwidth} 
\centering \includegraphics[width=0.95\textwidth]{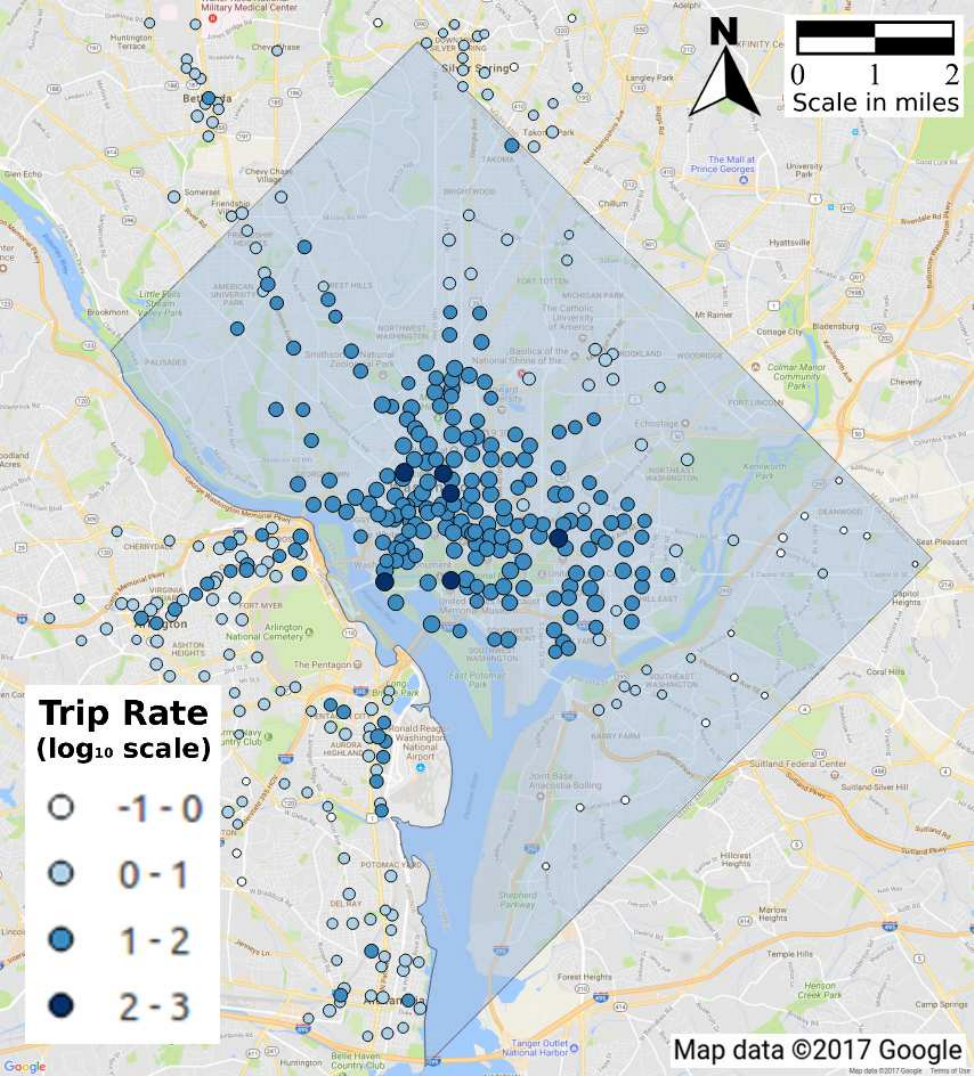} \caption{Trip Rates by Origins.}
\label{fig:BikeState_5Y_Geo_grp_osid} 
\end{subfigure}

\begin{subfigure}{.49\textwidth} 
\centering \includegraphics[width=0.95\textwidth]{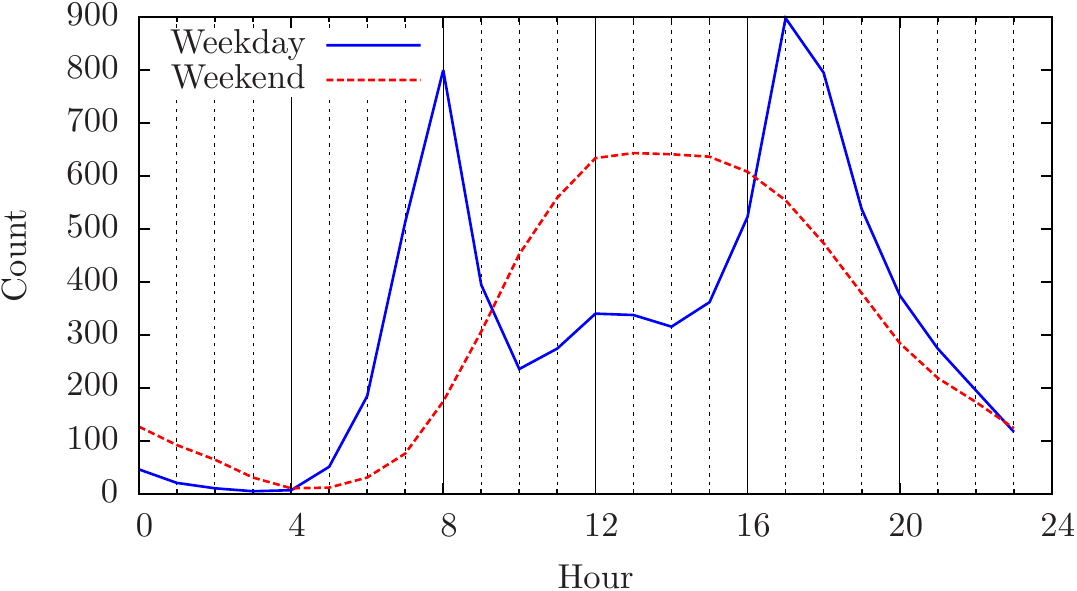} \caption{Trip Rates by Weekday and Weekend.}
\label{fig:BikeState_5Y_hod_Week} 
\end{subfigure}
\begin{subfigure}{.49\textwidth} 
\centering \includegraphics[width=0.95\textwidth]{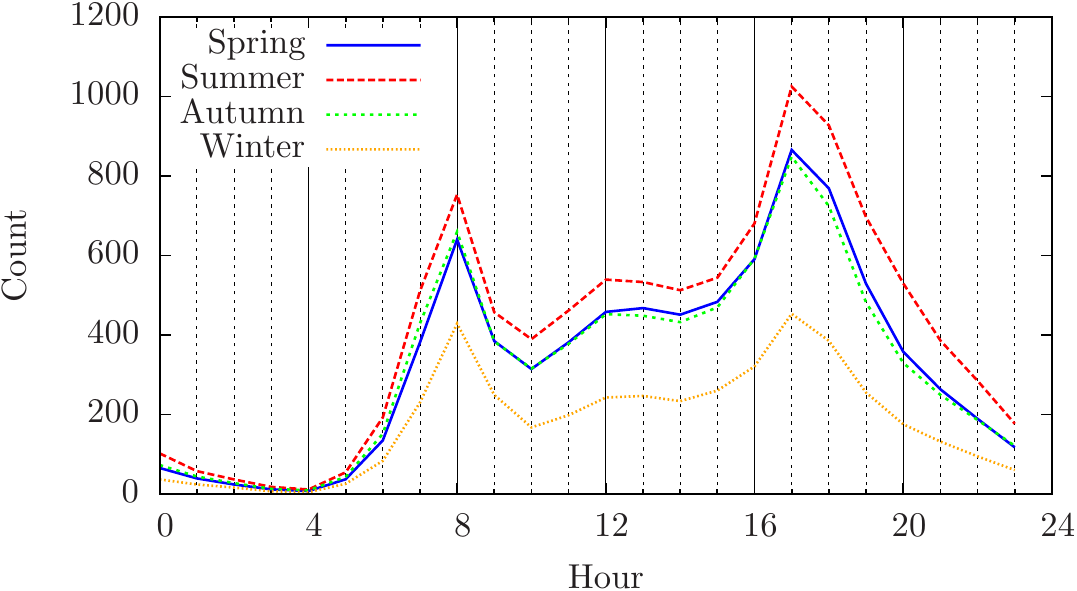} \caption{Trip Rates by Seasons.}
\label{fig:BikeState_5Y_hod_Season} 
\end{subfigure}

\setlength{\belowcaptionskip}{-5.6pt}
\caption{Cycling Infrastructure and Basic Temporal and Spatial Characteristics of Bike Trips in the D.C. Area.}
\label{fig:BikeState_5Y_hod}
\end{figure}

In this section, we conduct a comprehensive data analysis to uncover underlying patterns and characteristics of the system dynamics of the bikeshare network in the Washington DC area. We first show the basic spatial and temporal characteristics of the bikeshare network. Fig. \ref{fig:dc_bikelanes} gives the District of Columbia (DC) area and its cycling infrastructure. The total length of bike lanes and the percentage of residents regularly biking to work have respectively increased from 30.1 to 69 miles and from 1.68\% to 4.54\%, between 2007 and 2013, according to \citet{DDOT2014BikeFact}. Fig.~\ref{fig:BikeState_5Y_Geo_grp_osid} gives the spatial distributions of daily averaged trip counts by origin stations. We color each station $s$ according to the values of $\operatorname{log}_{10}(C_s/\operatorname{max}(T_{LD}, TH_{LD}))$, where $C_s \in \{C_{O(s)}, C_{D(s)}\}$ is the total number of trips using the station as either trip origin or destination, $T_{LD}$ is the number of operation days of the station, and $TH_{LD}$ is defined as the threshold value of $T_{LD}$. By default, $TH_{L}=50$ is used to prevent Fig.~\ref{fig:BikeState_5Y_Geo_grp_osid} from showing statistical noise which has very small values of $T_{LD}$. It is found that the use frequency of station varies greatly among various bike stations. As shown in Fig.~\ref{fig:BikeState_5Y_Geo_grp_osid}, the difference of the use counts among stations could be in several orders of magnitude in terms of the number of bike trips using them as origin or destination, demonstrating that most of the total flows are concentrated at a few stations. As displayed by the combined Figs.~\ref{fig:BikeState_5Y_Geo_grp_osid} and \ref{fig:dc_bikelanes}, the demand of bike stations has a strong spatial correlation with the bike lanes distribution. The higher is the spacial density of bike lanes, the higher demand for bike stations by users. It indicates that cycling infrastructure development is able to encourage and attract more users to use the bikeshare in the region.

Figs. \ref{fig:BikeState_5Y_hod_Week} and~\ref{fig:BikeState_5Y_hod_Season} show the temporal distributions of bikeshare trips averaged by the number of associated days, respectively by weekday/weekend and seasons. Two sharp AM and PM peaks emerge in the commuting pattern of weekdays, while only one single gentle peak appears in that of weekends, see Fig.~\ref{fig:BikeState_5Y_hod_Week}. The bikeshare trip distributions show a primary utilitarian pattern \citep{miranda2013classification}. It is consistent with the 2016 CaBi survey \citep{bikeshare2016capital}, where 65\% of respondents confirmed that commuting was a primary trip purpose for their bikeshare usage. In Fig.~\ref{fig:BikeState_5Y_hod_Season}, twelve months are classified into four seasons respectively as Spring (March, April, May), Summer (June, July, August), Autumn (September, October, November), and Winter (December, January, February) in terms of monthly average temperature in the region, which are respectively 4.67$^{\circ}$C, 14$^{\circ}$C, 25.67$^{\circ}$C, and 16$^{\circ}$C \citep{DC2017Weather}. The number of trips in Summer is about 2.33 times of that in Winter. The significantly lowered counts of biking trips in Winter are most likely resulted from the unfavorable weather and road conditions associated with cold temperatures \citep{gebhart2014impact,el2017effects}.

\subsection{Trip Demand and Flow} \label{sec:TripDemand}

\begin{figure} [t]
\centering

\begin{subfigure}{.49\textwidth} 
\centering \includegraphics[width=0.95\textwidth]{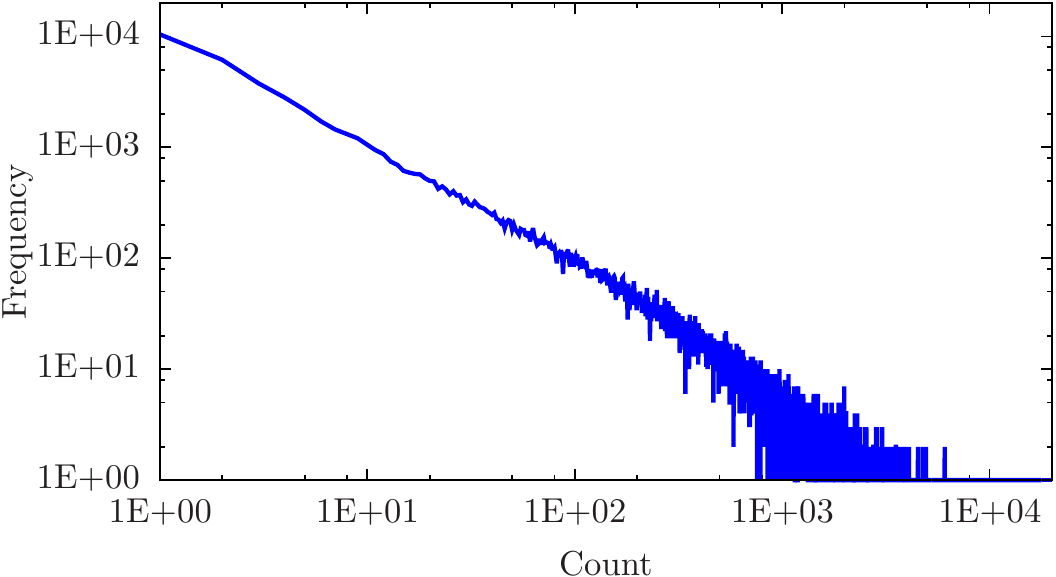} \caption{Flow Count Distribution.}
\label{fig:BikeState_5Y_grp_OD_DiffOD_Histogram} 
\end{subfigure}
\begin{subfigure}{.49\textwidth} 
\centering \includegraphics[width=0.95\textwidth]{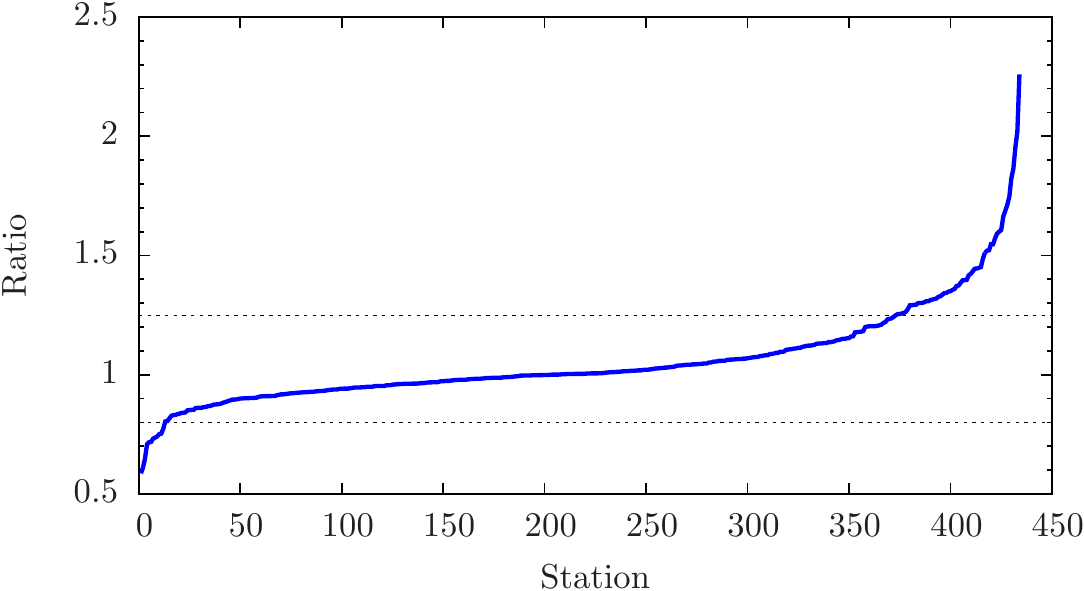} \caption{Demand-Supply (D-S) Ratio.}
\label{fig:BikeState_5Y_Geo_grp_odCount} 
\end{subfigure}

\caption{Flow Count Distribution and Demand-Supply Ratio.}
\label{fig:BikeState_5Y_UserType_hod}
\end{figure}

The bikeshare network is a self-organized network formed by the demand and supply of bike trips between stations. Let $C_{OD(s_o,s_d)}$ be the number of trips from station $s_o$ to station $s_d$. Fig. \ref{fig:BikeState_5Y_grp_OD_DiffOD_Histogram} gives the distribution of trip counts between stations in the bikeshare network. The trip counts follow a scale-free power-law distribution, which is a common pattern shown in other human mobility examples \citep{Xie2015,gonzalez2008understanding}, indicating a strong heterogeneity of human movements in this area. The formation of a scale-free network is an important feature. In such a network, maintenance or improvement of a small set of important O-D paths would bring benefits to bikeshare users on a large amount of trips.

For a station, let $C_{O}$ and $C_{D}$ respectively be the number of the trips taking it as the trip origin and destination, its approximate demand-supply (D-S) ratio $R_{DS}$ is defined as 
\begin{equation} \label{eq:r_ds}
R_{DS}=(C_{O}+C_\epsilon)/(C_{D}+C_\epsilon),
\end{equation}
where $C_\epsilon \ge 0$ is a constant to stabilize the result.  By default, $C_\epsilon=1000$. 

Based on $R_{DS}$, a station is classified as one in the demand-supply balance if 
\begin{equation} \label{eq:r_ds_balance}
R_{DS} \in [1-R_\delta, 1/(1-R_\delta)],
\end{equation}
where $R_\delta \in (0, 1)$ is a constant in the definition of the balance range. By default, $R_\delta=0.2$.

In Eq. \ref{eq:r_ds}, if $C_\epsilon=0$, we have the basic D-S ratio $R_{DS}^{(0)}=C_O/C_D$. If the sample sizes $C_{O}$ and $C_{D}$ are both very small, $R_{DS}^{(0)}$ could be a quite inaccurate estimation of the actual D-S ratio. In this case, however, the actual D-S ratio would be insignificant since it would require too much time to make a station reach the completely empty or full state. From Eq. \ref{eq:r_ds}, $R_{DS}$ is a value between 1 and $R_{DS}^{(0)}$. If $C_D \ll C_\epsilon$ and $C_O \ll C_\epsilon$, $R_{DS}$ would be closed to 1 and likely in the balance range. To have a $R_{DS}$ outside of the balance range, $C_D$ and/or $C_O$ should be sufficiently large, by taking $C_\epsilon$ as a reference value. When $C_D$ and $C_O$ are much larger than $C_\epsilon$, $R_{DS} \to R_{DS}^{(0)}$.

The balance range is determined by $R_\delta$ (see Eq. \ref{eq:r_ds_balance}). As $R_\delta$ approaches 0 or 1, respectively almost none or all values of $R_{DS}$ would be in the balance range. To evaluate if a selected $R_\delta$ value is suitable in practice, we could apply it to $R_{DS(s)}$ for all $s \in S$ and check the portion of stations outside of the balance.
Fig. \ref{fig:BikeState_5Y_Geo_grp_odCount} gives $R_{DS(s)}$ for all $s \in S$ in a sorted form. Here the default value $R_\delta=0.2$ is used. It shows that most stations have the $R_{DS}$ in the demand-supply balance range, i.e. with a demand-supply ratio in the range of $[0.8, 1.25]$; and only a small number of stations have the $R_{DS}$ away from the balance, i.e. $R_{DS}<0.8$ or $R_{DS}>1.25$. 

\begin{figure} [t]
\centering

\begin{subfigure}{.32\textwidth} \centering \includegraphics[width=.95\linewidth]{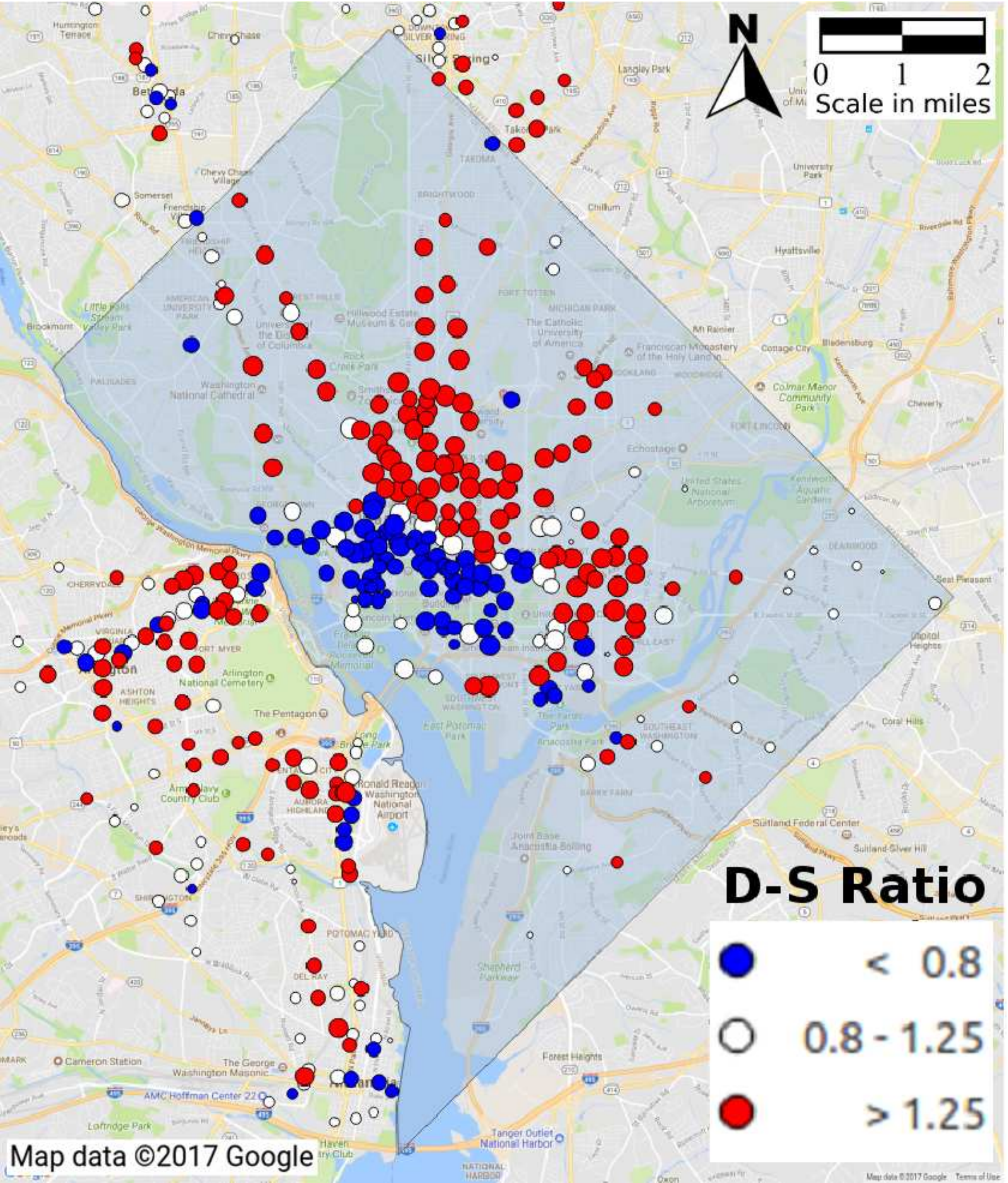} \caption{$R_{DS}$ in Weekday AM Peak} \label{fig:BikeState_5Y_Geo_grp_odCounts_WeekDayAM} \end{subfigure}
\begin{subfigure}{.32\textwidth} \centering \includegraphics[width=.95\linewidth]{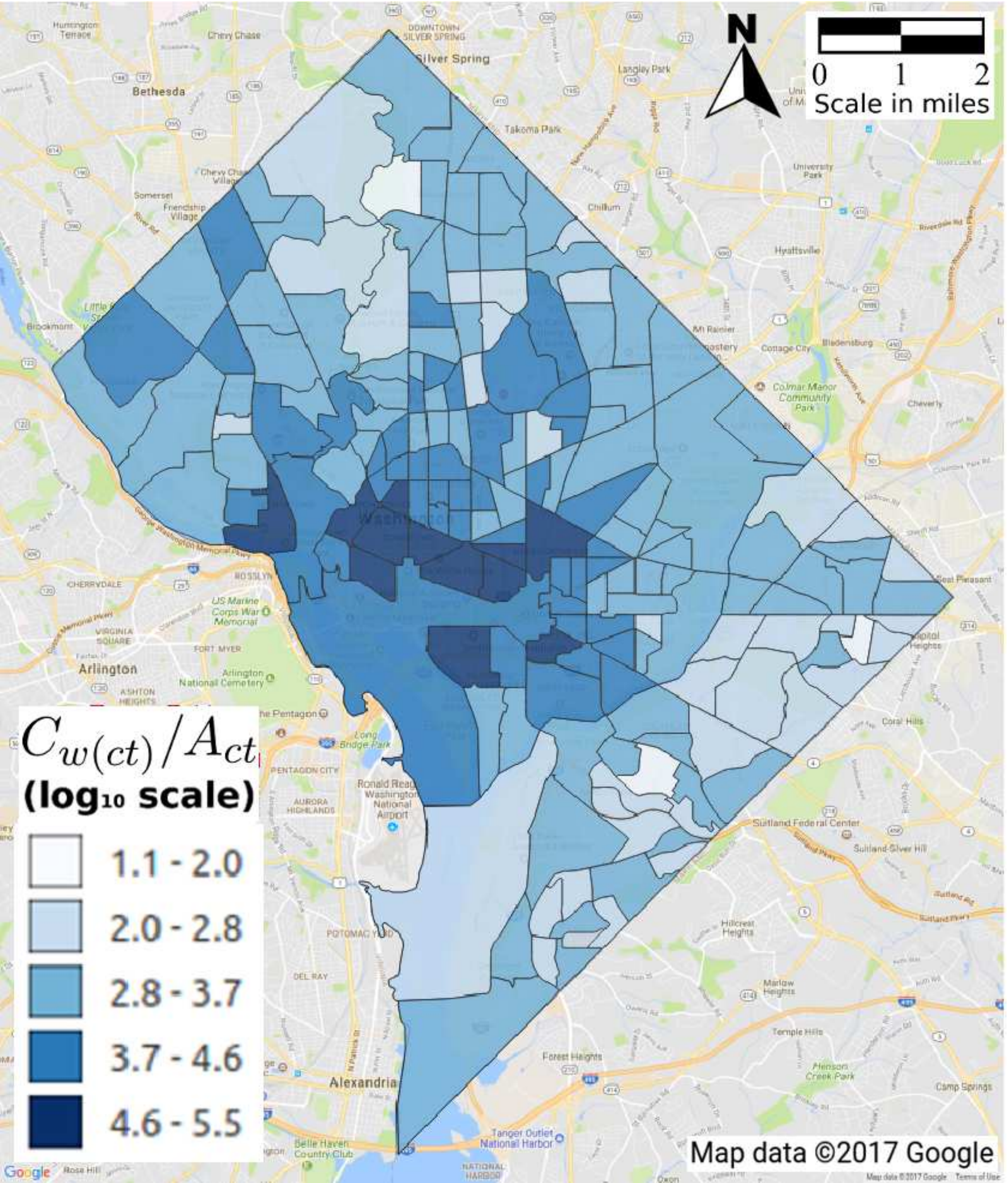} \caption{LODES Data by Workplace} \label{fig:Census_LODES2014_DC_grp_w_geocode_ct_Geo_Tract10} \end{subfigure}
\begin{subfigure}{.32\textwidth} \centering \includegraphics[width=.95\linewidth]{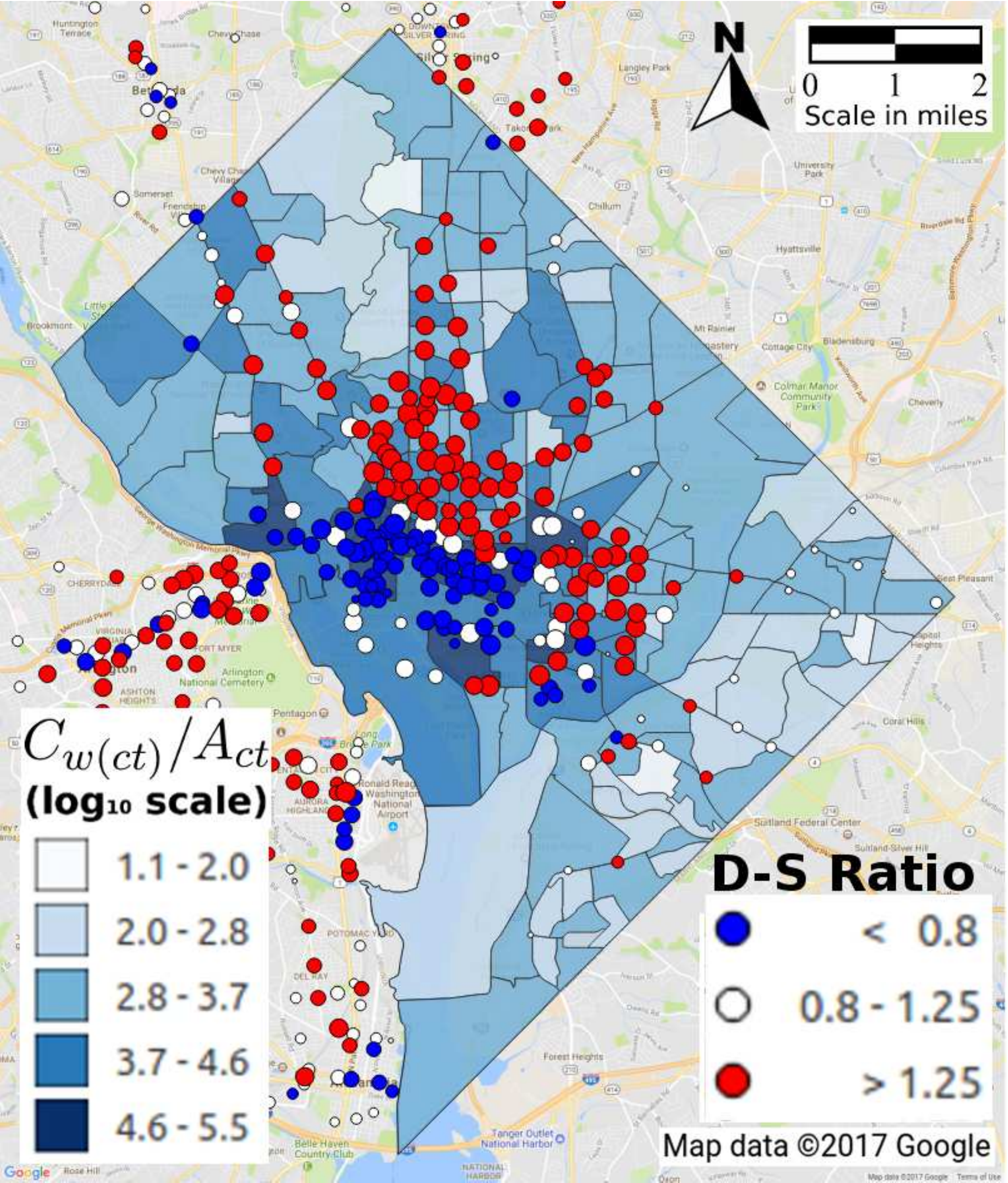} \caption{Overlap of (\ref{fig:BikeState_5Y_Geo_grp_odCounts_WeekDayAM}) and (\ref{fig:Census_LODES2014_DC_grp_w_geocode_ct_Geo_Tract10})} 
\label{fig:BikeState_5Y_Geo_grp_odCounts_WeekDayAM_Lodes14w} \end{subfigure}

\begin{subfigure}{.32\textwidth} \centering \includegraphics[width=.95\linewidth]{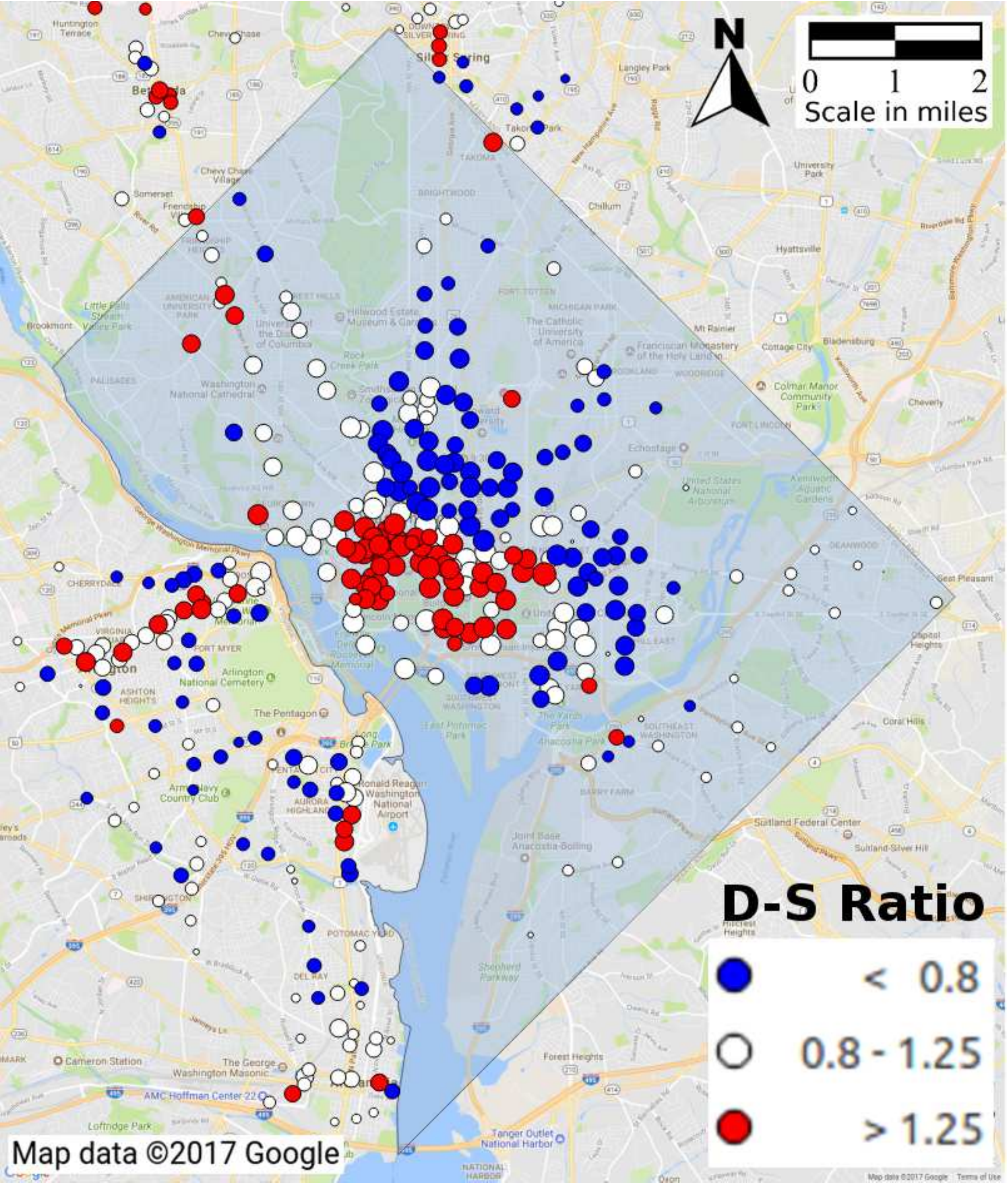} \caption{$R_{DS}$ in Weekday PM Peak} \label{fig:BikeState_5Y_Geo_grp_odCounts_WeekDayPM} \end{subfigure}
\begin{subfigure}{.32\textwidth} \centering \includegraphics[width=.95\linewidth]{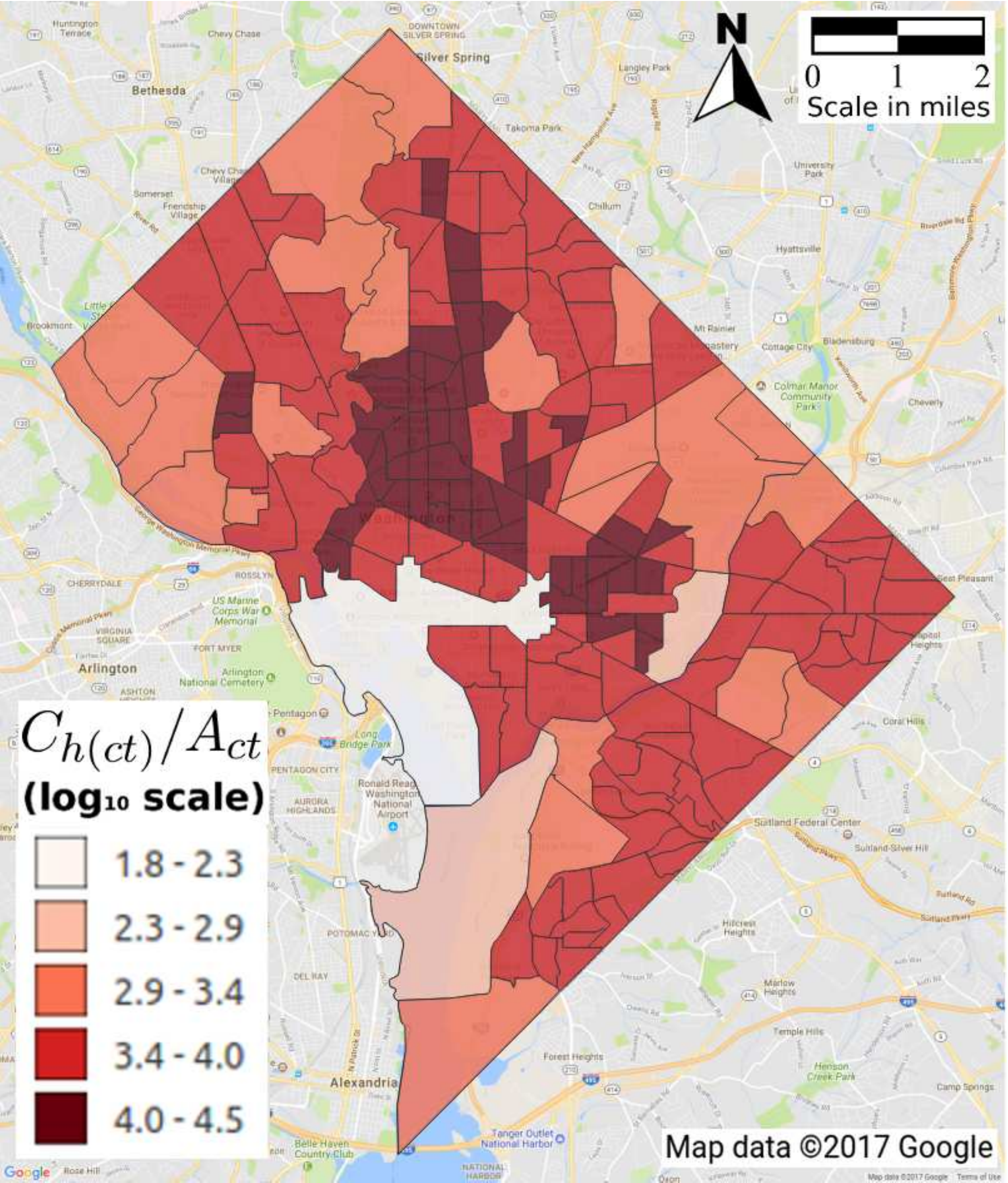} \caption{LODES Data by Residence} \label{fig:Census_LODES2014_DC_grp_h_geocode_ct_Geo_Tract10} \end{subfigure}
\begin{subfigure}{.32\textwidth} \centering \includegraphics[width=.95\linewidth]{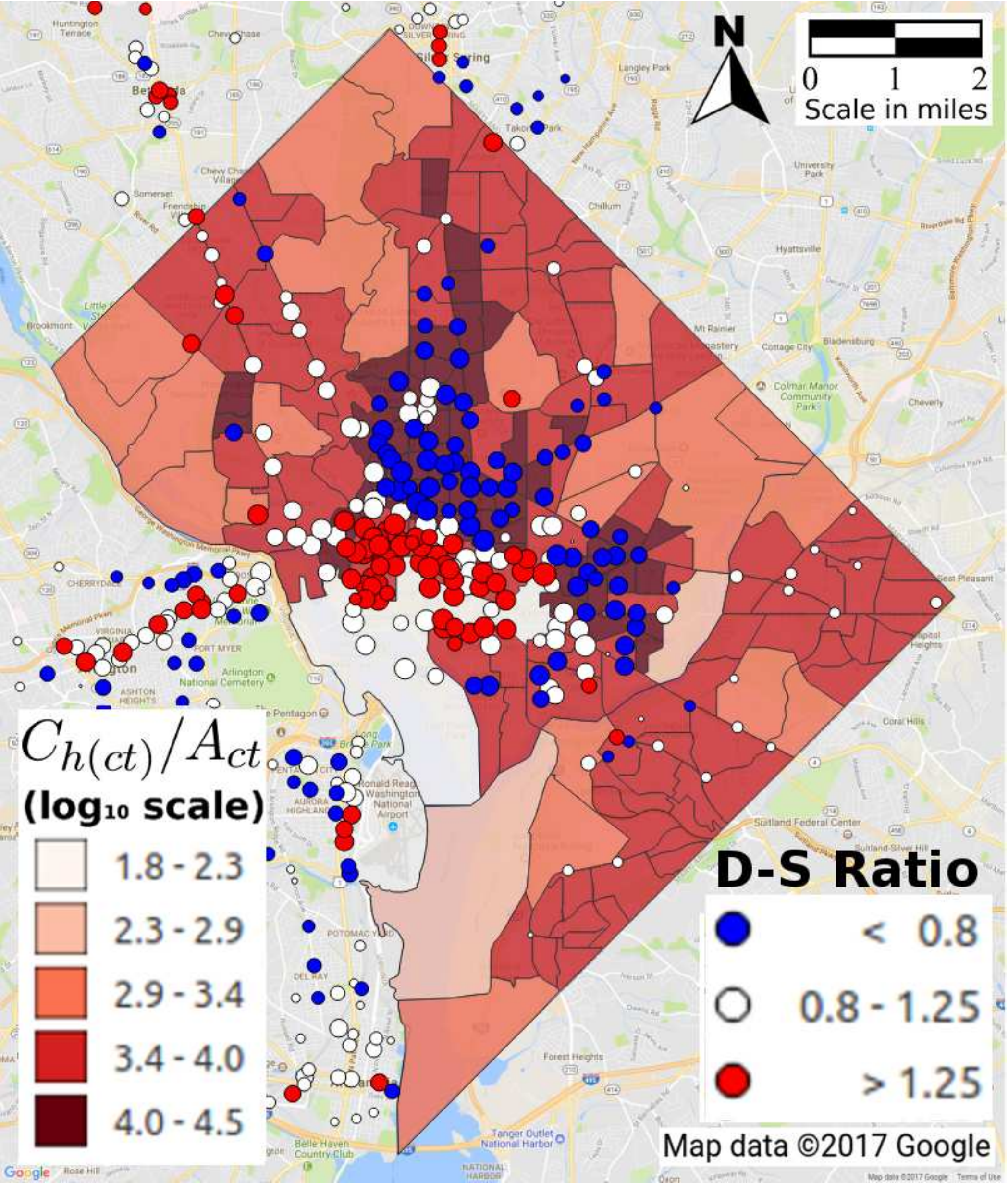} \caption{Overlap of (\ref{fig:BikeState_5Y_Geo_grp_odCounts_WeekDayPM}) and (\ref{fig:Census_LODES2014_DC_grp_h_geocode_ct_Geo_Tract10})} 
\label{fig:BikeState_5Y_Geo_grp_odCounts_WeekDayPM_Lodes14h} \end{subfigure}

\setlength{\belowcaptionskip}{-2.5pt}
\caption{Distributions of Stations Classified in Colors by Demand-Supply Ratios ($R_\delta=0.2$) from CaBi and Distributions of Commuters (in $\operatorname{log}_{10}$) by Workplace and Residence from LODES.}
\label{fig:BikeState_5Y_Geo_grp_osid_ToD_DoW}
\end{figure}

To further analyze the balancing situations of stations in different times of day (ToD), we focus on the stations corresponding to the two main trip-count peaks in Fig. \ref{fig:BikeState_5Y_hod_Week}. We classify the stations into three groups in terms of the range of demand-supply ratio, $R_{DS}$ (with $R_\delta=0.2$) and show their spatial locations in Figs. \ref{fig:BikeState_5Y_Geo_grp_odCounts_WeekDayAM} and \ref{fig:BikeState_5Y_Geo_grp_odCounts_WeekDayPM} in different colors, being $R_{DS}<0.8$ (blue), $R_{DS}\in [0.8, 1.25]$ (white), or $R_{DS}>1.25$ (red), where the spatial distributions of the stations in Figs. \ref{fig:BikeState_5Y_Geo_grp_odCounts_WeekDayAM} and \ref{fig:BikeState_5Y_Geo_grp_odCounts_WeekDayPM} are respectively corresponding to the two main trip-count peaks in Fig. \ref{fig:BikeState_5Y_hod_Week}.
During both of the two ranges of trip-count peak time (7-10 AM and 17-20 PM), the bikeshare stations are found highly spatial clusterable in terms of the color that represents the level of demand-supply ratio $R_{DS}$ (see Figs. \ref{fig:BikeState_5Y_Geo_grp_odCounts_WeekDayAM} and \ref{fig:BikeState_5Y_Geo_grp_odCounts_WeekDayPM}). Interestingly, the spatial patterns of the demand-supply imbalance show a well-observed inverse relationship between the stations used respectively in the AM and PM peak times. This may be due to the commuting of bikeshare users between their home and work places, as they use bikes from home to work around 7-10 AM and reversely from work to home around 17-20 PM. If $R_{DS}$ is too large or too small for a station, its bike docks will likely be all empty or full after some service time, i.e. hit the station's service limit where the station cannot provide renting (for the empty-docks status) or returning service (for the full-docks status) anymore~\citep{fishman2016bikeshare,bikeshare2016capital}. In this case, rebalancing strategies \citep{de2016bike} are needed to balance the demand and supply at some critical bikeshare stations. 

Two other methods, which respectively measure the occupancy rate \citep{o2014mining} and the normalized hourly pickup and return rates \citep{vogel2011understanding}, have also been proposed for identifying the demand-supply imbalance of bikeshare. Although the occupancy rate can directly show the status of docks as is, $R_{DS}$ is the underlying cause of imbalance. Beyond functioning as an alternative measure of the imbalance, $R_{DS}$ has an extra usage in simulation-based studies at an advantage for predictive analytics, which is important as potential impacts can be tested and evaluated accordingly to acquire better rebalancing strategies. It is worthy to note that using the normalized hourly pickup and return rates separately ~\citep{vogel2011understanding} actually removes the information corresponding to $R_{DS}$, thus is complementary with the method using $R_{DS}$ for measuring the imbalance of bikeshare.

To gain more insights of the underlying driving force for the demand-supply patterns, we use LODES data in our analysis. LODES covers workplace and residence information on wage and salary jobs in private sectors and state and local governments. In Figs. \ref{fig:Census_LODES2014_DC_grp_w_geocode_ct_Geo_Tract10} and \ref{fig:Census_LODES2014_DC_grp_h_geocode_ct_Geo_Tract10}, each census tract $ct$ in DC is colored according to the values of $\operatorname{log}_{10}(C_{w(ct)}/A_{ct})$ and  $\operatorname{log}_{10}(C_{h(ct)}/A_{ct})$ respectively, where $A_{ct}$ is the land area of $ct$ in square miles,  $C_{w(ct)}$ and $C_{h(ct)}$ are the worker counts in $ct$ grouped respectively by their workplace and residence. In Geographic Information System (GIS), workplaces are mostly clustered and located at the mixed-use neighborhoods in the north and south of National Mall, while residing homes are mainly spatially clustered and located at the neighborhoods in the north and east of the main workplace region. Most of the workers are commuters between the two workspace and residence regions during weekdays. 

To reveal the relationship between commuting behavior and demand-supply ratio, we give 
Figs. \ref{fig:BikeState_5Y_Geo_grp_odCounts_WeekDayAM_Lodes14w} and \ref{fig:BikeState_5Y_Geo_grp_odCounts_WeekDayPM_Lodes14h} respectively showing the overlapped views of Figs. \ref{fig:BikeState_5Y_Geo_grp_odCounts_WeekDayAM} and \ref{fig:Census_LODES2014_DC_grp_w_geocode_ct_Geo_Tract10}, and of Figs. \ref{fig:BikeState_5Y_Geo_grp_odCounts_WeekDayPM} and \ref{fig:Census_LODES2014_DC_grp_h_geocode_ct_Geo_Tract10}. For bike stations with $R_{DS}<0.8$ (in blue color), they are mostly located in high-density workplace and residence regions during the AM and PM peak periods, respectively. Figs. \ref{fig:BikeState_5Y_Geo_grp_odCounts_WeekDayAM_Lodes14w} and \ref{fig:BikeState_5Y_Geo_grp_odCounts_WeekDayPM_Lodes14h} indicate that there is a strong spatial correlation between commuting behavior and demand-supply ratio. It confirms that the commuting between workplace and residence is one of main factors affecting the demand-supply balance of the bikeshare stations.

\subsection{Operating Activities} \label{sec:OperationalActivity}

Operations are important for maintaining daily functions of BSS. Here, we analyze and try to measure the efficiency of operating activities of bikeshare through data mining. We take the valet and corral service \citep{de2016bike,CaBi2017Corrals} as an example of case study on operating activities. If a bikeshare station has valet or corral service, operating staffs will take care of bike returning at the station or keep the station from fully occupied by removing bikes from docks and storing them to a corralled place. Corral service can lift serving capacity of a bikeshare station from a limited to a rather high level during its operational period. In addition, the service can guarantee users to return their bikes at their expected stations, without wasting time (or paying extra fee if the bikeshare charging system is usage-time-based) to frustratingly find an empty dock in neighbor stations. It is especially useful for high-demand stations of bikeshare. In practice, corral service can be provided by bikeshare operators regularly in high-demand seasons or specially on some high-attendance events. We analyze the both conditions in our case study.  

According to social media and CaBi public reports, CaBi launched seasonally regular corral service for weekdays since the year of 2015. At the beginning, corral service was provided only at two stations --- the Stations 31205 and 31227, which both started from May 14, 2015, but respectively ended on November 16 and December 18, 2015. In 2016, the two stations continued to provide the service between April 4 and December 23. In addition, four more stations started to provide corral service in 2016, where Station 31233 provided the service between June 6 and December 23, Stations 31259, 31243 and 31620 all started the service from Jun 8, but respectively ended on November 9, November 9, and October 14. The time-of-day servicing periods were [7AM, 11AM] and [8AM, 12PM] respectively for 2015 and 2016, meaning that the operating cost is at least $20$ hours of work per week by staff operators for each station.

\afterpage{
\begin{figure} [h]
\centering

\begin{subfigure}{.49\textwidth} 
\centering \includegraphics[width=0.95\textwidth]{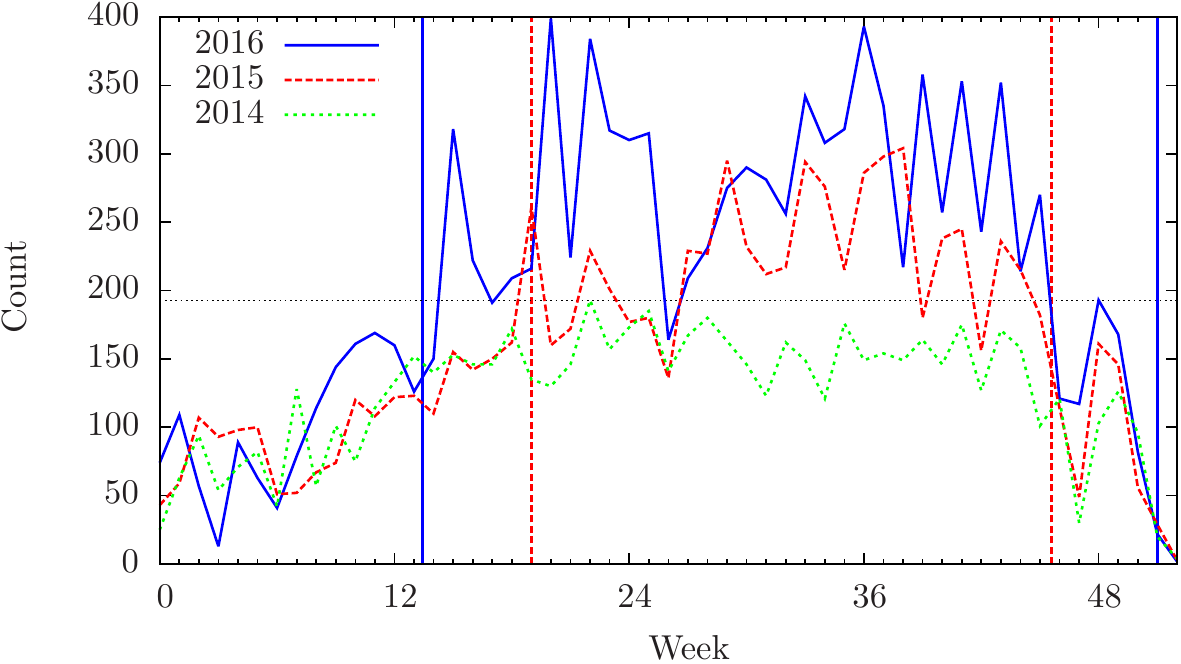} \caption{Station 31205.}
\label{fig:BikeState_5Y_grp_Week_AMCorrals_31205} 
\end{subfigure}
\begin{subfigure}{.49\textwidth} 
\centering \includegraphics[width=0.95\textwidth]{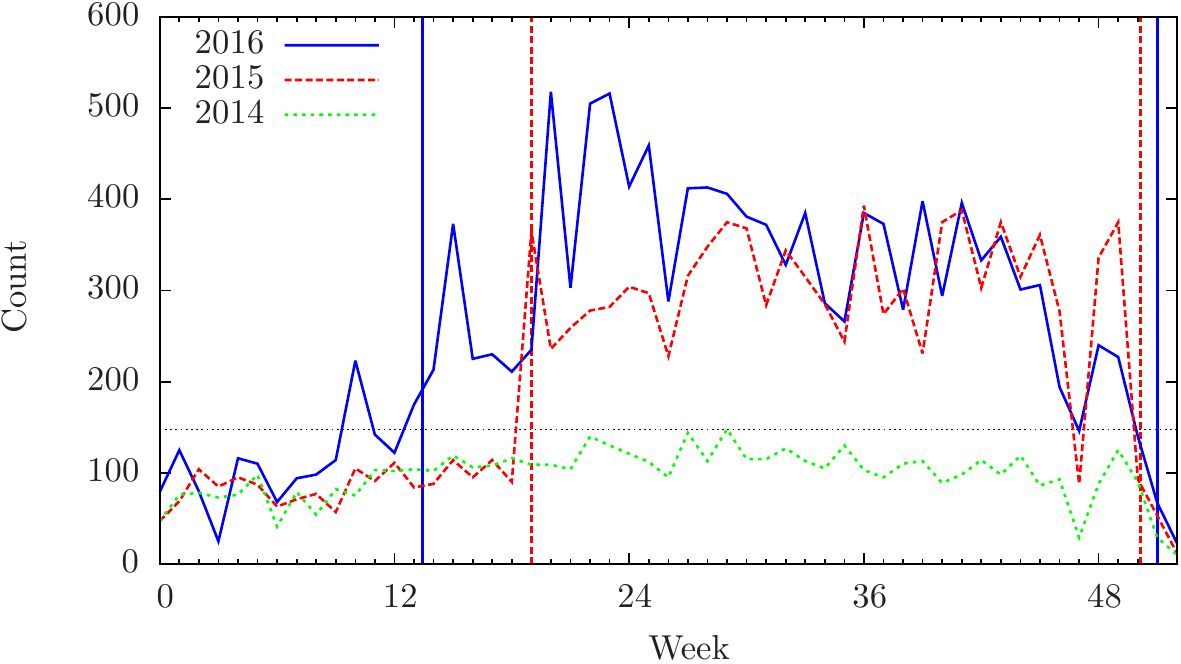} \caption{Station 31227.}
\label{fig:BikeState_5Y_grp_Week_AMCorrals_31227} 
\end{subfigure}

\begin{subfigure}{.49\textwidth} 
\centering \includegraphics[width=0.95\textwidth]{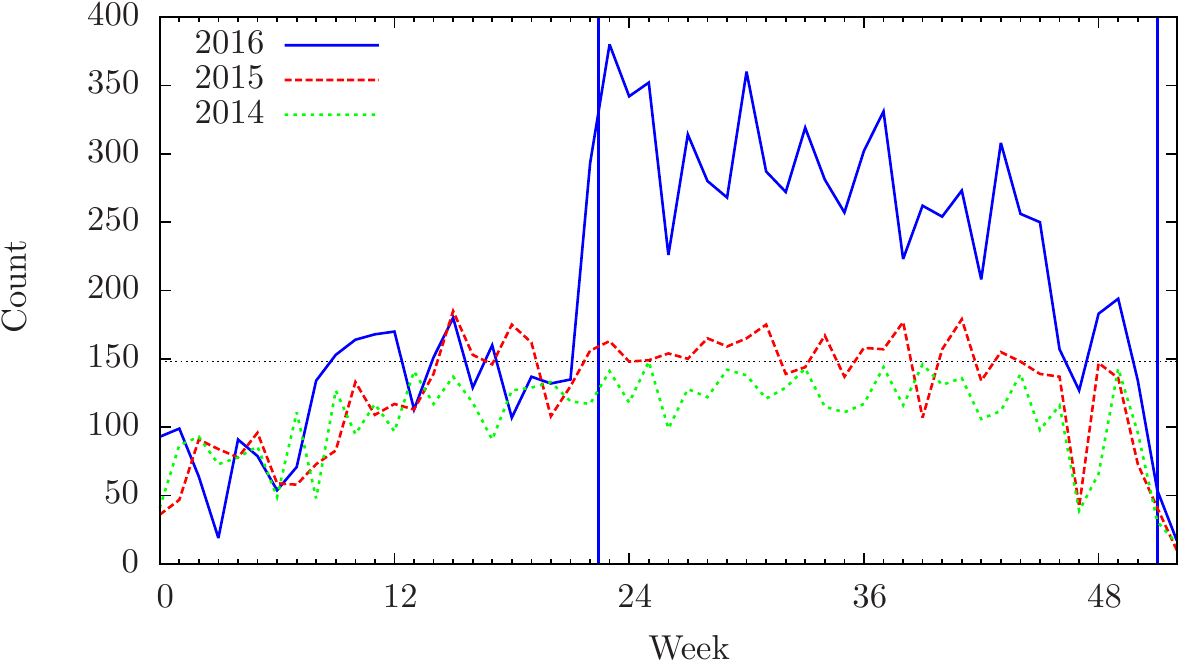} \caption{Station 31233.}
\label{fig:BikeState_5Y_grp_Week_AMCorrals_31233} 
\end{subfigure}
\begin{subfigure}{.49\textwidth} 
\centering \includegraphics[width=0.95\textwidth]{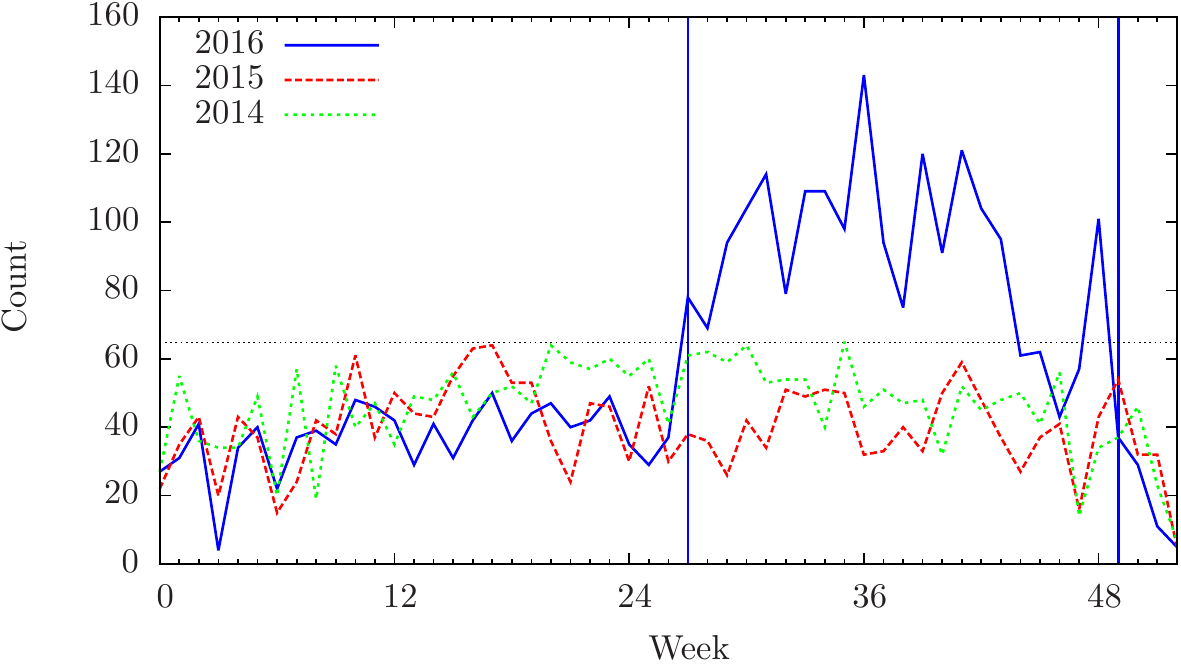} \caption{Station 31259.}
\label{fig:BikeState_5Y_grp_Week_AMCorrals_31259} 
\end{subfigure}

\begin{subfigure}{.49\textwidth} 
\centering \includegraphics[width=0.95\textwidth]{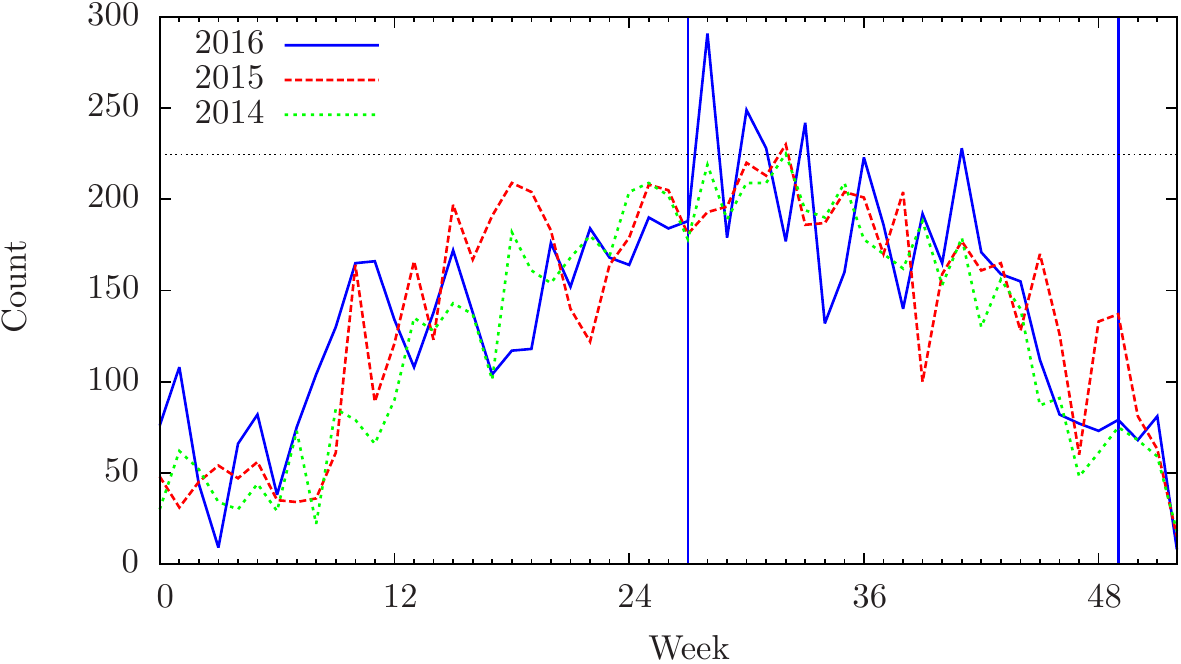} \caption{Station 31243.}
\label{fig:BikeState_5Y_grp_Week_AMCorrals_31243} 
\end{subfigure}
\begin{subfigure}{.49\textwidth} 
\centering \includegraphics[width=0.95\textwidth]{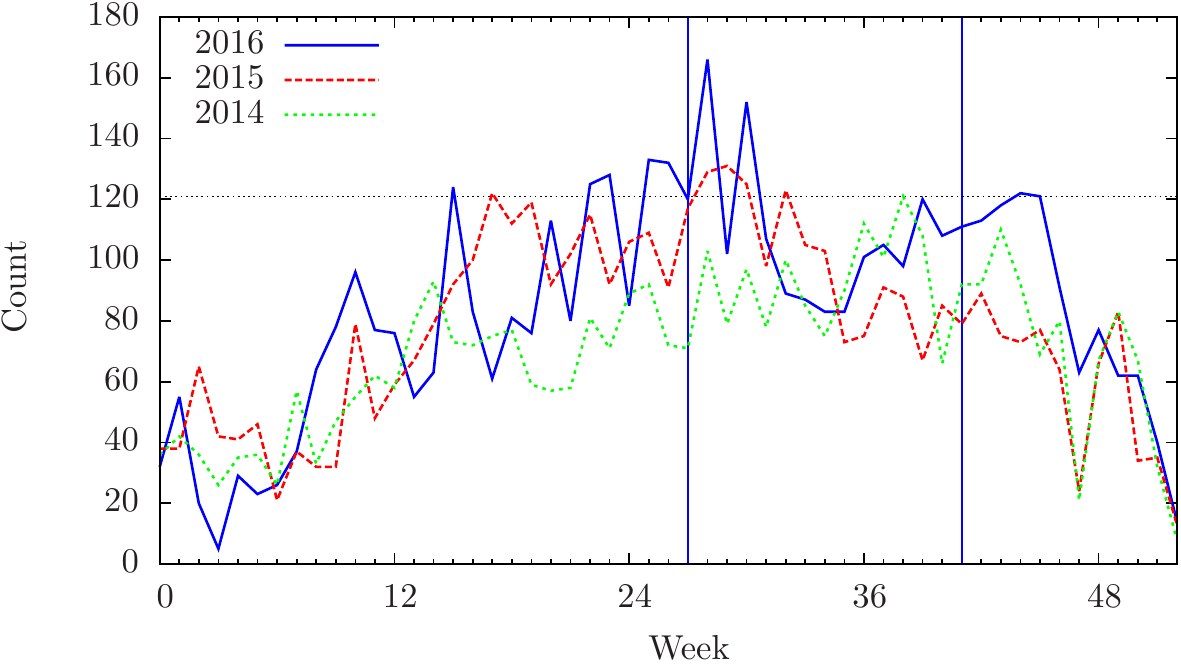} \caption{Station 31620.}
\label{fig:BikeState_5Y_grp_Week_AMCorrals_31620} 
\end{subfigure}
\caption{Results of Regular Bike Corral Service from Data: Comparisons of Weekly Drop-off Counts during [8AM, 12PM] of Weekdays among 2014, 2015, and 2016 at Six Stations Respectively. For Each Station, its Start and End Dates of Corral Service are Marked with a Pair of Vertical Lines for Each Service-Available Year (Dash for 2015 and Solid for 2016), and its Peak of Weekly Drop-off Counts of 2014 is Marked with a Horizon Line.}
\label{fig:BikeState_5Y_AMCorral_Regular}
\end{figure}
}

Fig. \ref{fig:BikeState_5Y_AMCorral_Regular} shows the comparisons of weekly drop-off counts during [8AM, 12PM] of Weekdays among 2014 (green), 2015 (red), and 2016 (blue) respectively at each of the six stations where seasonally regular bike corral service was provided by CaBi in 2015 and 2016. In the figure, the start and end dates of corral service are all marked using vertical lines. Notice that neither station in 2014 nor any of the last four CaBi stations in 2015 was provided corral service by CaBi. These results of no corral service provide us baselines to compare bikeshare capacity for studying effects of corral service. For each station without corral service, let $\hat{C}_D^{w}$ be its maximum weekly drop-off count and $\bar{C}_D^{w}$ its maximum weekly capacity, then $\hat{C}_D^{w}$ is an approximation to the lower bound of $\bar{C}_D^{w}$. As shown in Fig. \ref{fig:BikeState_5Y_AMCorral_Regular}, during the periods without corral service, all the weekly drop-off counts are below $\hat{C}_D^{w} + \delta$, where $\hat{C}_D^{w}$ is the peak weekly drop-off count in 2014 (see the horizon lines in Fig. \ref{fig:BikeState_5Y_AMCorral_Regular}) and $\delta$ is a small value representing the noises in data. It turned out that the peak demand in 2014 was sufficiently high to yield the results of $\hat{C}_D^{w}$ approaching $\bar{C}_D^{w}$. More importantly, Fig. \ref{fig:BikeState_5Y_AMCorral_Regular} reveals that at each of the six stations, the value of $\hat{C}_D^{w}$ with corral service (in 2015 or 2016) is higher than that without the service, reflecting that corral service has the capability of raising capacity of bikeshare stations. Concerning the six CaBi stations in Fig. \ref{fig:BikeState_5Y_AMCorral_Regular}, it is worth noting that corral service led to significant increases in weekly drop-off counts at four of the stations, but induced only a little change at the other two stations. In Figs. \ref{fig:BikeState_5Y_grp_Week_AMCorrals_31205} through \ref{fig:BikeState_5Y_grp_Week_AMCorrals_31259}, the increase in weekly counts is notable for most of the weeks with corral service. While in Figs. \ref{fig:BikeState_5Y_grp_Week_AMCorrals_31243} and \ref{fig:BikeState_5Y_grp_Week_AMCorrals_31620}, weekly drop-off counts are found raised only for a few of the weeks with corral service, and the extent of increase in weekly drop-off counts is rather small in comparison with those shown in Figs. \ref{fig:BikeState_5Y_grp_Week_AMCorrals_31205} through \ref{fig:BikeState_5Y_grp_Week_AMCorrals_31259}. The results provide us a straightforward MoE on corral service at different stations. To augment the overall QoS and operational efficiency, operators of BSS would need optimize their operating activities. As an example, Fig. \ref{fig:BikeState_5Y_AMCorral_Regular} shows that data analysis is able to provide bikeshare operators evidence-based supports to help them reallocate their resources such as redistribute corral service among different stations for a better operational efficiency. 

Although the system data does not provide the details of redistribution efforts at these stations, we can gain the insights by comparing the redistribution efforts between different years. Here, as an example, we compare the redistribution efforts during [8AM, 12PM] from the start to the end dates of corral service in 2016 with those during the same time-of-day and date-of-year period in 2014. The comparison is performed with the following method. At each station $s$, we collect all the trips that have arrived the station during the considered time period in each year as $L_C$. For each $l_c \in L_C$, the next trip $l_n$ using the same bike is extracted. If the origin station of $l_n$ is the same as $s$, the bike is considered as in normal use, otherwise as in maintenance. In this context, the maintenance activity would most likely be the redistribution effort moving the bike from $s$ to the origin station of $l_n$. Table \ref{tab:carral_redistribution} summarizes the results in the two years of 2014 and 2016, where $|L_C|$, $|L_{C,N}|$, and $|L_{C,M}|$ are respectively the numbers of total trips arriving the station, the trips in normal use departing at the station, and the maintenance trips at the station, and $R_{C,M}=|L_{C,M}|/|L_C|$. 
It is clear that coral service has enabled the bike stations to serve users with more trips. As shown in Table \ref{tab:carral_redistribution}, although the number of departing trips, $|L_{C,N}|$, remained similar in the two years, the number of arriving trips, $|L_C|$, increased significantly from 2014 to 2016 due to corral service (which was used in 2016 but not in 2014). Table \ref{tab:carral_redistribution} also shows that the number of bikes in maintenance (mainly redistribution) trips, $|L_{C,M}|$, remarkably increased in 2016, compared to 2014. It is consistent with the fact that the number of arriving trips, $|L_C|$, is mostly much larger than that of departing trips, $|L_{C,N}|$, at the stations. Essentially, corral service at a station could function as a temporary bike depot during the service period, but the corralled bikes should be cleared/redistributed from the station before the end of service. 

Although the redistribution operation might be costly, it could increase the availability and usage of bikes. The corralled bikes could be relocated to the stations in supply shortage, therefore total bike ridership could be increased, leading to an operational gain that would offset redistribution cost. Some of the increased ridership in bikeshare could contribute to reducing VMT, which would offset the impact to environment from using rebalancing trucks in the redistribution operation. In addition, with corral service at the stations, high capacity trucks could be utilized to redistribute a large number of bikes at once, which would reduce the average redistribution cost and the impact to environment per bike. Another cost-effective operational option could be extending the service to the PM peak period. In this case, corralled bikes would be picked up by users (as indicated by the changes of the time-of-day $R_{DS}$ in Fig. \ref{fig:BikeState_5Y_Geo_grp_osid_ToD_DoW}), rather than call for any redistributions by system operators. Notice that, in one hand, the extension of corral service could support normal bike trips for more revenue and reduce redistribution efforts for less costs; in another hand, the extension of service time would need several extra work hours of staffs and also reduce bicycle availability as some bikes would stay in corrals. An optimization should be considered by system operators in order to reach the best solution in corral service operations.

\begin{table*}
\centering  \caption{Bike Use and Redistribution Efforts during Regular Corral Service in 2016 and Same Time Periods in 2014.}
  \label{tab:carral_redistribution}
\begin{tabular}{|l|c|c|c|c|c|c|c|c|} \hline 
\multirow{2}{*}{Station} & \multicolumn{4}{|c|}{2014} & \multicolumn{4}{|c|}{2016} \tabularnewline \cline{2-9} 
 & $|L_C|$ & $|L_{C,N}|$ & $|L_{C,M}|$ & $|R_{C,M}|$ & $|L_C|$ & $|L_{C,N}|$ & $|L_{C,M}|$ & $|R_{C,M}|$   \\ \hline 
31205 & 5684 & 4581 & 1103 & 19.4\% & 9884 & 4196 & 5688 & 57.5\%   \\ \hline 
31227 & 4488 & 3085 & 1403 & 31.3\% & 12695 & 3372 & 9323 & 73.4\%   \\ \hline 
31233 & 3776 & 2838 & 938 & 24.8\% & 8151 & 3181 & 4970 & 61.0\%   \\ \hline 
31259 & 1273 & 826 & 447 & 35.1\% & 2212 & 543 & 1669 & 75.5\%   \\ \hline 
31243 & 2480 & 2120 & 360 & 14.5\% & 2554 & 1397 & 1157 & 45.3\%   \\ \hline 
31620 & 2010 & 1717 & 293 & 14.6\% & 2566 & 1835 & 731 & 28.5\%   \\ \hline 
Average & 3285.2 & 2527.8 & 757.3 & 23.1\% & 6343.7 & 2420.7 & 3923.0 & 61.8\%   \\ \hline
\end{tabular}  
\end{table*}

\afterpage{
\begin{figure} [H]
\centering

\begin{subfigure}{.49\textwidth} 
\centering \includegraphics[width=0.95\textwidth]{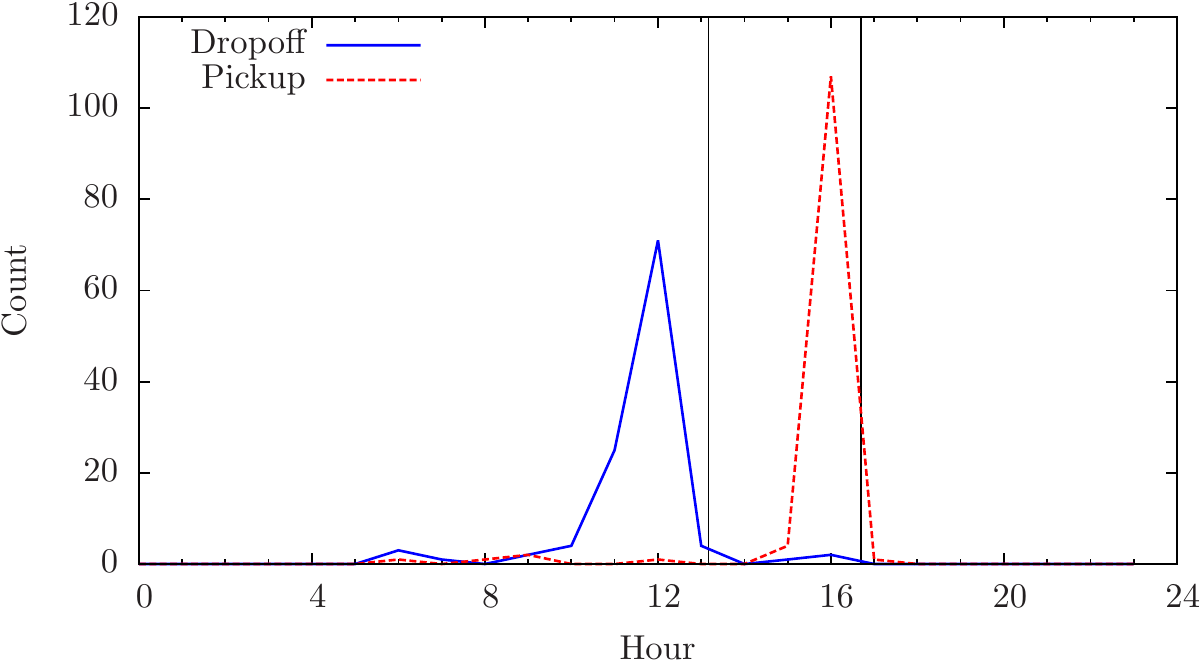} \caption{2012-10-10.}
\label{fig:BikeState_5Y_grp_Hour_31209_2012-10-10} 
\end{subfigure}
\begin{subfigure}{.49\textwidth} 
\centering \includegraphics[width=0.95\textwidth]{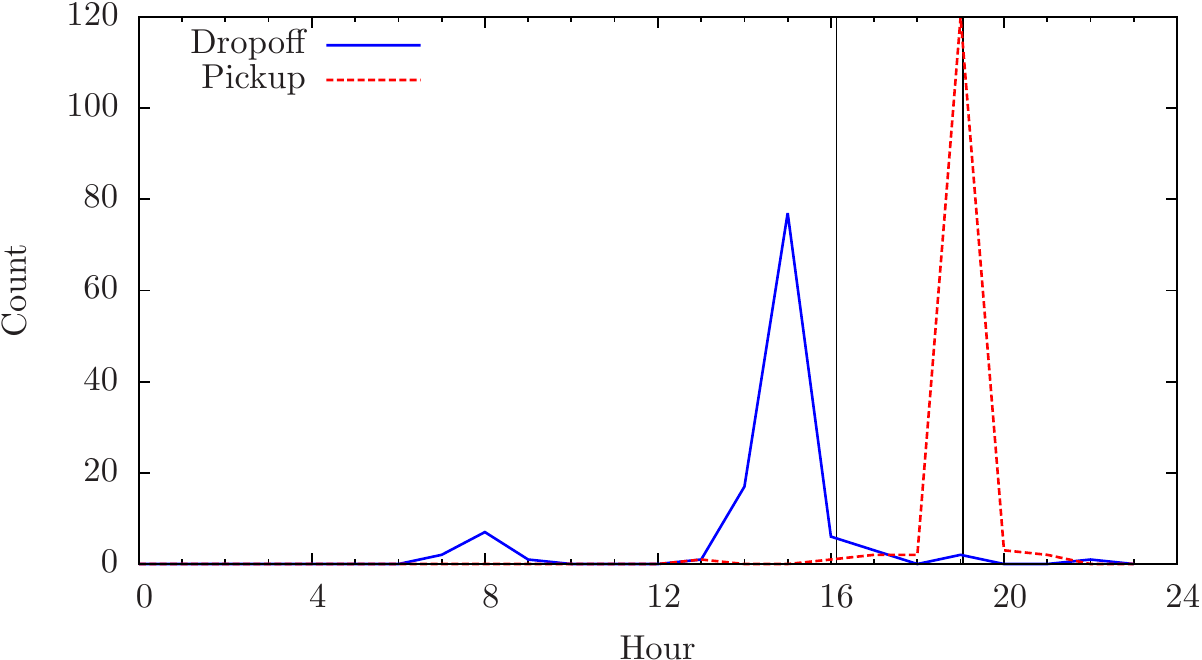} \caption{2012-10-11.}
\label{fig:BikeState_5Y_grp_Hour_31209_2012-10-11} 
\end{subfigure}

\begin{subfigure}{.49\textwidth} 
\centering \includegraphics[width=0.95\textwidth]{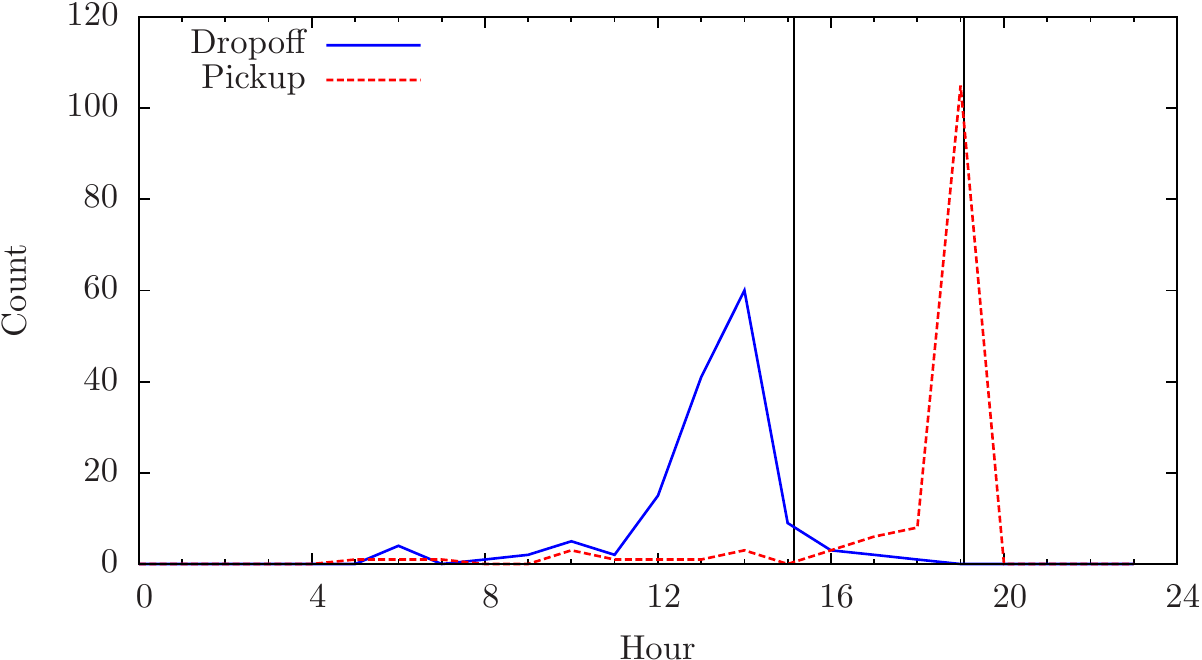} \caption{2014-10-03.}
\label{fig:BikeState_5Y_grp_Hour_31209_2014-10-03} 
\end{subfigure}
\begin{subfigure}{.49\textwidth} 
\centering \includegraphics[width=0.95\textwidth]{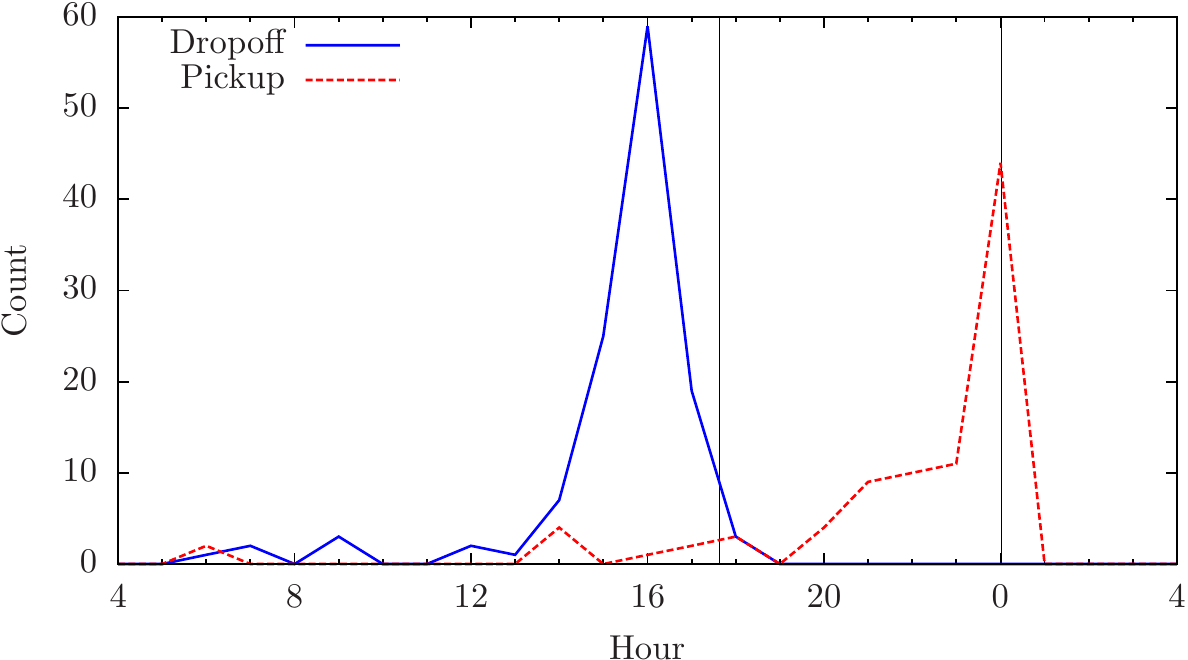} \caption{2014-10-04.}
\label{fig:BikeState_5Y_grp_Hour_31209_2014-10-04} 
\end{subfigure}

\begin{subfigure}{.49\textwidth} 
\centering \includegraphics[width=0.95\textwidth]{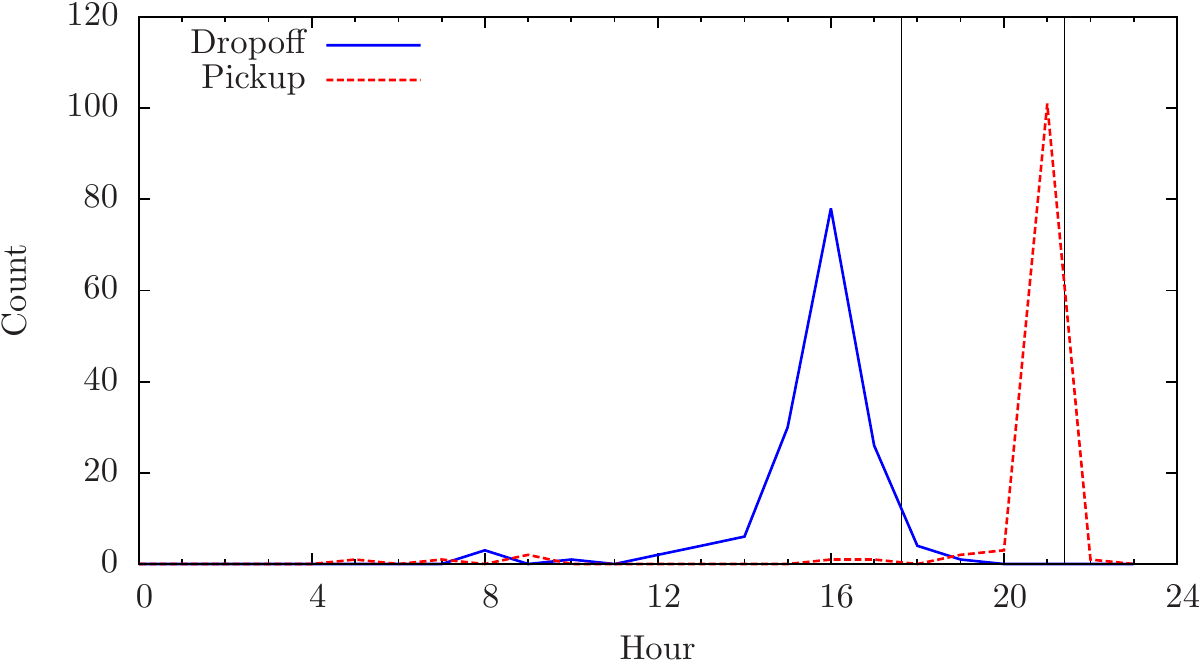} \caption{2016-10-07.}
\label{fig:BikeState_5Y_grp_Hour_31209_2016-10-07} 
\end{subfigure}
\begin{subfigure}{.49\textwidth} 
\centering \includegraphics[width=0.95\textwidth]{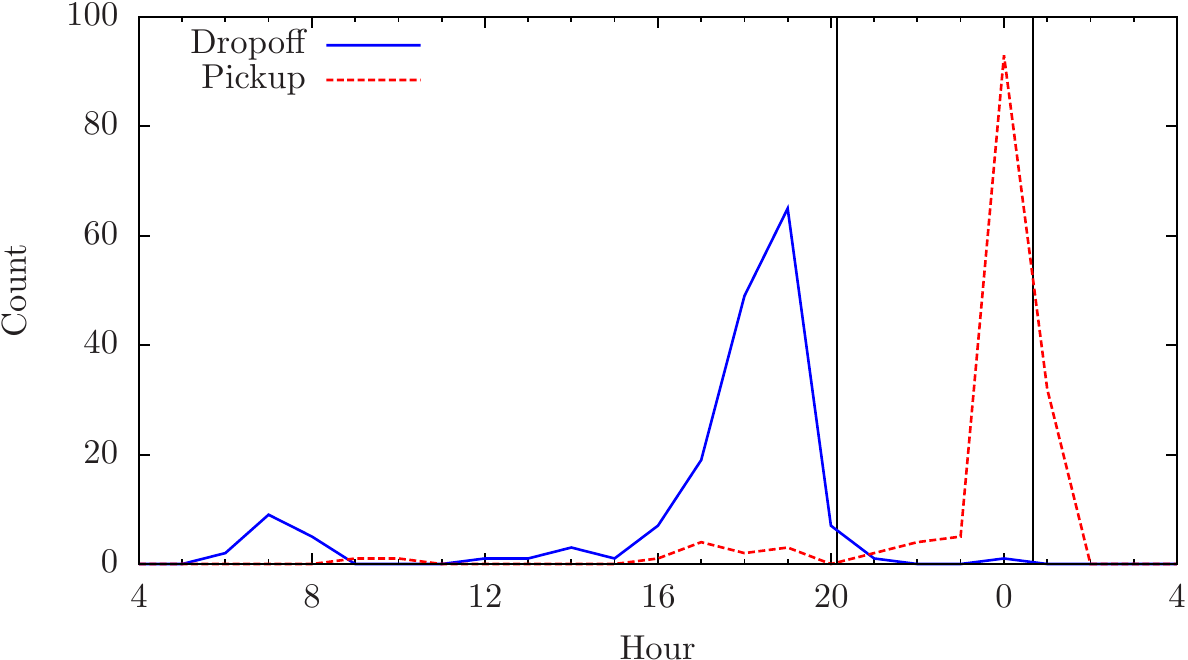} \caption{2016-10-13.}
\label{fig:BikeState_5Y_grp_Hour_31209_2016-10-13} 
\end{subfigure}

\caption{Results of Bike Corral Service for Events from Data: Drop-offs and Pickups at the Nearest Station 31209 during the Six Events of Washington Nationals Playing at Nationals Park in 2012, 2014 and 2016 National League Division Series (NLDS). For Each Event, Two Vertical Lines are Used to Mark the Official Start and End Times of Baseball Game.}
\label{fig:BikeState_5Y_AMCorral_Events}
\end{figure}
}

CaBi provided bike corral service for some high-attendance events \citep{CaBi2017Corrals}. In DC area, the baseball games of Washington Nationals held at Nationals Park have often attracted high attendances. Let us take this event as an example to analyze corral service. CaBi Station 31209 is chosen for this analysis, as it is the nearest station to Nationals Park. The data shows that the top six days in the ranking of the highest daily drop-off counts at Station 31209 are respectively 2012-10-10, 2012-10-11, 2014-10-03, 2014-10-04, 2016-10-07 and 2016-10-13, coincident with the dates when Washington Nationals played their games in the National League Division Series (NLDS) (\url{https://en.wikipedia.org/wiki/National_League_Division_Series}) at Nationals Park in 2012, 2014 and 2016. Fig. \ref{fig:BikeState_5Y_AMCorral_Events} gives hourly drop-off and pickup counts at Station 31209 around the time when the six events were held at Nationals Park, where the start and end times are respectively marked with two vertical lines. As shown in the figure, both drop-off and pickup hourly counts were very low at Station 31209 during normal time, but they both suddenly increased to reach rather high values around the start and end times of the games. 
Concerning the corral service for each event, the operational time with respect to work of staffs is $T_E^O=T_E^B+T_E^D+T_E^A$, where $T_E^B$ is the period to handle drop-off demand before the event, $T_E^D$ is the duration of the event, and $T_E^A$ is the period to handle pickup demand after the event. For the six events, $T_E^D$ is in the range of between 2.92 hours (Fig. \ref{fig:BikeState_5Y_grp_Hour_31209_2012-10-11}) and 6.38 hours (Fig. \ref{fig:BikeState_5Y_grp_Hour_31209_2014-10-04}). The longer $T_E^D$ is, the more costly the corral service is. 
Fig. \ref{fig:BikeState_5Y_AMCorral_Events} indicates that corral service significantly increased capacity of Station 31209 for the high-attendance events. As an operating activity, corral service can be very useful and effective for attracting more car users to adopt biking transportation mode when participating these events, which not only augments operating gains of bikeshare, but also relieves the difficulty and problems in transportation due to regional traffic congestion and parking limits.

\subsection{Use and Idle Time} \label{sec:UseTime}

Now we analyze the use and idle times of bikes. For all bikes in $B$, the total operational time $T_{L}$ contains three components, i.e., 
\begin{equation}
T_{L}=T_{U}+T_{N}=T_{U}+T_{I}+T_{M},
\end{equation}
\begin{equation}
T_{U}=\sum_{b\in B}T_{U(b)},~~~~T_{I}=\sum_{b\in B}T_{I(b)},~~~~T_{M}=\sum_{b\in B}T_{M(b)},
\end{equation}
where $T_{U}$ is the total {\em use time} of all bikes ridden by users between stations, $T_{I}$ is the total {\em idle time} of all bikes sitting at docking stations, $T_{M}$ is the total {\em maintenance time} of all bikes taken away from the system for maintenance, and $T_{U(b)}$, $T_{I(b)}$ and $T_{M(b)}$ are respectively the use, idle, and maintenance times of bike $b \in B$. The {\em non-use time} $T_{N}$ is $T_{N}=T_{I}+T_{M}$.  The {\em use time ratio} $R_U$ is defined as $R_U=T_{U}/T_{L}$. Higher $R_U$ represents a better utilization, meaning that the BSS functions more actively as a transportation mode for users and therefore generates more revenue to support the development of itself.

For bike $b$, its use time can be directly obtained from individual trips, i.e.,
\begin{equation}
T_{U(b)}=\sum_{l \in L_b} t_{U(l)}=\sum_{l \in L_b} (t_{d(l)}-t_{o(l)}),
\end{equation}
where $L_b$ is the set of trips by bike $b$, and $t_{U(l)}=t_{d(l)}-t_{o(l)}$ is the duration of trip $l \in L_b$. 
Let the trips in $L_b$ be ordered by $t_{o(l)}$, and $l$ and $l'$ be the $k$th and $(k-1)$th trips in $L_b$, for $k \in [1, |L_b|]$. For bike $b$, we can calculate the non-use time between the trips as 
\begin{equation}
T_{N(b)}=\sum_{l \in L_b} t_{N(l)}=\sum_{l \in L_b} (t_{o(l)}-t_{d(l')}),
\end{equation}
where $t_{N(l)}=t_{o(l)}-t_{d(l')}$ is the trip-level non-use time between trips $l'$ and $l$. 

The trips used for or associated with any maintenance have all been removed from the data source by CaBi. Thus it is impossible for us to precisely separate idle time $t_{I(l)}$ and maintenance time $t_{M(l)}$ from each $t_{N(l)}$ at a trip level. To extract reasonable samples of $t_{I(l)}$, we use the following conditions: (a) $s_{o(l)} \equiv s_{d(l')}$, and (b) $t_{o(l)}-t_{d(l')}<TH_I$, where $TH_I$ is a threshold value. By default, $TH_I=\infty$. Both of the assumptions are rationale and meet practical conditions. The condition of (a) can be satisfied to exclude some major maintenance activities, which is reasonable in practice, for example, {\em rebalancing} maintenance \citep{de2016bike} will always move bikes to different bike stations, and other maintenance activities may also return a bike to a different station in $S$ in a high probability. The condition of (b) is used in consideration of some other maintenance activities that might cost a significant amount of time to get accomplished, such as bike repairing.

\begin{figure} [ht]
\centering

\begin{subfigure}{.49\textwidth} 
\centering \includegraphics[width=0.95\textwidth]{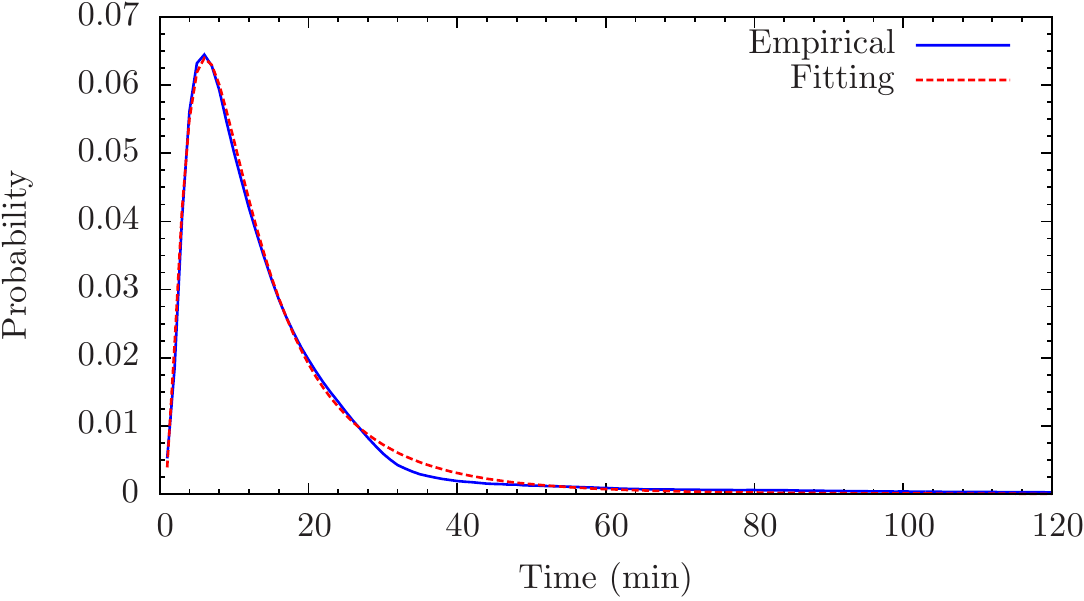} \caption{Use Time Distribution: PDF}
\label{fig:BikeState_5Y_grp_Time_PDF} 
\end{subfigure}
\begin{subfigure}{.49\textwidth} 
\centering \includegraphics[width=0.95\textwidth]{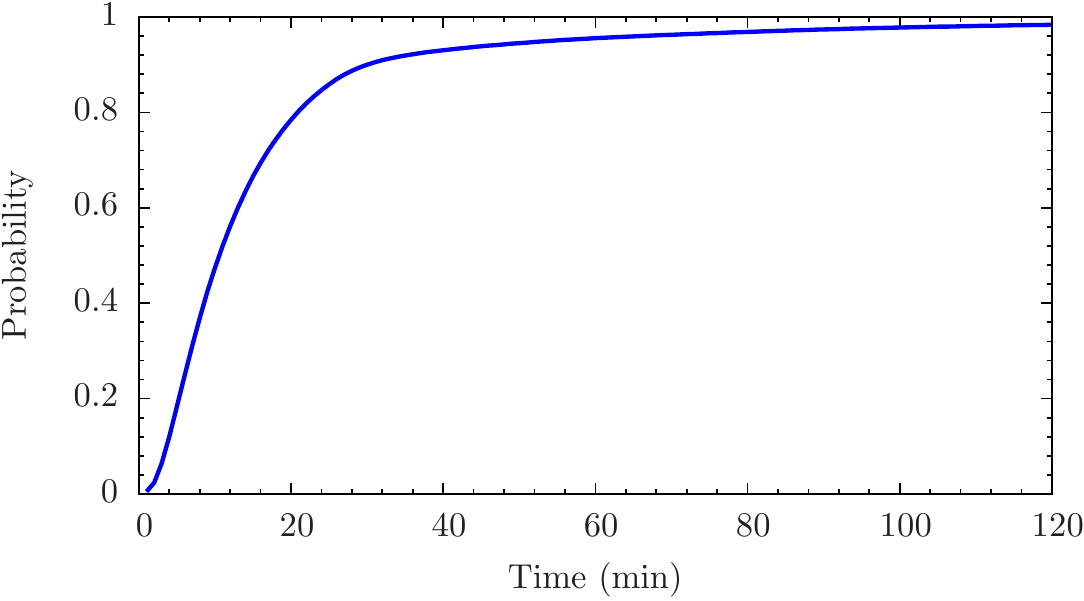} \caption{Use Time Distribution: CDF}
\label{fig:BikeState_5Y_grp_Time_PDF_CDF} 
\end{subfigure}

\begin{subfigure}{.49\textwidth} 
\centering \includegraphics[width=0.95\textwidth]{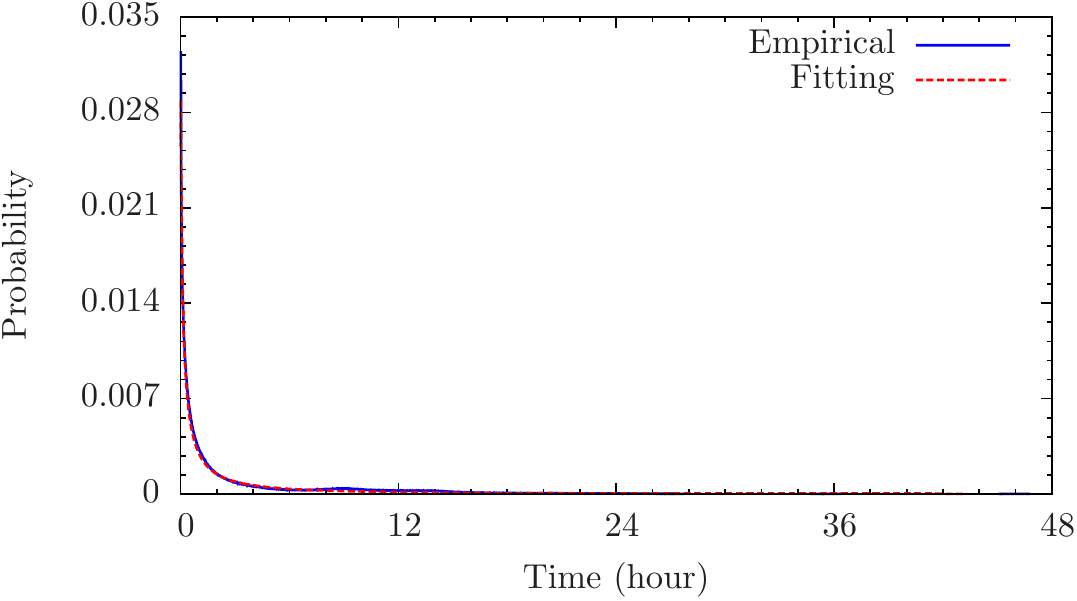} \caption{Idle Time Distribution: PDF}
\label{fig:BikeState_5Y_grp_idletime_PDF} 
\end{subfigure}
\begin{subfigure}{.49\textwidth} 
\centering \includegraphics[width=0.95\textwidth]{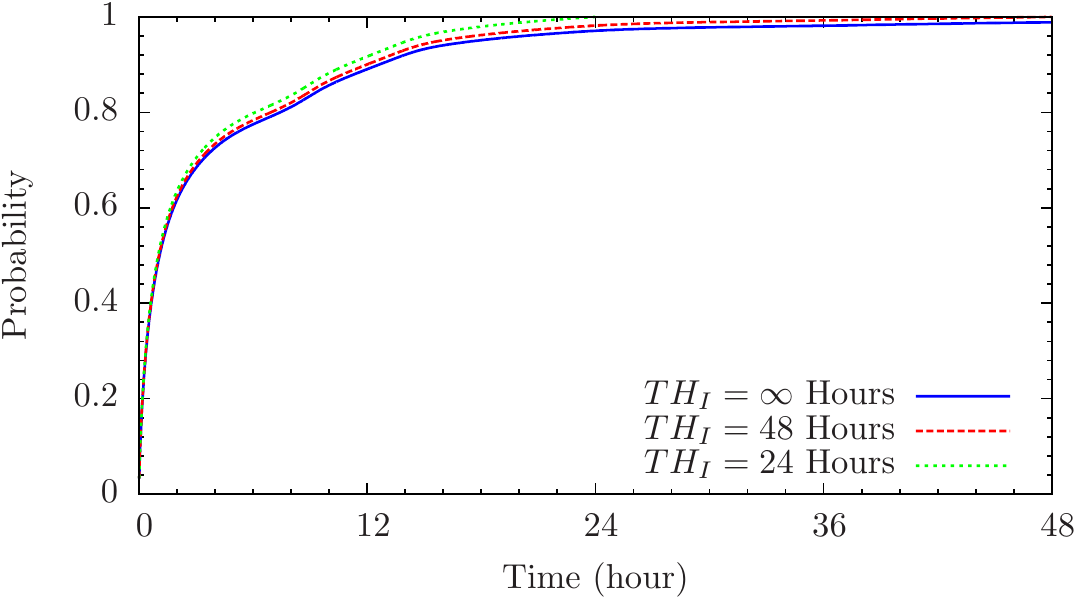} \caption{Idle Time Distribution: CDF}
\label{fig:BikeState_5Y_grp_idletime_PDF_CDF} 
\end{subfigure}

\caption{Empirical Distributions of Use and Idle Times.}
\label{fig:BikeState_5Y_User_Idle_Time_PDF_CDF}
\end{figure}

Figs. \ref{fig:BikeState_5Y_grp_Time_PDF} and \ref{fig:BikeState_5Y_grp_idletime_PDF} show the empirical probability density functions (PDF) respectively for the use time $t_{U(l)}$ of each trip and for the idle time $t_{I(l)}$ before each trip, $\forall l \in L$. Correspondingly, Fig. \ref{fig:BikeState_5Y_grp_Time_PDF_CDF} and Fig. \ref{fig:BikeState_5Y_grp_idletime_PDF_CDF} give the cumulative distribution functions (CDF). As shown in Figs. \ref{fig:BikeState_5Y_grp_Time_PDF} and \ref{fig:BikeState_5Y_grp_idletime_PDF}, the two empirical PDFs can be well fitted with the lognormal distribution, i.e.,
\begin{equation}
f(x | \mu, \sigma)=(x\sigma\sqrt{2\pi})^{-1}\operatorname{exp}({-\frac{\left(\ln x-\mu\right)^2}{2\sigma^2}}),
\end{equation}
where the location and shape parameters $\mu$ and $\sigma$ are estimated using the cooperative group optimization (CGO) \citep{xie2014cooperative} to minimize the least squares. For the two PDFs, the parameters ($\mu$, $\sigma$) are respectively (2.3911, 0.7641) and (4.2167, 2.2145), and the root-mean-square errors (RMSE) are respectively 1.89E-4 and 4.47E-5. The lognormal model allows us to describe an empirical PDF in two parameters, to help modeling the stochastic use time in simulations, and to robustly estimate the mean value as ${\displaystyle \exp(\mu +\sigma ^{2}/2)}$ if there are any outliers in the data. 

As shown in Fig. \ref{fig:BikeState_5Y_grp_Time_PDF_CDF}, the median of use time is 10.57 minutes, 90.01\% of CaBi bike trips have a use time below the 30-minutes limit to avoid additional charges, and only around 1.64\% of trips have a use time over 2 hours. As shown in Fig. \ref{fig:BikeState_5Y_grp_idletime_PDF_CDF}, the median of idle time is 1.09 hours, and 2.90\% of CaBi bikes have an idle time over 24 hours, if $TH_I=\infty$ is considered. There is only a minor and negligible difference in the median of idle time between the cases using $TH_I=\infty$ and $TH_I \in \{24, 48\}$ hours. 
From the whole data, the total use and non-use time, i.e., $T_{U}$ and $T_{N}$, are respectively 467.02 and 13563.77 bikes$\cdot$years. The use time ratio $R_U$ is 3.33\%, meaning that there is quite a large room to improve the operational efficiency of BSS, and the improvement could be significant. The rebalancing maintenance contributes to an increase in idle time, but it also increases use time by reducing the imbalance between demand and supply at stations as described in Section \ref{sec:TripDemand}. Advanced algorithms may be implemented to optimize a practical rebalancing strategy for better tackling the imbalance issue, which can enable a BSS to serve more users while to reduce cost of maintenance. Subject to solving the imbalance between demand and supply of bikeshare, the $R_{DS}$ values can be used as fast-and-frugal heuristic information \citep{gigerenzer1999simple} to identify the imbalance regions at different times of day, as shown in Fig.~\ref{fig:BikeState_5Y_Geo_grp_osid_ToD_DoW}.

\subsection{Trip Purpose} \label{sec:UserTripTypes}

In general, bike users take two basic types of trips in terms of trip purpose, which is either utilitarian or recreational~\citep{miranda2013classification}. If the primary purpose of a trip is utilitarian, for example, commuting for work, the bike user would more likely prefer a shorter travel time to reach the destination and the user normally does not take any intermediate stop~\citep{conley2016view}. A commuting trip has typical AM/PM peaks in volume during weekday and has a lower volume during weekend. Instead, if the primary purpose of a trip is recreational, for example, riding in parkland, the user would pay more attention to the attributes pertinent to comfort rather than travel time. Recreational trips often have a higher volume during weekend, and most of them would be taken around an area with an open or recreational space.

CaBi data of bike trips has no direct record to identify purpose of trips by users. We will study if it is possible to estimate purpose of trips by users from data analysis on typical features of trips in terms of the types of trips and user. We classify trips to two types based on their forms of origin and destination stations, $s_o$ and $s_d$. One type is called {\em O-O trip} \citep{zhao2015exploring}, or ``loop'' trip \citep{Noland2017}, where $s_o=s_d$. Another type is called {\em O-D trip}, where $s_o \neq s_d$. In the collected data, O-O trips account for 4.03\% of trips in total. All commuting trips are O-D trips. In addition, two user types are defined in the CaBi data, which are {\em casual users} and {\em member users} \citep{buck2013bikeshare,Noland2017,Wergin2017} respectively. The trips taken by casual users account for 20.64\% of trips in total. 

\begin{figure} [p]
\centering

\begin{subfigure}{.49\textwidth} 
\centering \includegraphics[width=0.95\textwidth]{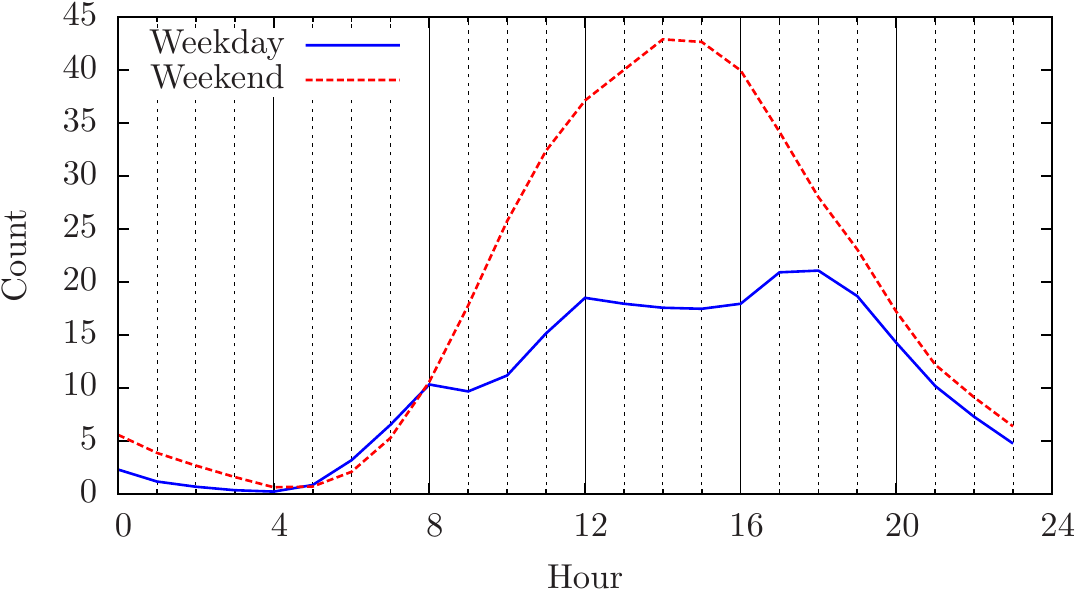} \caption{O-O Trips.}
\label{fig:BikeState_5Y_SameOD_hod_Week} 
\end{subfigure}
\begin{subfigure}{.49\textwidth} 
\centering \includegraphics[width=0.95\textwidth]{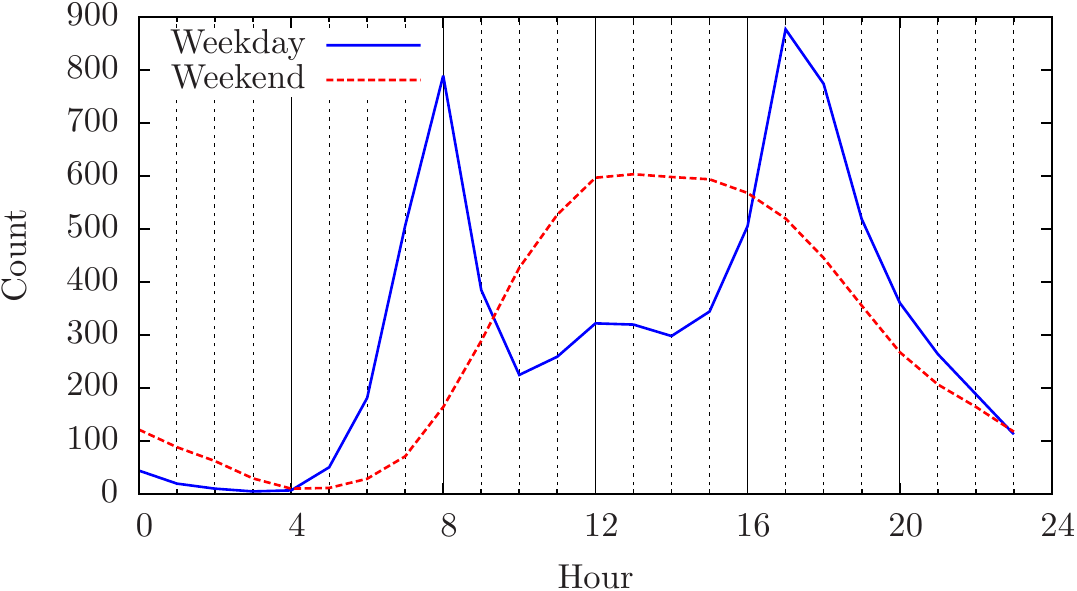} \caption{O-D Trips.}
\label{fig:BikeState_5Y_DiffOD_hod_Week} 
\end{subfigure}

\begin{subfigure}{.49\textwidth} 
\centering \includegraphics[width=0.95\textwidth]{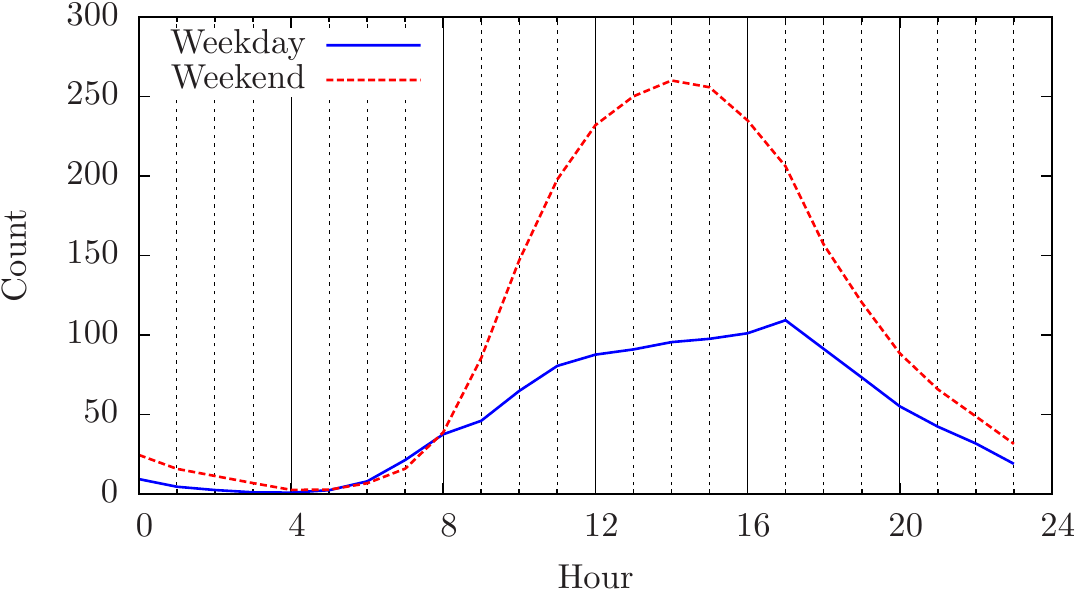} \caption{Casual Users.}
\label{fig:BikeState_5Y_CasualUser_hod_Week} 
\end{subfigure}
\begin{subfigure}{.49\textwidth} 
\centering \includegraphics[width=0.95\textwidth]{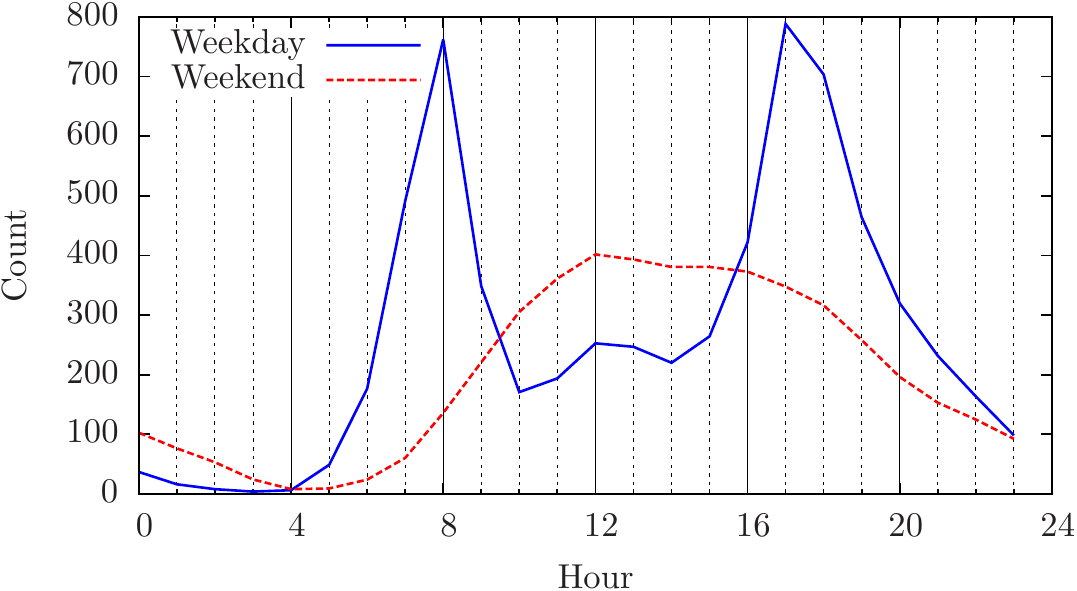} \caption{Member Users.}
\label{fig:BikeState_5Y_MemberUser_hod_Week} 
\end{subfigure}

\caption{Average Hour-of-Day Trip Counts by Trip and User Types.}
\label{fig:BikeState_5Y_Trip_User_Type_hod}
\end{figure}

\begin{figure} [p]
\centering

\begin{subfigure}{.49\textwidth} 
\centering \includegraphics[width=0.95\textwidth]{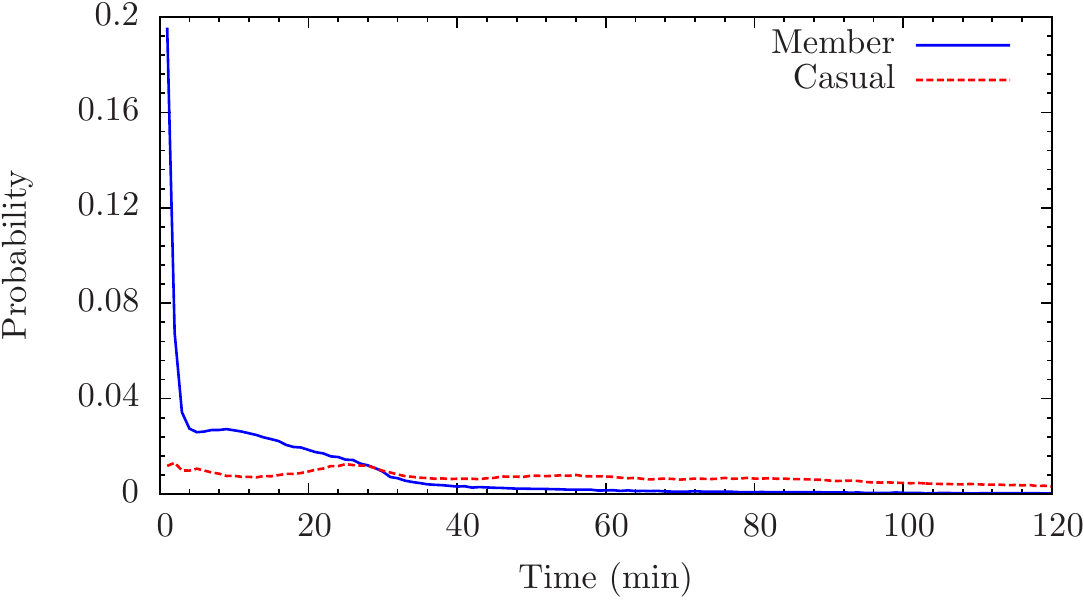} \caption{Empirical PDF for O-O Trips.}
\label{fig:BikeState_5Y_OD_grp_Time_SameOD_UserType_PDF} 
\end{subfigure}
\begin{subfigure}{.49\textwidth} 
\centering \includegraphics[width=0.95\textwidth]{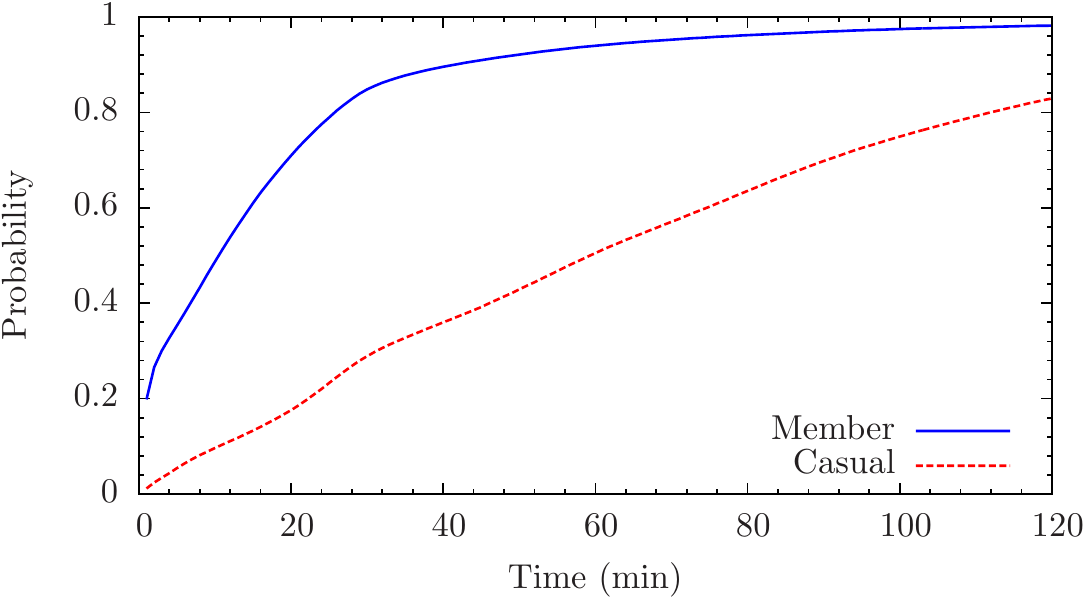} \caption{Empirical CDF for O-O Trips.}
\label{fig:BikeState_5Y_OD_grp_Time_SameOD_UserType_PDF_CDF} 
\end{subfigure}

\begin{subfigure}{.49\textwidth} 
\centering \includegraphics[width=0.95\textwidth]{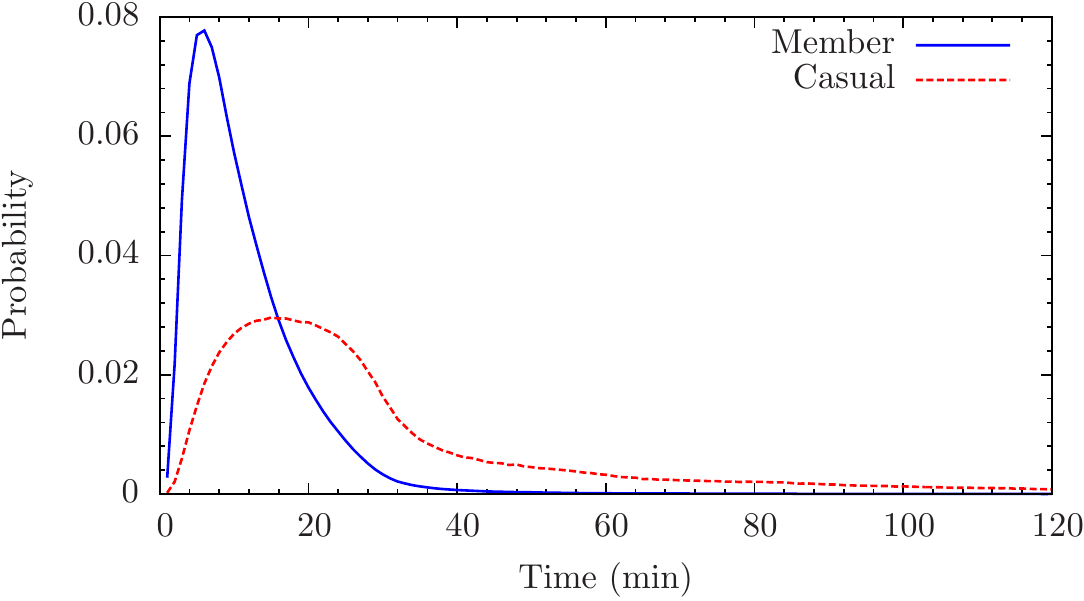} \caption{Empirical PDF for O-D Trips.}
\label{fig:BikeState_5Y_OD_grp_Time_DiffOD_UserType_PDF} 
\end{subfigure}
\begin{subfigure}{.49\textwidth} 
\centering \includegraphics[width=0.95\textwidth]{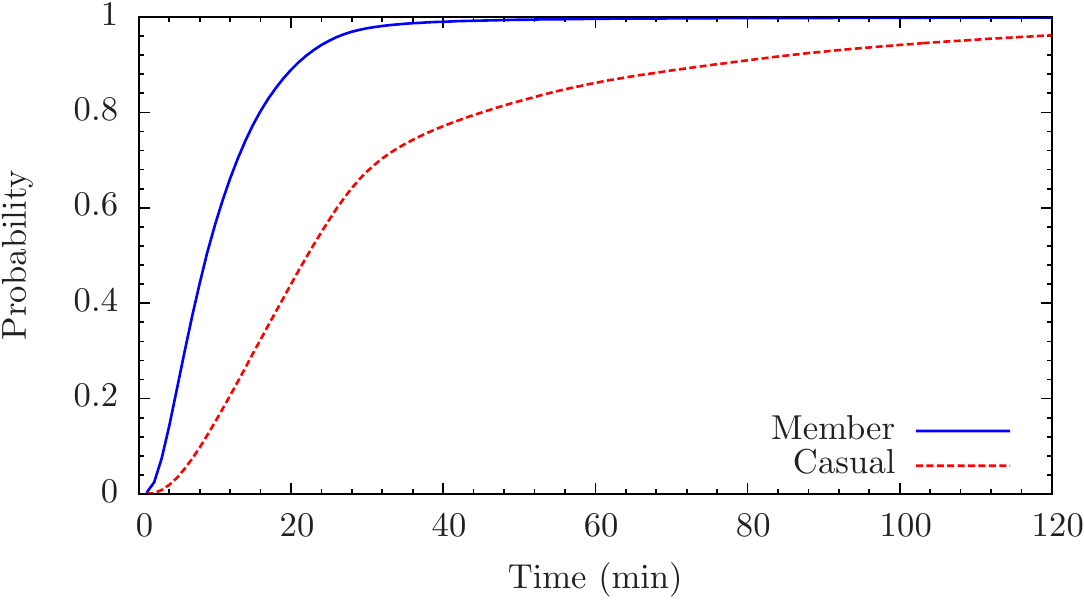} \caption{Empirical CDF for O-D Trips.}
\label{fig:BikeState_5Y_OD_grp_Time_DiffOD_UserType_PDF_CDF} 
\end{subfigure}

\caption{Empirical Distributions of Use Time by Trip and User Types.}
\label{fig:BikeState_5Y_UserTime_Trip_User_Type_PDF_CDF}
\end{figure}

Fig. \ref{fig:BikeState_5Y_Trip_User_Type_hod} shows the comparisons of the temporal distributions of hourly averaged trip counts between weekday and weekend by different types of trips and users respectively. Both the O-O trip users and casual users take more trips in weekend than in weekday, expressing a typical recreational pattern \citep{miranda2013classification} as shown in Fig.~\ref{fig:BikeState_5Y_SameOD_hod_Week} and Fig.~\ref{fig:BikeState_5Y_CasualUser_hod_Week}. Consistent with previous study, the pattern is a fairly general for recreational trips. As indicated by the study on the bikesharing data in New York \citep{Noland2017}, casual users are more likely to take recreational trips, including loop trips. The O-D trip users and member users both have two peak trip periods at AM and PM in weekday but a single peak trip period in weekend, revealing a typical commuting pattern as shown in Fig.~\ref{fig:BikeState_5Y_DiffOD_hod_Week} and Fig.~\ref{fig:BikeState_5Y_MemberUser_hod_Week}.

Fig. \ref{fig:BikeState_5Y_UserTime_Trip_User_Type_PDF_CDF} gives the comparison of the distributions (i.e. empirical PDFs and CDFs) in bikeshare use time between casual and member users by different trip types. In Fig. \ref{fig:BikeState_5Y_OD_grp_Time_SameOD_UserType_PDF}, at a very small value of use time ($t_{U} \le 2$ minutes), the distribution of O-O trips has a sharp peak for member users (the blue curve). This is because member users would return rented bikes to the docking stations under an unsatisfied condition. For member users, the use time distribution of O-D trips shows a higher peak value than that of the O-O trips (see the comparison between the blue curves in Figs. \ref{fig:BikeState_5Y_OD_grp_Time_SameOD_UserType_PDF} and \ref{fig:BikeState_5Y_OD_grp_Time_DiffOD_UserType_PDF}). For O-D trips, member users are likely to have a shorter use time than casual users, as indicated by the comparison in Fig.~\ref{fig:BikeState_5Y_OD_grp_Time_DiffOD_UserType_PDF}. 
These features in use time by different users are clearly shown by Figs. \ref{fig:BikeState_5Y_OD_grp_Time_SameOD_UserType_PDF_CDF} and \ref{fig:BikeState_5Y_OD_grp_Time_DiffOD_UserType_PDF_CDF}, where the CDF curve by member users is at the left side of that by casual users for both O-O and O-D trips, and the CDF curve of O-D trips is at the left side of that of O-O trips for both member and casual users. Comparing member and casual users, we find that the median use time for member and casual users are respectively 10.50 and 59.31 minutes for O-O trips, and respectively 8.90 and 22.15 minutes for O-D trips. The trips below 30 minutes for the O-O and O-D trips respectively account for 84.84\% and 97.64\% for member users, whereas respectively 28.90\% and 67.63\% for casual users. Combining the result from the comparisons above with that from Fig. \ref{fig:BikeState_5Y_Trip_User_Type_hod}, we can conclude that member users prefer to use bikeshare for utilitarian trips, while casual users prefer recreational trips.  

\subsection{O-D Flows in Bikesharing Network} \label{sec:odFlows}

\begin{figure} [htb]
\centering

\begin{subfigure}{.32\textwidth} \centering \includegraphics[width=.95\linewidth]{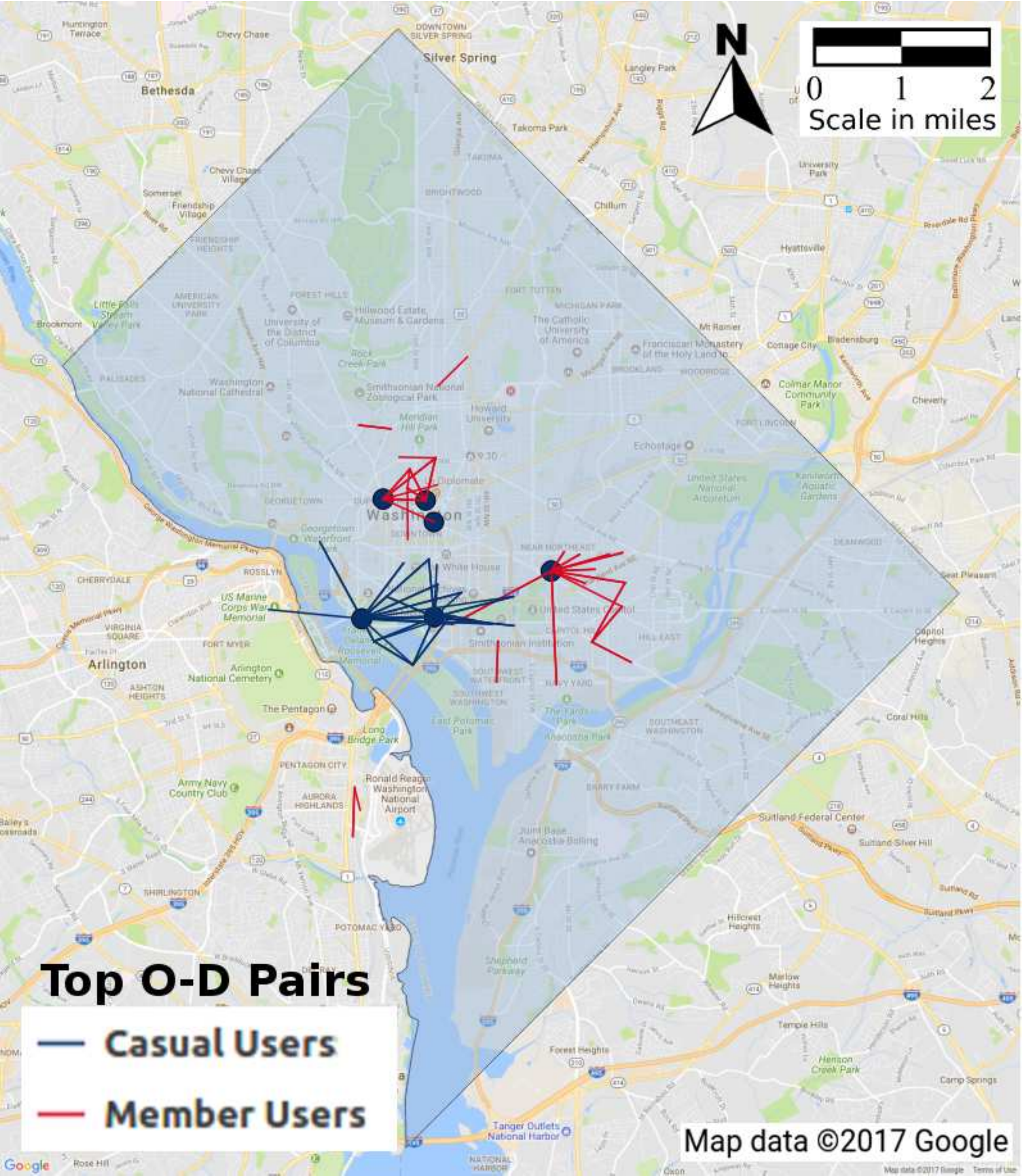} \caption{Top 50 O-D Pairs for Casual (Blue) and Member (Red) Users} \label{fig:BikeState_5Y_Geo_grp_osid_DailyRate_Top50_AllUsers_Top6_2} \end{subfigure}
\begin{subfigure}{.32\textwidth} \centering \includegraphics[width=.95\linewidth]{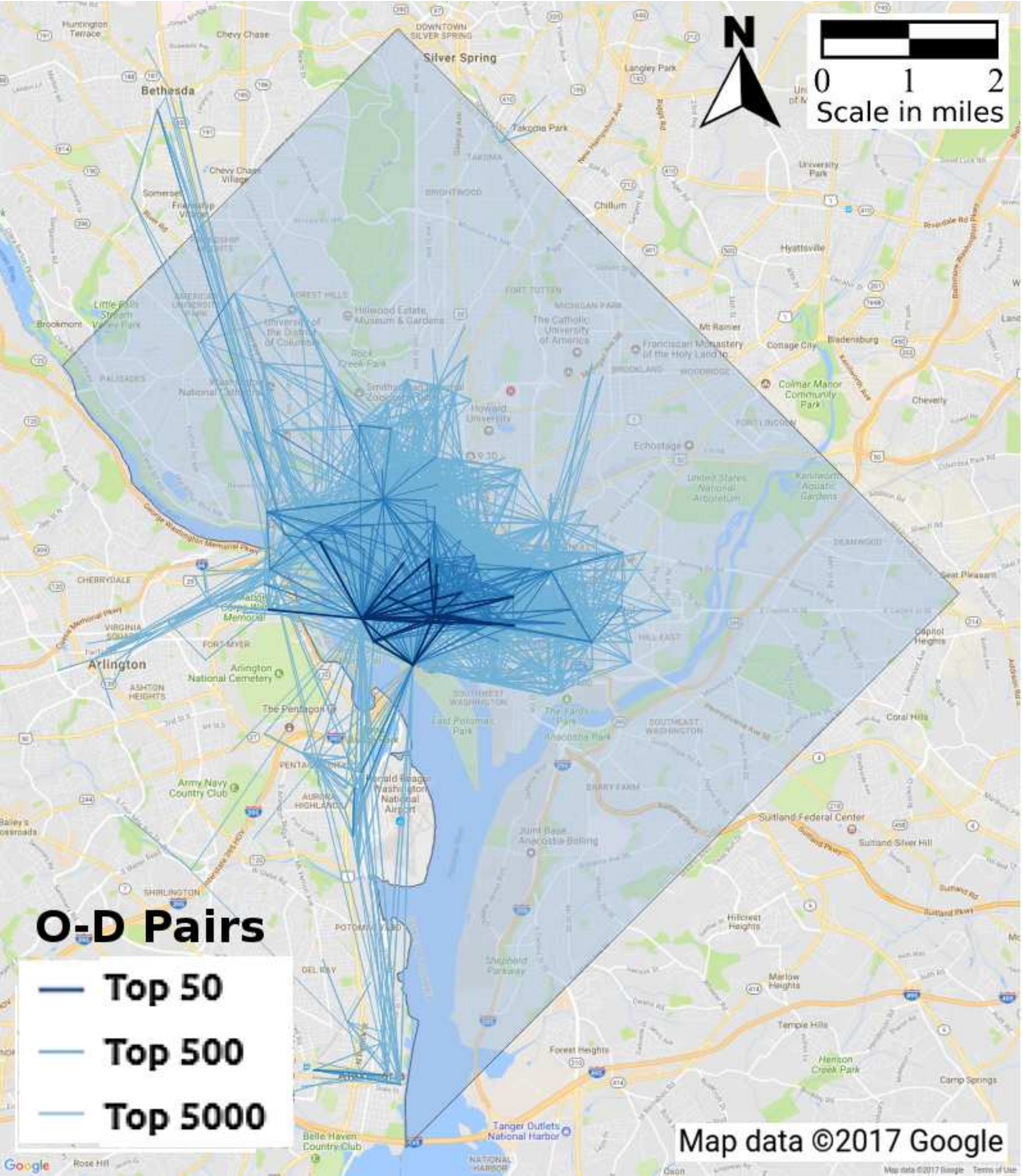} \caption{Top 50, 500, and 5000 O-D Pairs for Casual Users} \label{fig:BikeState_5Y_Geo_grp_OD_DiffOD_CasualUser_Sorted_50,500,5000_4} \end{subfigure}
\begin{subfigure}{.32\textwidth} \centering \includegraphics[width=.95\linewidth]{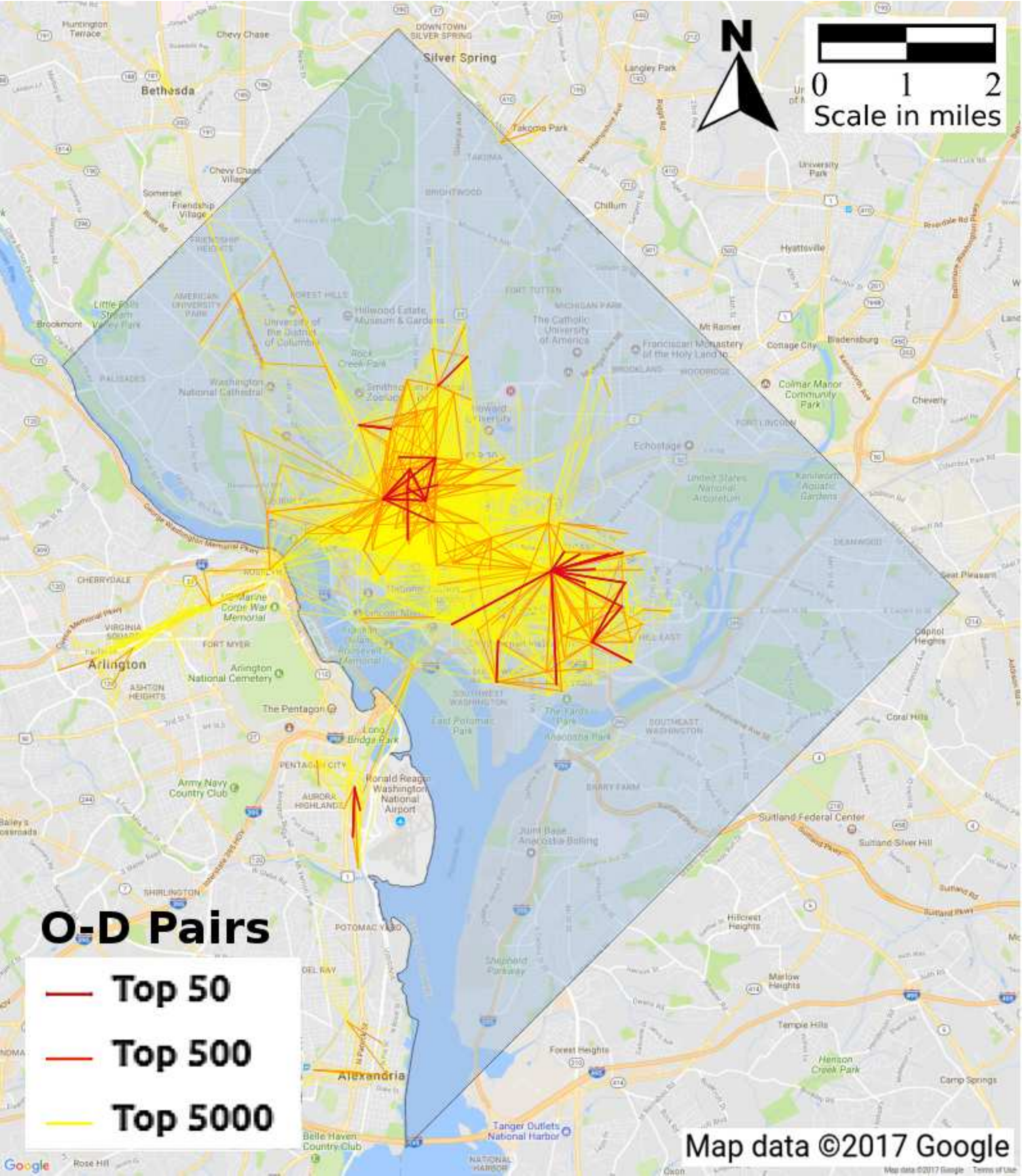} \caption{Top 50, 500, and 5000 O-D Pairs for Member Users} 
\label{fig:BikeState_5Y_Geo_grp_OD_DiffOD_MemberUser_Sorted_50,500,5000_4} \end{subfigure}

\caption{Top O-D Pairs in the Ranking of the Highest O-D Flows by Casual and Member Users.}
\label{fig:BikeState_5Y_CasualUser_TopRoutes}
\end{figure}

To better understand a bikesharing network, we analyze the top O-D pairs in the ranking of the highest O-D flows respectively by casual and member users (see Fig. \ref{fig:BikeState_5Y_CasualUser_TopRoutes}). As shown in Fig. \ref{fig:BikeState_5Y_Geo_grp_osid_DailyRate_Top50_AllUsers_Top6_2}, the top 50 O-D pairs form one main cluster by casual users in and around a famous recreational area --- the National Mall and Memorial Parks, while form two clusters by member users respectively at the east and north neighborhoods of central business district. The top 6 origin stations in the ranking of the highest O-D flows are separated into three clusters (see the black dots in Fig. \ref{fig:BikeState_5Y_Geo_grp_osid_DailyRate_Top50_AllUsers_Top6_2}). Among the 6 origin stations, one at the east cluster is the Union Station, three at the north cluster are near the triangle of Dupont Circle, Logan Circle and Thomas Circle Park, and two are in National Mall area. To analyze how a bikeshare network structure grows with the increase in the number of O-D links, we show the top 50, 500, and 5000 highest-ranking O-D pairs in O-D flows respectively by casual users (see Fig. \ref{fig:BikeState_5Y_Geo_grp_OD_DiffOD_CasualUser_Sorted_50,500,5000_4}) and by member users (see Fig.\ref{fig:BikeState_5Y_Geo_grp_OD_DiffOD_MemberUser_Sorted_50,500,5000_4}).  Fig.~\ref{fig:BikeState_5Y_Geo_grp_OD_DiffOD_MemberUser_Sorted_50,500,5000_4} indicates that the trip network formed by member users covers the neighborhoods for commuters --- the areas that feature high densities of workplaces (see Fig.~\ref{fig:Census_LODES2014_DC_grp_w_geocode_ct_Geo_Tract10}) or homes (see Fig.~\ref{fig:Census_LODES2014_DC_grp_h_geocode_ct_Geo_Tract10}). 
In contrast, casual users take more long-distance trips (see Fig. \ref{fig:BikeState_5Y_Geo_grp_OD_DiffOD_CasualUser_Sorted_50,500,5000_4}). 
From Figs. \ref{fig:BikeState_5Y_Geo_grp_OD_DiffOD_CasualUser_Sorted_50,500,5000_4} and \ref{fig:BikeState_5Y_Geo_grp_OD_DiffOD_MemberUser_Sorted_50,500,5000_4}, community structure can be clearly identified, where three clusters appear showing densely connected links of O-D pairs. The formation of such a community structure is quite common in real-world self-organized networks \citep{girvan2002community}. The polycentricity at a metropolitan scale is an interesting feature of modern urban landscapes \citep{anas1998urban}.
Most users prefer to bike on a short-time trip (as shown in \ref{fig:BikeState_5Y_grp_Time_PDF}), thus most regions far from core areas of existing bikeshare clusters are not reached by users, even though some of these regions have sufficiently high densities of workplace or residence for generating a large trip demand (see Figs. \ref{fig:Census_LODES2014_DC_grp_w_geocode_ct_Geo_Tract10} and \ref{fig:Census_LODES2014_DC_grp_h_geocode_ct_Geo_Tract10} on the densities of workplace and residence respectively). To increase bikeshare ridership of these regions, it is important to foster the formation of new clusters with densely connected O-D links in the bikesharing network, which may be considered by system operators while new bikeshare stations need to be added or by city planners while building environment need to be improved.

\begin{figure} [p]
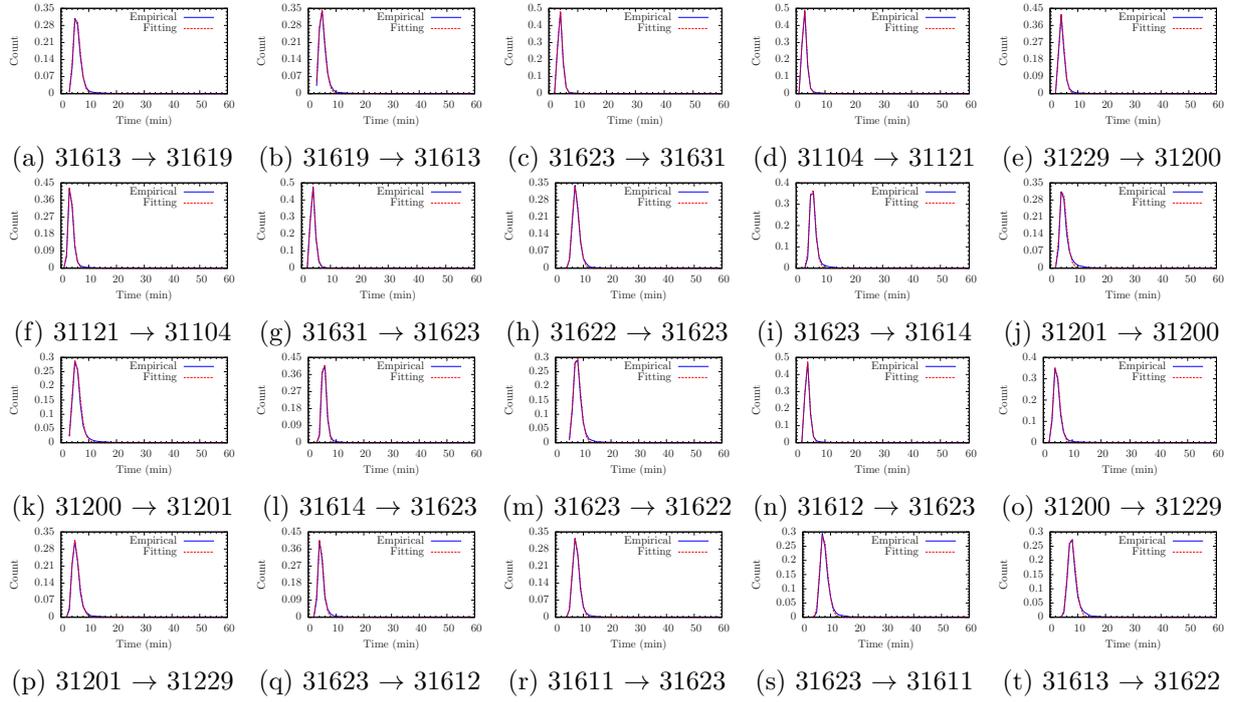

\centering

\begin{subfigure}{.19\textwidth} \centering \includegraphics[width=.95\linewidth]{img/{{BikeState_5Y_grp_Time_31613_31619_MemberUser_PDF_Fitting}}} \caption{31613 $\rightarrow$ 31619} \label{fig:BikeState_5Y_grp_Time_31613_31619_MemberUser_PDF_Fitting} \end{subfigure}
\begin{subfigure}{.19\textwidth} \centering \includegraphics[width=.95\linewidth]{img/{{BikeState_5Y_grp_Time_31619_31613_MemberUser_PDF_Fitting}}} \caption{31619 $\rightarrow$ 31613} \label{fig:BikeState_5Y_grp_Time_31619_31613_MemberUser_PDF_Fitting} \end{subfigure}
\begin{subfigure}{.19\textwidth} \centering \includegraphics[width=.95\linewidth]{img/{{BikeState_5Y_grp_Time_31623_31631_MemberUser_PDF_Fitting}}} \caption{31623 $\rightarrow$ 31631} \label{fig:BikeState_5Y_grp_Time_31623_31631_MemberUser_PDF_Fitting} \end{subfigure}
\begin{subfigure}{.19\textwidth} \centering \includegraphics[width=.95\linewidth]{img/{{BikeState_5Y_grp_Time_31104_31121_MemberUser_PDF_Fitting}}} \caption{31104 $\rightarrow$ 31121} \label{fig:BikeState_5Y_grp_Time_31104_31121_MemberUser_PDF_Fitting} \end{subfigure}
\begin{subfigure}{.19\textwidth} \centering \includegraphics[width=.95\linewidth]{img/{{BikeState_5Y_grp_Time_31229_31200_MemberUser_PDF_Fitting}}} \caption{31229 $\rightarrow$ 31200} \label{fig:BikeState_5Y_grp_Time_31229_31200_MemberUser_PDF_Fitting} \end{subfigure}

\begin{subfigure}{.19\textwidth} \centering \includegraphics[width=.95\linewidth]{img/{{BikeState_5Y_grp_Time_31121_31104_MemberUser_PDF_Fitting}}} \caption{31121 $\rightarrow$ 31104} \label{fig:BikeState_5Y_grp_Time_31121_31104_MemberUser_PDF_Fitting} \end{subfigure}
\begin{subfigure}{.19\textwidth} \centering \includegraphics[width=.95\linewidth]{img/{{BikeState_5Y_grp_Time_31631_31623_MemberUser_PDF_Fitting}}} \caption{31631 $\rightarrow$ 31623} \label{fig:BikeState_5Y_grp_Time_31631_31623_MemberUser_PDF_Fitting} \end{subfigure}
\begin{subfigure}{.19\textwidth} \centering \includegraphics[width=.95\linewidth]{img/{{BikeState_5Y_grp_Time_31622_31623_MemberUser_PDF_Fitting}}} \caption{31622 $\rightarrow$ 31623} \label{fig:BikeState_5Y_grp_Time_31622_31623_MemberUser_PDF_Fitting} \end{subfigure}
\begin{subfigure}{.19\textwidth} \centering \includegraphics[width=.95\linewidth]{img/{{BikeState_5Y_grp_Time_31623_31614_MemberUser_PDF_Fitting}}} \caption{31623 $\rightarrow$ 31614} \label{fig:BikeState_5Y_grp_Time_31623_31614_MemberUser_PDF_Fitting} \end{subfigure}
\begin{subfigure}{.19\textwidth} \centering \includegraphics[width=.95\linewidth]{img/{{BikeState_5Y_grp_Time_31201_31200_MemberUser_PDF_Fitting}}} \caption{31201 $\rightarrow$ 31200} \label{fig:BikeState_5Y_grp_Time_31201_31200_MemberUser_PDF_Fitting} \end{subfigure}

\begin{subfigure}{.19\textwidth} \centering \includegraphics[width=.95\linewidth]{img/{{BikeState_5Y_grp_Time_31200_31201_MemberUser_PDF_Fitting}}} \caption{31200 $\rightarrow$ 31201} \label{fig:BikeState_5Y_grp_Time_31200_31201_MemberUser_PDF_Fitting} \end{subfigure}
\begin{subfigure}{.19\textwidth} \centering \includegraphics[width=.95\linewidth]{img/{{BikeState_5Y_grp_Time_31614_31623_MemberUser_PDF_Fitting}}} \caption{31614 $\rightarrow$ 31623} \label{fig:BikeState_5Y_grp_Time_31614_31623_MemberUser_PDF_Fitting} \end{subfigure}
\begin{subfigure}{.19\textwidth} \centering \includegraphics[width=.95\linewidth]{img/{{BikeState_5Y_grp_Time_31623_31622_MemberUser_PDF_Fitting}}} \caption{31623 $\rightarrow$ 31622} \label{fig:BikeState_5Y_grp_Time_31623_31622_MemberUser_PDF_Fitting} \end{subfigure}
\begin{subfigure}{.19\textwidth} \centering \includegraphics[width=.95\linewidth]{img/{{BikeState_5Y_grp_Time_31612_31623_MemberUser_PDF_Fitting}}} \caption{31612 $\rightarrow$ 31623} \label{fig:BikeState_5Y_grp_Time_31612_31623_MemberUser_PDF_Fitting} \end{subfigure}
\begin{subfigure}{.19\textwidth} \centering \includegraphics[width=.95\linewidth]{img/{{BikeState_5Y_grp_Time_31200_31229_MemberUser_PDF_Fitting}}} \caption{31200 $\rightarrow$ 31229} \label{fig:BikeState_5Y_grp_Time_31200_31229_MemberUser_PDF_Fitting} \end{subfigure}

\begin{subfigure}{.19\textwidth} \centering \includegraphics[width=.95\linewidth]{img/{{BikeState_5Y_grp_Time_31201_31229_MemberUser_PDF_Fitting}}} \caption{31201 $\rightarrow$ 31229} \label{fig:BikeState_5Y_grp_Time_31201_31229_MemberUser_PDF_Fitting} \end{subfigure}
\begin{subfigure}{.19\textwidth} \centering \includegraphics[width=.95\linewidth]{img/{{BikeState_5Y_grp_Time_31623_31612_MemberUser_PDF_Fitting}}} \caption{31623 $\rightarrow$ 31612} \label{fig:BikeState_5Y_grp_Time_31623_31612_MemberUser_PDF_Fitting} \end{subfigure}
\begin{subfigure}{.19\textwidth} \centering \includegraphics[width=.95\linewidth]{img/{{BikeState_5Y_grp_Time_31611_31623_MemberUser_PDF_Fitting}}} \caption{31611 $\rightarrow$ 31623} \label{fig:BikeState_5Y_grp_Time_31611_31623_MemberUser_PDF_Fitting} \end{subfigure}
\begin{subfigure}{.19\textwidth} \centering \includegraphics[width=.95\linewidth]{img/{{BikeState_5Y_grp_Time_31623_31611_MemberUser_PDF_Fitting}}} \caption{31623 $\rightarrow$ 31611} \label{fig:BikeState_5Y_grp_Time_31623_31611_MemberUser_PDF_Fitting} \end{subfigure}
\begin{subfigure}{.19\textwidth} \centering \includegraphics[width=.95\linewidth]{img/{{BikeState_5Y_grp_Time_31613_31622_MemberUser_PDF_Fitting}}} \caption{31613 $\rightarrow$ 31622} \label{fig:BikeState_5Y_grp_Time_31613_31622_MemberUser_PDF_Fitting} \end{subfigure}

\caption{Empirical PDFs of Bikeshare Use Time for the Top 20 Highest-Ranking O-D Pairs in O-D Flows by Member Users, where Empirical Data (Blue Solid Line) are Fitted with Lognormal Distributions (Red Dash Line).}
\label{fig:BikeState_5Y_DiffOD_Time_MemberUser_PDF}
\end{figure}

\begin{figure} [p]
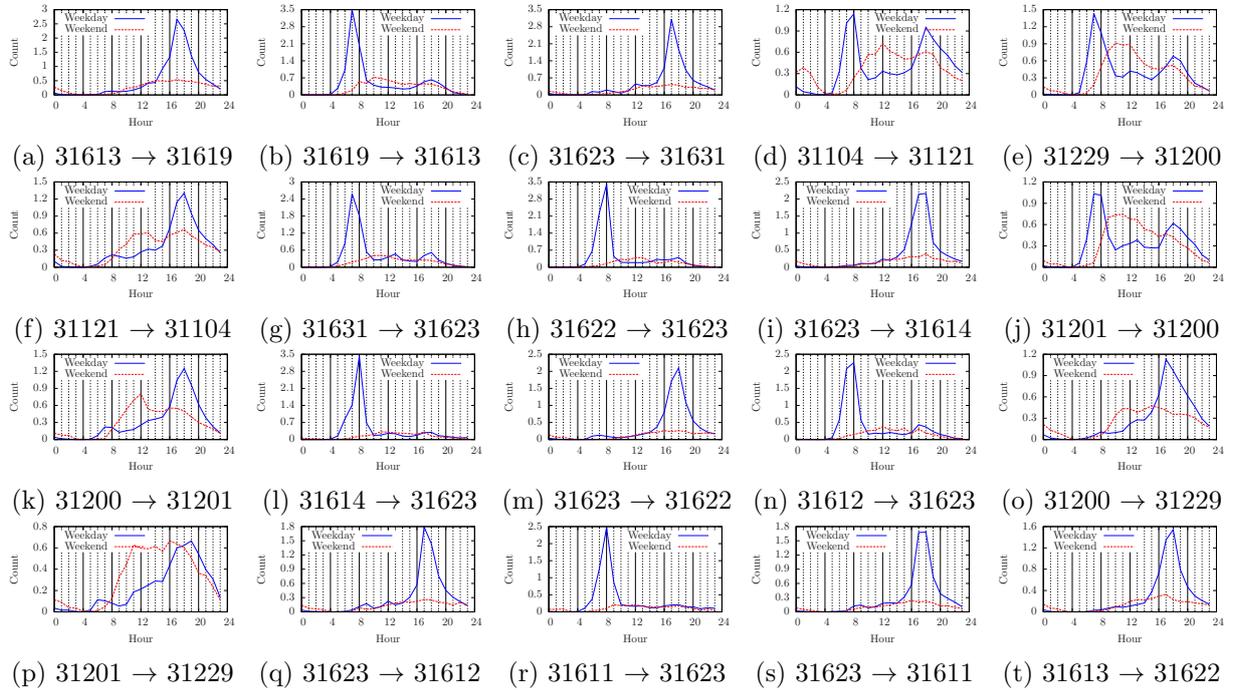

\centering

\begin{subfigure}{.19\textwidth} \centering \includegraphics[width=.95\linewidth]{img/{{BikeState_5Y_31613_31619_MemberUser_hod_Week}}} \caption{31613 $\rightarrow$ 31619} \label{fig:BikeState_5Y_31613_31619_MemberUser_hod_Week} \end{subfigure}
\begin{subfigure}{.19\textwidth} \centering \includegraphics[width=.95\linewidth]{img/{{BikeState_5Y_31619_31613_MemberUser_hod_Week}}} \caption{31619 $\rightarrow$ 31613} \label{fig:BikeState_5Y_31619_31613_MemberUser_hod_Week} \end{subfigure}
\begin{subfigure}{.19\textwidth} \centering \includegraphics[width=.95\linewidth]{img/{{BikeState_5Y_31623_31631_MemberUser_hod_Week}}} \caption{31623 $\rightarrow$ 31631} \label{fig:BikeState_5Y_31623_31631_MemberUser_hod_Week} \end{subfigure}
\begin{subfigure}{.19\textwidth} \centering \includegraphics[width=.95\linewidth]{img/{{BikeState_5Y_31104_31121_MemberUser_hod_Week}}} \caption{31104 $\rightarrow$ 31121} \label{fig:BikeState_5Y_31104_31121_MemberUser_hod_Week} \end{subfigure}
\begin{subfigure}{.19\textwidth} \centering \includegraphics[width=.95\linewidth]{img/{{BikeState_5Y_31229_31200_MemberUser_hod_Week}}} \caption{31229 $\rightarrow$ 31200} \label{fig:BikeState_5Y_31229_31200_MemberUser_hod_Week} \end{subfigure}

\begin{subfigure}{.19\textwidth} \centering \includegraphics[width=.95\linewidth]{img/{{BikeState_5Y_31121_31104_MemberUser_hod_Week}}} \caption{31121 $\rightarrow$ 31104} \label{fig:BikeState_5Y_31121_31104_MemberUser_hod_Week} \end{subfigure}
\begin{subfigure}{.19\textwidth} \centering \includegraphics[width=.95\linewidth]{img/{{BikeState_5Y_31631_31623_MemberUser_hod_Week}}} \caption{31631 $\rightarrow$ 31623} \label{fig:BikeState_5Y_31631_31623_MemberUser_hod_Week} \end{subfigure}
\begin{subfigure}{.19\textwidth} \centering \includegraphics[width=.95\linewidth]{img/{{BikeState_5Y_31622_31623_MemberUser_hod_Week}}} \caption{31622 $\rightarrow$ 31623} \label{fig:BikeState_5Y_31622_31623_MemberUser_hod_Week} \end{subfigure}
\begin{subfigure}{.19\textwidth} \centering \includegraphics[width=.95\linewidth]{img/{{BikeState_5Y_31623_31614_MemberUser_hod_Week}}} \caption{31623 $\rightarrow$ 31614} \label{fig:BikeState_5Y_31623_31614_MemberUser_hod_Week} \end{subfigure}
\begin{subfigure}{.19\textwidth} \centering \includegraphics[width=.95\linewidth]{img/{{BikeState_5Y_31201_31200_MemberUser_hod_Week}}} \caption{31201 $\rightarrow$ 31200} \label{fig:BikeState_5Y_31201_31200_MemberUser_hod_Week} \end{subfigure}

\begin{subfigure}{.19\textwidth} \centering \includegraphics[width=.95\linewidth]{img/{{BikeState_5Y_31200_31201_MemberUser_hod_Week}}} \caption{31200 $\rightarrow$ 31201} \label{fig:BikeState_5Y_31200_31201_MemberUser_hod_Week} \end{subfigure}
\begin{subfigure}{.19\textwidth} \centering \includegraphics[width=.95\linewidth]{img/{{BikeState_5Y_31614_31623_MemberUser_hod_Week}}} \caption{31614 $\rightarrow$ 31623} \label{fig:BikeState_5Y_31614_31623_MemberUser_hod_Week} \end{subfigure}
\begin{subfigure}{.19\textwidth} \centering \includegraphics[width=.95\linewidth]{img/{{BikeState_5Y_31623_31622_MemberUser_hod_Week}}} \caption{31623 $\rightarrow$ 31622} \label{fig:BikeState_5Y_31623_31622_MemberUser_hod_Week} \end{subfigure}
\begin{subfigure}{.19\textwidth} \centering \includegraphics[width=.95\linewidth]{img/{{BikeState_5Y_31612_31623_MemberUser_hod_Week}}} \caption{31612 $\rightarrow$ 31623} \label{fig:BikeState_5Y_31612_31623_MemberUser_hod_Week} \end{subfigure}
\begin{subfigure}{.19\textwidth} \centering \includegraphics[width=.95\linewidth]{img/{{BikeState_5Y_31200_31229_MemberUser_hod_Week}}} \caption{31200 $\rightarrow$ 31229} \label{fig:BikeState_5Y_31200_31229_MemberUser_hod_Week} \end{subfigure}

\begin{subfigure}{.19\textwidth} \centering \includegraphics[width=.95\linewidth]{img/{{BikeState_5Y_31201_31229_MemberUser_hod_Week}}} \caption{31201 $\rightarrow$ 31229} \label{fig:BikeState_5Y_31201_31229_MemberUser_hod_Week} \end{subfigure}
\begin{subfigure}{.19\textwidth} \centering \includegraphics[width=.95\linewidth]{img/{{BikeState_5Y_31623_31612_MemberUser_hod_Week}}} \caption{31623 $\rightarrow$ 31612} \label{fig:BikeState_5Y_31623_31612_MemberUser_hod_Week} \end{subfigure}
\begin{subfigure}{.19\textwidth} \centering \includegraphics[width=.95\linewidth]{img/{{BikeState_5Y_31611_31623_MemberUser_hod_Week}}} \caption{31611 $\rightarrow$ 31623} \label{fig:BikeState_5Y_31611_31623_MemberUser_hod_Week} \end{subfigure}
\begin{subfigure}{.19\textwidth} \centering \includegraphics[width=.95\linewidth]{img/{{BikeState_5Y_31623_31611_MemberUser_hod_Week}}} \caption{31623 $\rightarrow$ 31611} \label{fig:BikeState_5Y_31623_31611_MemberUser_hod_Week} \end{subfigure}
\begin{subfigure}{.19\textwidth} \centering \includegraphics[width=.95\linewidth]{img/{{BikeState_5Y_31613_31622_MemberUser_hod_Week}}} \caption{31613 $\rightarrow$ 31622} \label{fig:BikeState_5Y_31613_31622_MemberUser_hod_Week} \end{subfigure}

\caption{Comparison of the Trip Counts between Weekdays (Blue) and Weekends (Red) for the Top 20 Highest-Ranking O-D Pairs in O-D Flows by Member Users. The Stations 31121, 31200, 31613, and 31623 are Respectively Adjacent to Four Transportation Hubs (Metrorail and Railway Stations), i.e., Woodley Park, Dupont Circle, Eastern Market, and Union Station.}
\label{fig:BikeState_5Y_DiffOD_MemberUser_hod_Week}
\end{figure}

Fig.~\ref {fig:BikeState_5Y_DiffOD_Time_MemberUser_PDF} shows the empirical PDFs of use time for the top 20 O-D pairs in the ranking of the highest O-D flows by member users. Fig.~\ref{fig:BikeState_5Y_DiffOD_MemberUser_hod_Week} gives the comparison of average daily counts of the trips between weekdays (blue) and weekends (red) for the top 20 O-D pairs. Although the distributions of use time by member users for all the top 20 O-D pairs express a similar pattern on a utilitarian purpose --- with a single sharp peak that can be fitted into the lognormal distribution (see Fig.~\ref {fig:BikeState_5Y_DiffOD_Time_MemberUser_PDF}), their daily trip counts show diverse patterns (see Fig.~\ref{fig:BikeState_5Y_DiffOD_MemberUser_hod_Week}, especially the blue curves for weekdays on the same trip purpose). These patterns give us additional information on the land use near origin and destination stations (such as workplace or high-density residence) as well as more detailed trip purpose of an O-D pair $(s_o, s_d)$ (such as for commuting or for non-commuting). 

Here we present a few decision rules based on associated knowledge. First, a dominant AM or PM peak of trip counts in weekday indicates that the trip purpose between the O-D pair is for commuting. Let R, W, and T respectively be a residence, workplace, and transportation hub. Basic commuting trip segments include H $\rightarrow$ W, H $\rightarrow$ T, T $\rightarrow$ T, and T $\rightarrow$ W during the AM period, and W $\rightarrow$ H, W $\rightarrow$ T, T $\rightarrow$ T, and T $\rightarrow$ H during the PM period. If one and only one station (either $s_o$ or $s_d$) is T, we can define the following {\em commuting-related} rule ($R_C$): $s_o$ is H if $s_d$ is T and $s_d$ is W if $s_o$ is T given a dominant AM peak, whereas $s_o$ is W if $s_d$ is T and $s_d$ is H if $s_o$ is T given a dominant PM peak. 

Let us take the O-D pairs in Fig.~\ref{fig:BikeState_5Y_DiffOD_MemberUser_hod_Week} as an example to illustrate the usage of the $R_C$ rule. The Stations 31121, 31200, 31613, and 31623 are known respectively adjacent to four transportation hubs (Metrorail and railway stations) of Woodley Park, Dupont Circle, Eastern Market, and Union Station.  Based on the $R_C$ rule and the patterns shown in Fig.~\ref{fig:BikeState_5Y_DiffOD_MemberUser_hod_Week}, we can identify that the eight stations, i.e. Stations 31201, 31229, 31611, 31612, 31614, 31619, 31622, and 31631, are located in primary residence neighborhoods. This is consistent with the result in Fig. \ref{fig:Census_LODES2014_DC_grp_h_geocode_ct_Geo_Tract10}, where the LODES data by residence confirmed that seven and one of the eight stations are respectively located in the high-density residential census tracts with $\operatorname{log}_{10}(C_{h(ct)}/A_{ct})$ in $[4.0,4.5]$ and $[3.4, 4.0]$.  Among the top 20 O-D pairs in the ranking of the highest O-D flows by member users, 17 are the O-D links between the four T stations and the eight H stations, i.e. all the top O-D pairs in Fig.~\ref{fig:BikeState_5Y_DiffOD_MemberUser_hod_Week} are T $\rightarrow$ H or H $\rightarrow$ T except for Figs. \ref{fig:BikeState_5Y_31104_31121_MemberUser_hod_Week}, \ref{fig:BikeState_5Y_31121_31104_MemberUser_hod_Week} and \ref{fig:BikeState_5Y_31201_31229_MemberUser_hod_Week}. It indicates that bikesharing plays an important role in providing first- and last-mile connections between residence places and transportation hubs.

We can also define decision rules to identify the land use related to non-commuting trips. In the {nightlife-related} rule ($R_N$), $s_o$ or $s_d$ is likely located near a nightlife area, if there is a nontrivial trip rate during 0--4 AM in weekend. For example, in Fig. \ref{fig:BikeState_5Y_31104_31121_MemberUser_hod_Week}, the station 31104 is at Adams Morgan, which is a major night life area with many bars and restaurants. 

\subsection{Mobility} \label{sec:Mobility}

\begin{figure} [htb]
\centering

\begin{subfigure}{.49\textwidth} 
\centering \includegraphics[width=0.95\textwidth]{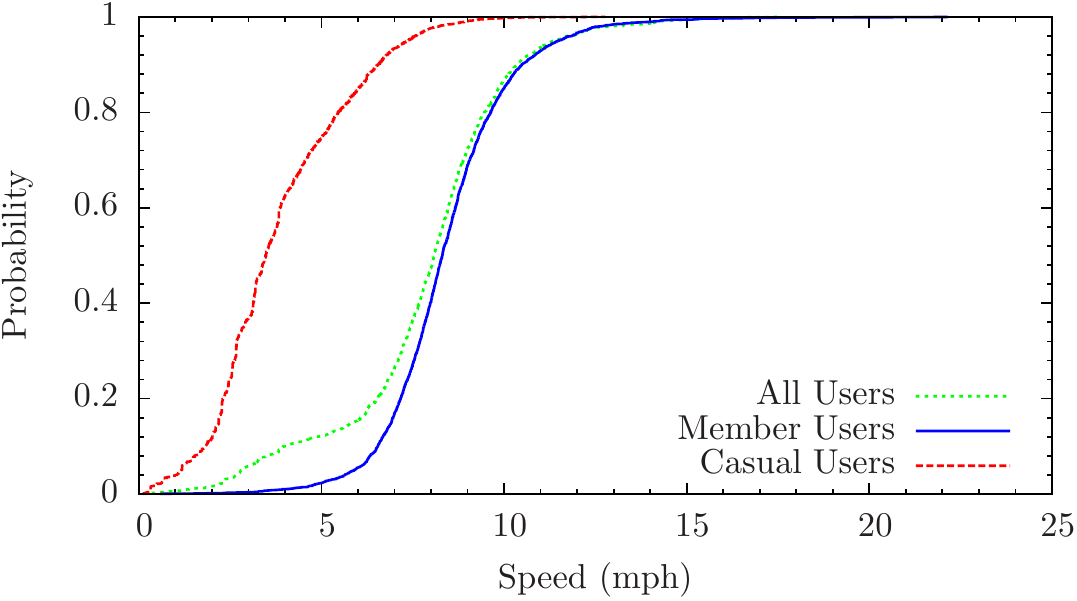} \caption{Full System in Terms of User Types.}
\label{fig:BikeState_5Y_med_osid_dsid_DiffOD_medSpeed_UserType_T1000_PDF_CDF}
\end{subfigure}
\begin{subfigure}{.49\textwidth} 
\centering \includegraphics[width=0.95\textwidth]{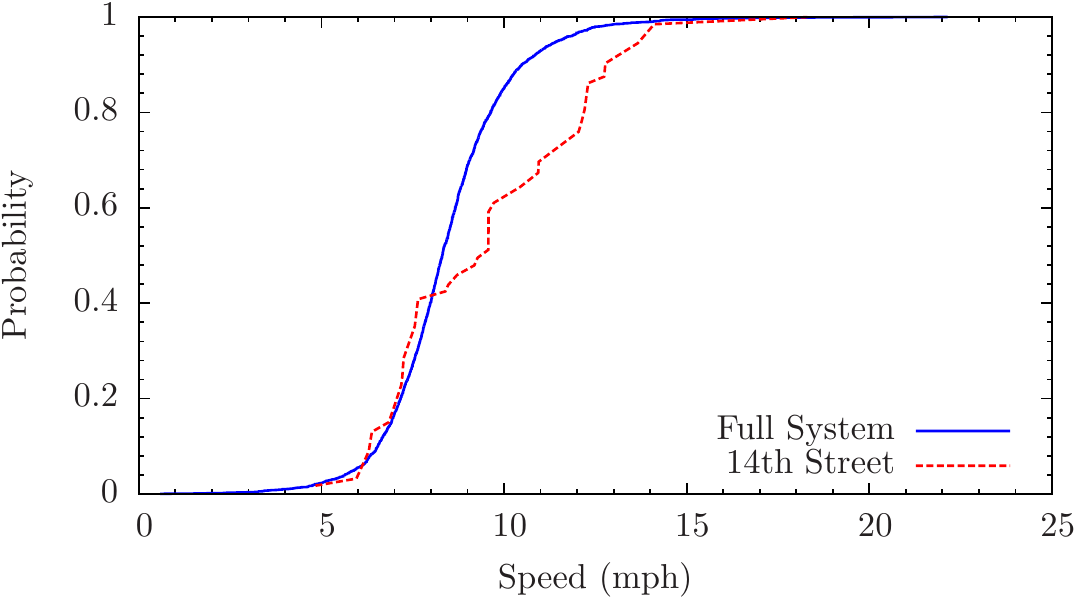} \caption{Member Users in Different Regions.}
\label{fig:BikeState_5Y_med_osid_dsid_Regions_medSpeed_Member_T1000_PDF_CDF}
\end{subfigure}

\caption{CDF Distributions of Median Travel Speed from O-D Trips.} \label{fig:BikeState_5Y_med_osid_dsid_medSpeed_T1000_PDF_CDF}
\end{figure}

On mobility, travel or trip speed is important. For each bicycle trip, its bikeshare use time provides an upper bound of travel time. Particularly, for the utilitarian O-D trips, the bikeshare use time approaches the travel time if there is no intermediate stop. Therefore, we can estimate the lower bound of trip speed for member users using the bikeshare use time and the shortest path length between each O-D pair, without requiring extensive GPS tracking data \citep{Wergin2017}. Fig. \ref{fig:BikeState_5Y_med_osid_dsid_DiffOD_medSpeed_UserType_T1000_PDF_CDF} gives the distribution of median speed $v_{med}$ for all the O-D pairs with more than 1000 trips taken by bikeshare users. For each O-D pair, we compute $v_{med}=L_R/T_{U,med}$, where $L_R$ is the distance of the recommended bicycling route between the O-D pair retrieved with Google Maps Distance Matrix API, $T_{U,med}$ is the median bikeshare use time for the O-D pair. For casual, member, and all users, the medians values of $v_{med}$ are respectively 3.48, 8.31, and 8.08 mph. Comparison between member and all users shows a significant difference in the 10th percentile values of $v_{med}$, which are respectively 6.54 and 4.00 mph.

Concerning the traditional vehicle flow, traffic congestion has been a serious problem \citep{schrank20152015}. Let the free flow, average, and 95th percentile travel times respectively be $TT_{O}$, $\overline{TT}$, and $TT_{95}$, the {\em Travel Time Index} (TTI) and {\em Planning Time Index} (PTI) are respectively
\begin{equation}
\text{TTI}=\overline{TT}/TT_{O},~~~~\text{PTI}=TT_{95}/TT_{O},
\end{equation}
as defined in the Urban Congestion Reports (UCR) by \cite{FHWA2017Congestion}. According to UCR, during the last quarter of 2016, DC had TTI=1.48 and PTI=3.08, and the average duration of daily congestion lasted for nearly 8 hours. According to \cite{DDOT2017Mobility}, some road segments have severe congestion with TTI $>$ 3.5 during peak time. Notice that the maximum lawful speed is 25 mph on most streets in DC. Thus, an extremely high TTI means a rather slow traffic speed. 

Congestion affects both travel time and reliability of vehicles~\citep{martin2014evaluating,stinson2004frequency}. In general, congestion factors do not have effects on bike users as much as on car and bus transit users. With the current median speed of about 8 mph, bicycling is certainly a viable commuting option in urban transportation, as compared to other surface transportation modes. In particular, if travel time reliability is important for users, bicycling might be a better choice during peak time. Compared to bicycling, although driving might save travel time in terms of the free flow speeds, the variation in vehicle travel time is rather large in rush hours with a high TTI and PTI for vehicles, where a delay can even reach several times of the travel time savings~\citep{asensio2008commuters}. According to the 2016 member survey, 89\% of respondents cited access and speed as their primary reasons for joining CaBi \citep{bikeshare2016capital}.

The speed for CaBi members is still far below than the average cycling speed that could reach 14.6 mph \citep{moritz1997survey}. It is likely due to the high density of intersections in urban area imposing unnecessary delays for cyclists. At this point, smart multimodal intersection control \citep{portilla2016model} can help to gain  a better biking mobility while without any significant interruption to existing vehicle flows. The improvement in bike mobility might further promote more bike usage as a part of the green intermodal transportation system, and accelerate the traffic mode shift towards reducing VMT~\citep{fishman2014barriers} and mitigating traffic congestion as well as vehicle emissions~\citep{schrank20152015,hamilton2017bicycle}.

It is important to understand the dependence of bike mobility on the network infrastructure. For urban planners and policymakers, such knowledge could provide critical supports to upgrade the bike infrastructure in multimodal urban transportation networks. To gain some insights, a case study is performed on a dedicated bike lane with bikeshare stations. The bike lane is on the 14th Street between Station 31407 at Colorado Ave and Station 31241 at Thomas Circle. There are 12 CaBi stations in total(i.e., 31101, 31105, 31119, 31123, 31124, 31202, 31203, 31241, 31401, 31402, 31406, 31407) on the bike lane, and the O-D trips among these stations are considered. Only member users are considered for evaluating the bike mobility of utilitarian trips. It is reasonable to assume that nearly all these bikeshare users will follow the route on the 14th Street, since it is the shortest route and it is a dedicated bike lane. Fig. \ref{fig:BikeState_5Y_med_osid_dsid_Regions_medSpeed_Member_T1000_PDF_CDF} shows the difference in bike mobility between the full system and the selected region on the 14th Street. Compared with the full system, there are more percentiles of users at the higher travel speeds on the 14th Street. For a travel speed over 10 mph, the percentages of users are respectively 37.68\% and 14.88\% on the 14th Street and in the full system. The result indicates that bike mobility could be significantly improved on dedicated bike lanes.

\subsection{Safety} \label{sec:Safety}

Concerning Safety, we focus on the bike crash data in DC. According to the combined data of the traffic safety statistics of 2013-2015 from \cite{DDOT2015SafetyFact} and the crash data from Open Data DC, there were 707.4 bike crashes per year in average between 2012-2016, where 46.8 bicyclists were fatal or major injured and 436.6 were minor injured. These crashes involved all bike riders, although it was reported that bikeshare users are associated with a lower crash risk as compared to other cyclists \citep{martin2016bikesharing}. For CaBi, there were 132 reported crashes (including 50 hospital injures) in its 9.16 million trips between 2012 to July 2015 \citep{martin2016bikesharing}.

\begin{figure} [h]
\centering

\begin{subfigure}{.328\textwidth} \centering \includegraphics[width=.95\linewidth]{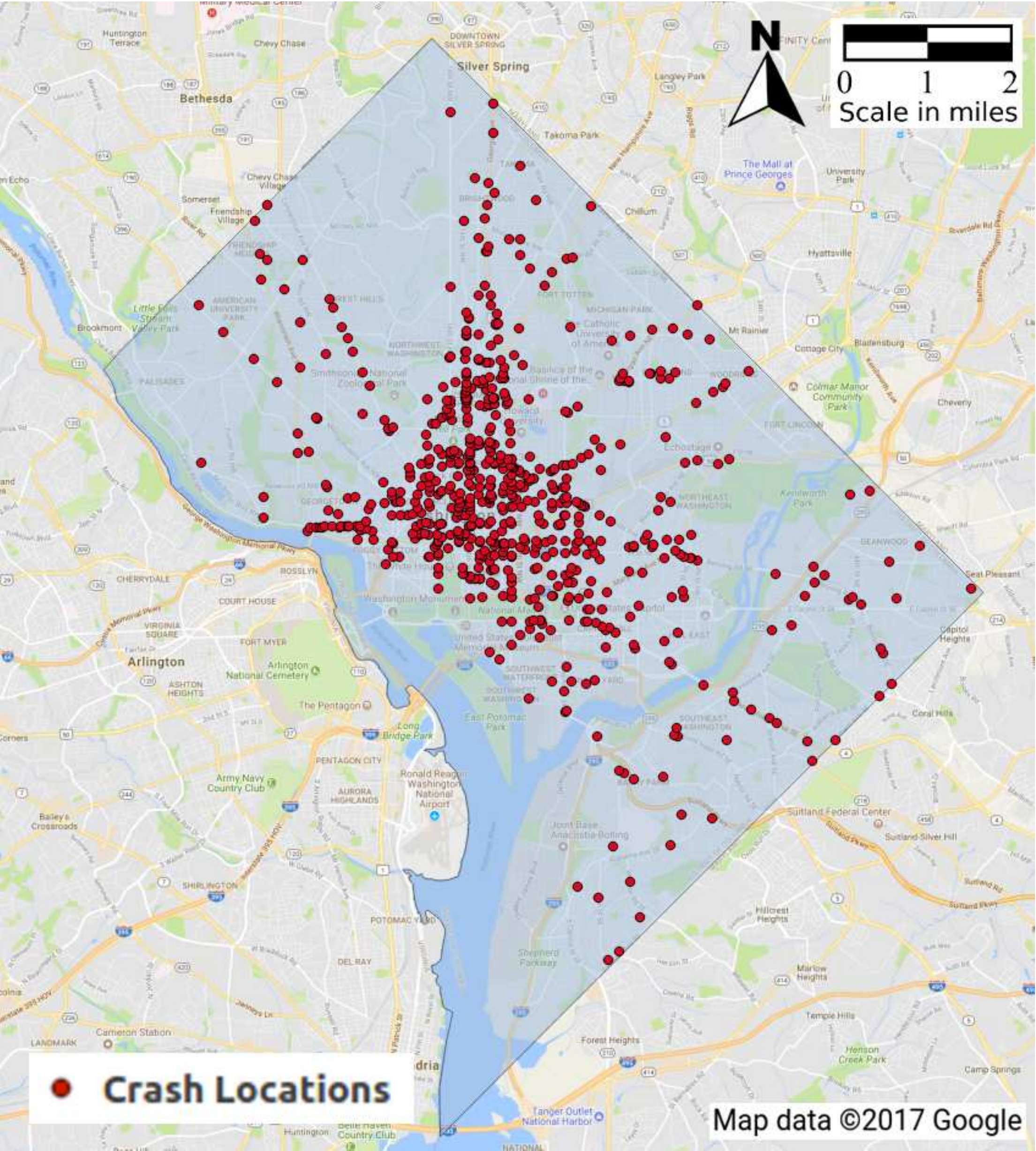} \caption{DC Bike Crashes in 2016} \label{fig:dc_crashes2_grp_Geo_TOTAL_BICYCLES_Y2016} \end{subfigure}
\begin{subfigure}{.32\textwidth} \centering \includegraphics[width=.95\linewidth]{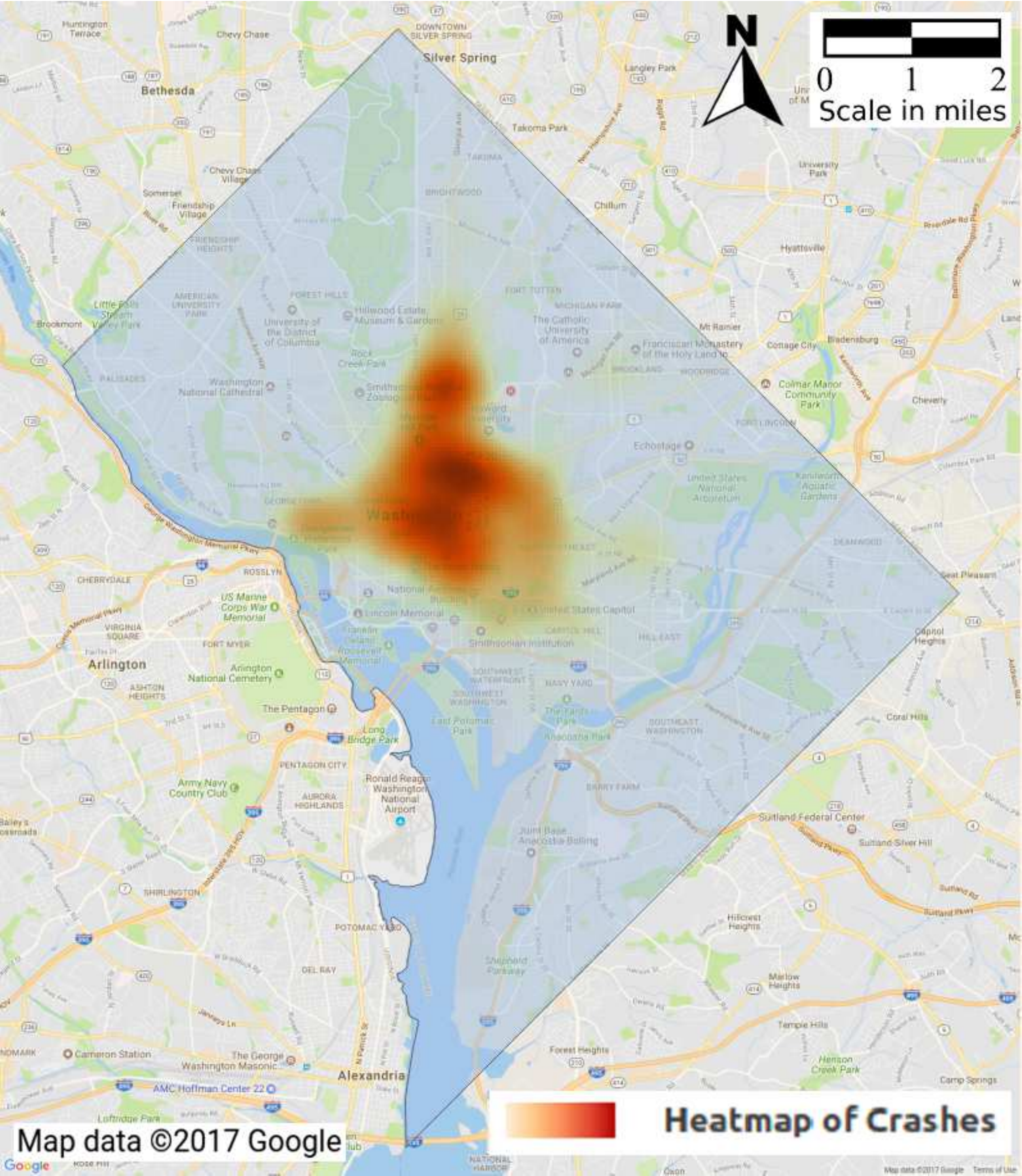} \caption{Crash Heatmap} \label{fig:dc_crashes2_grp_Geo_TOTAL_BICYCLES_Y2016_w_Heatmap} \end{subfigure}
\begin{subfigure}{.32\textwidth} \centering \includegraphics[width=.95\linewidth]{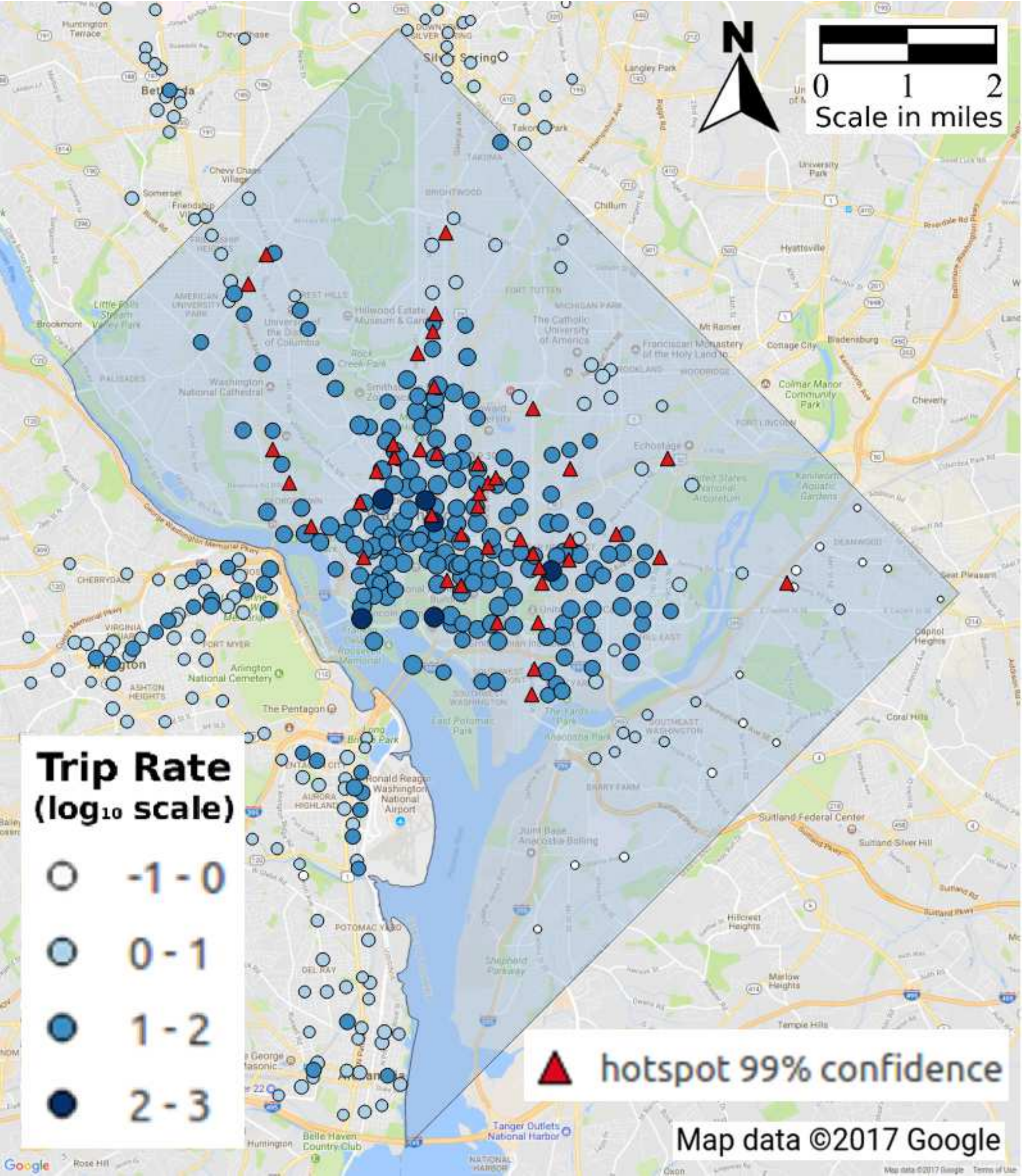} \caption{Demands and Crash Hot Spots} \label{fig:BikeState_5Y_Geo_grp_osid_DailyRate_BikeAccidentHotSpots} \end{subfigure}

\caption{Spatial Distribution of Bike Crashes in Washington DC.} \label{fig:Bike_Accident_16}
\end{figure}

In DC, there were 768 bicyclists in total involved in traffic crashes in 2016. Fig. \ref{fig:dc_crashes2_grp_Geo_TOTAL_BICYCLES_Y2016} shows the crash locations. The broad spatial distribution of crashes certainly gives a warning that bike safety requires a significant improvement. This is an important issue, as more people are adopting this active transportation mode nowadays. Fig.~\ref{fig:dc_crashes2_grp_Geo_TOTAL_BICYCLES_Y2016} also illustrates that the whole bike network is well connected in DC, since crashes are rare events in the total trips but they distributed widely in the map of DC. The heatmap of crashes is shown in Fig. \ref{fig:dc_crashes2_grp_Geo_TOTAL_BICYCLES_Y2016_w_Heatmap}, which provides us a visual summary of the most frequent crash regions.
Fig. \ref{fig:BikeState_5Y_Geo_grp_osid_DailyRate_BikeAccidentHotSpots} gives the significant crash hot spots in statistics, which are obtained using the method of Z scores from the Getis-Ord Gi* statistic with 99\% confidence. By including the trip demand information from Fig. \ref{fig:BikeState_5Y_Geo_grp_osid} into Fig. \ref{fig:BikeState_5Y_Geo_grp_osid_DailyRate_BikeAccidentHotSpots}, we find that the hot spots of bike crashes have a strong spatial correlation with trip demand. 

Safety has been a major barrier for people to adopt biking as a new form of mobility. Especially, bike users would have concerns when biking in traffic \citep{fishman2016bikeshare}. Besides education and enforcement, upgrading bike infrastructure and using technology can both help to reduce considerable risks or exposures and to lower the stress level \citep{lowry2016prioritizing} for cyclists on urban roads.

\section{Discussions}

\subsection{Implications for Data-Driven Decision Supports} \label{sec:ddds}

\begin{figure} [htbp]
\centering \includegraphics[width=0.99\textwidth]{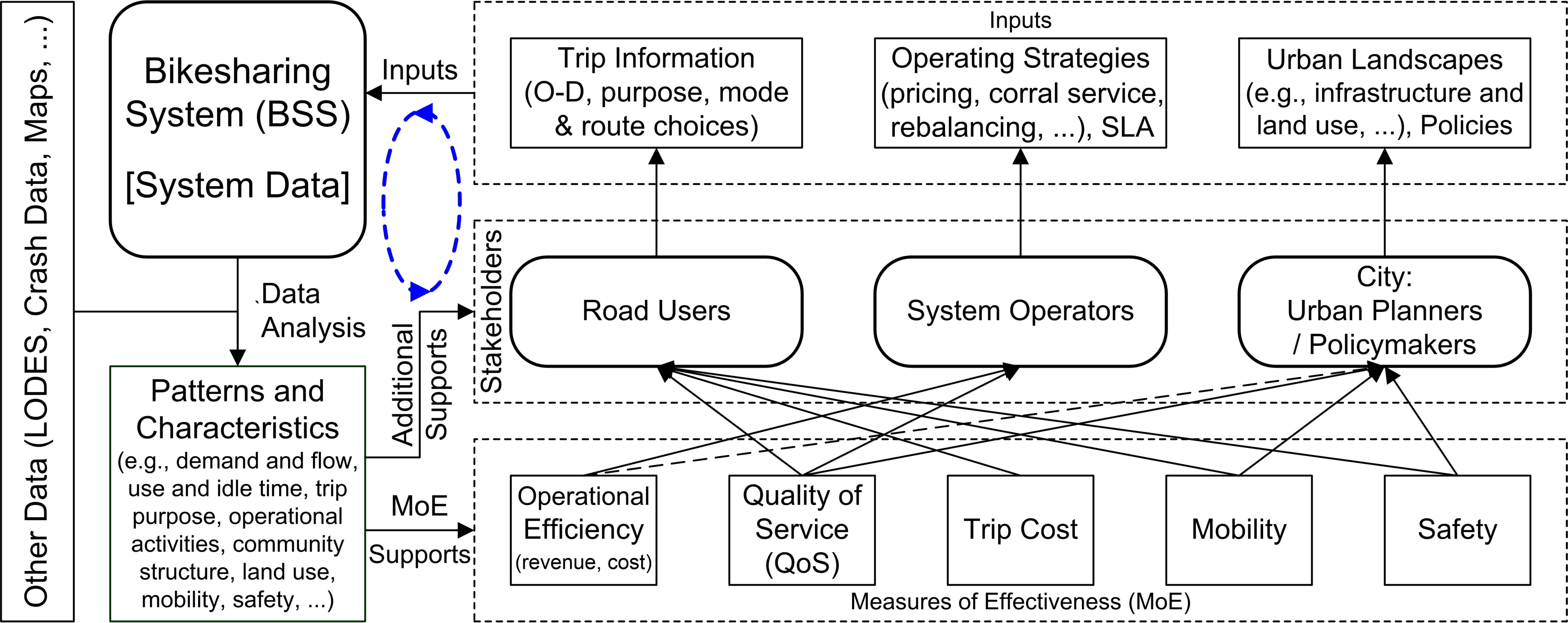} \caption{Data-Driven Decision Supports from Relations Between Bikesharing Systems and Stakeholders.}
\label{fig:BSSStakeholders}
\end{figure}

Fig. \ref{fig:BSSStakeholders} shows data-driven decision supports (DDDS) from the relations between BSS and key stakeholders for taking advantages of the mined travel patterns and characteristics from data, where the key stakeholders consist of road users, system operators, and city (urban planners and policymakers). The inputs from key stakeholders to BSS include trip information (such as O-D, purpose, mode and route choice), bikeshare operating strategies (such as corral service, rebalancing, and pricing) and service level agreements (SLA), and urban landscapes (such as land use and infrastructure) and policies. Data analysis is then performed upon the data sources integrating BSS and other related datasets, such as LODES, crash data, and data from Google Maps APIs. From the viewpoint of DDDS, the data analysis could provide supporting information to key stakeholders to help them achieve their goals. The process could be iterative where the inputs to BSS are updated over time, which would make key stakeholders benefit from the changes and also foster the transportation shift towards more sustainable urban mobility in a complex urban environment.

The patterns and characteristics extracted from data are able to provide supports for measures of effectiveness (MoE) to stakeholders through tracking key metrics of system on an ongoing basis. 
We consider essential MoEs in Fig. \ref{fig:BSSStakeholders} and discuss them mostly in the context of the CaBi system. Note that DDDS is extensible, and new MoEs could be incorporated into DDDS implementations according to specific goals while the relevant data are available. 
The use time ratio of bikeshare (Section \ref{sec:UseTime}) can be used by system operators for understanding or estimating their operational efficiency. The demand-supply imbalance (Section \ref{sec:TripDemand}) can function as an alert on the need of rebalancing to system operators for a better QoS of bikeshare. The use time  distribution in Fig. \ref{fig:BikeState_5Y_User_Idle_Time_PDF_CDF} indicates that most road users are sensitive to trip time as additional trip time costs them more usage-time-based fee. Concerning QoS, although SLA is provided by operators and has to be adopted by individual users, it is often negotiated between city and system operators \citep{de2016bike} if they are in public-private partnerships (which is in fact a usual case). The evaluations on mobility and safety in Sections \ref{sec:Mobility} and \ref{sec:Safety} have straightforward impacts on miscellaneous transportation decision makings by key stakeholders at different effecting weights. Notice that the metrics of mobility and safety represent aggregated experiences of road users, which can be improved by city planners or policymakers through upgrading urban landscapes or adopting new technologies to reduce number of bike crash incidents or encouraging safety culture among road users.

For system operators, a higher operational efficiency is corresponding to a higher revenue and/or a lower operating cost. The revenue of BSS system depends on the pricing scheme and ridership. The trip cost of road users depends on the pricing scheme and trip time. If BSS improves its operational efficiency by reducing its operating cost (e.g., through optimizing rebalancing activities \citep{shui2017dynamic,elhenawy2018heuristic}), some flexibility in adjustment of the pricing scheme would be available to attract more ridership from road users, and the revenue would be improved in return. According to \citet{DDOT2015CabiPlan}, although the cost recovery rate of CaBi is projected to increase from 75\% (FY2016) to 87\% (FY2021), its operating balance might stay in negative in the foreseeing future. 
Notice that BSS could bring us public benefits from social, environmental, economic, and health-related aspects that may not be shown in the operating balance~\citep{hamilton2017bicycle,DDOT2015CabiPlan,shaheen2013public,fishman2016bikeshare}. There are many usable business models as surveyed in \cite{shaheen2013public}.
In the case that BSS needs subsidy from the city, the city may also consider the operational efficiency as one of MoEs. Notice that one of important goals of the city is to achieve sustainable urban transportation, for which reducing private car usage through increasing bike ridership in road users is a highly viable way and tightly associated with the operational efficiency of BSS. The indirect yet tight relation between the city and BSS operational efficiency is indicated with a dash line in Figure \ref{fig:BSSStakeholders}.

To be concise, we do not show relatively loose relations among the components of DDDS in Fig. \ref{fig:BSSStakeholders}. For example, people's preferences for biking facilities may impact the quality of land use through changing the values of businesses \citep{buehler2015business,Poirier2018Bicycle} and residential properties \citep{Liu2017Impact}. The reduction in trip costs and the increase in mobility and safety would attract more road users to adopt BSS, which might potentially raise operational efficiency of BSS. System operators can enhance biking safety through adding protection or safety alert accessories to bikes, such as installing self-powered LED lights to provide more visibility during riding.

Some MoEs could involve conflicting goals. For instance, on operational efficiency, system operators might explore options to reduce QoS for saving the high maintenance and service costs \citep{demaio2009bike}, or pursue a higher payment of trip cost from users through increasing membership rates or usage fees \citep{DDOT2015CabiPlan}, which however might reduce total ridership. The total impacts of the operation on efficiency should be carefully evaluated given any conflicts are entailed in the operation. Another example is on usage of helmet. Bicycle helmet has been shown effective to protect bike users and improve road safety \citep{attewell2001bicycle}. Although the law in DC does not enforce the usage of bike helmet for the people above the age of 16, \cite{DDOT2015CabiPlan} has a target to increase the percentage of CaBi riders wearing helmet year by year. As shown in the references~\citep{shaheen2013public,fishman2016bikeshare}, bikeshare users demonstrate a strong reluctance to wear helmet. If without sufficient improvements in the practical conditions, the policies aiming to increase road safety through more helmet usage might conflict with that of augmenting total ridership and operational efficiency of BSS. For one more example, electric bicycles, or e-bikes, would help increasing the usage of bikesharing \citep{pucher2017cycling} as they could lower the activity burdens for seniors and long-distance travelers, could make cycling easier in hilly areas, also could significantly improve mobility for all bike users; however, on road safety, e-bike users have shown more likely to be involved in severe crashes than classical bike users \citep{schepers2014safety}. For achieving overall benefits among difference goals, constraints and relations should be considered and the total effect could be evaluated as data-driven decision support.

Beyond providing MoE supports, the patterns and characteristics can be used by stakeholders as additional decision supports. Effectiveness of some operating activities such as corral service, which is shown evaluable from the patterns and characteristics in drop-off/pickup counts at different stations or for different events (Section \ref{sec:OperationalActivity}), can be used by system operators to gauge the impact of their operating activities, to evaluate their operating cost, and to augment operational efficiency and QoS. The empirical PDFs of O-D trip time, which are shown able to be fitted by parametric models in Fig. \ref{fig:BikeState_5Y_DiffOD_Time_MemberUser_PDF} (Section \ref{sec:UseTime}), can be used not only by system operators for operational simulations, but also by road users for trip planning. The trip flow imbalance in Fig. \ref{fig:BikeState_5Y_Geo_grp_osid_ToD_DoW} (Section \ref{sec:TripDemand}) and the flow patterns in Fig. \ref{fig:BikeState_5Y_DiffOD_MemberUser_hod_Week} (Section \ref{sec:odFlows}), which are both shown associated with the information of land use, can be used by system operators to optimize operating efforts and by city to improve urban landscapes. Trip purpose, which is shown extractable in Section \ref{sec:UserTripTypes}, can greatly influences travel behavior and mode choice. The advantage of building biking infrastructure such as dedicated bike lanes in enhancing biking mobility, as shown in Section \ref{sec:Mobility}, can be used by system operators to plan or develop bikeshare networks and by city to develop or improve biking infrastructure. Significant improvement of biking mobility and safety is critical for road users to determine whether or not to adopt biking as their transportation mode, and it would highly affect if city could win a mode shift to have the VMT and traffic congestion reduced.

Fig. \ref{fig:BikeState_5Y_CasualUser_TopRoutes} shows the community structure emerged from the O-D flows of bikeshare users. It reveals a self-organized formation of polycentricity in urban spatial structure \citep{anas1998urban} as the result of the addition of a BSS into urban landscapes of city. Traditionally, traffic congestion is often considered as the most serious negative externality \citep{louf2014congestion} leading to the formation of polycentricity that helps to keep urban spatial structure efficient. In the bikesharing network, the primary negative externality however appears to be trip time, as most people would try to keep their biking trips done within 30 minutes in order to retain a low trip cost under the current price scheme of CaBi. Fig. \ref{fig:BikeState_5Y_CasualUser_TopRoutes} indicates that the existing CaBi coverage on communities is rather small, limited to only a small portion of DC area. It is also shown in Fig. \ref{fig:BikeState_5Y_Geo_grp_osid} that the trip rates of CaBi stations in quite a large area of DC are lower than the rate in the bikeshare-centric areas by a few orders of magnitude. This problem can considerably reduce the overall operational efficiency of the BSS, but simple solutions such as eliminating the low-use bikeshare stations \citep{garcia2012optimizing} may increase inequity in accessibility to bikeshare service. From the viewpoint of polycentricity, there are some potential solutions, such as, 1) foster the formation of new communities in the bikeshare low-use area, which can be taken into account by system operators when some bikeshare stations would be relocated or new pricing schemes would be implemented, or by city when urban landscapes would be updated; and 2) improve biking mobility to extend the biking distance accessible by road users at the same time limit, which can significantly expand biking communities to cover broader areas. In addition, if an adjustment of current pricing scheme could be made by system operators to encourage longer trips beyond the existing communities, it would help solving the problem. 
 
It is worthy to note that data itself is incomplete and is only one of the inputs providing supports to the complex decision making for the public good. To prevent potential risks from data-driven decision making \citep{lepri2017tyranny}, we must adopt a human-centric perspective, since people are the subjects of the decisions. Moreover, stakeholders might have different perspectives and weights on some key MoEs. For example, the pursue of operational efficiency by system operators might impact transportation equity given any disadvantages were induced to some groups of road users \citep{lee2017understanding}. Due to the conflicts and differences in perspectives, negotiations among different stakeholders should be included in the policy decision-making process to achieve widely acceptable solutions or service terms that could be written into SLA. 

DDDS adopts the way of iterative improvement process in real-world decision making and policy implemention \citep{mclaughlin1987learning}. 
As indicated (by the blue dash cycle) in Fig. \ref{fig:BSSStakeholders}, the DDDS can iteratively incorporate all the changes and inputs from the key stakeholders into BSS to update the function of BSS. The patterns and characteristics as the outputs of each updated BSS bring up the MoEs and supports to date and serve as the information sources to the key stakeholders, leading to updated decision makings by the key stakeholders and new changes in the inputs to BSS in the next iteration. The iterative process can link the inputs from the key stakeholders with the patterns and characteristics in a systematic way for BSS. This is important for BSS optimization in practice. For example, the changes may be high-cost (for example, implementing bike infrastructure), may significantly impact road users, or both; however, their potential effects are rather difficult to assure in advance, based on incomplete information or in complex urban environments. In this case, the changes could be exploited first in a small scale as a pilot, from which the data as the responses from road users on the changes could be collected and analyzed to gain patterns and characteristics providing evidence-based supports to evaluate the effect and efficiency of the performed changes. By gradually expanding the usage of the changes in the DDDS, the iterative improvements of BSS would allow city policymakers and BSS system operators progressively evaluate the impacts of changes, learn from experiences, and make their decisions of optimizing BSS at a lower risk. 

\subsection{Contributions on Travel Patterns and Characteristics}

Besides the MoEs and evidence-based decision supports in DDDS as discussed in Section \ref{sec:ddds}, some additional contributions adding to the body of knowledge on travel behaviors are offered by the systematic examination of new patterns and characteristics in BSS. Here we provide some brief discussions.

In Section \ref{sec:TripDemand}, we have studied the cause of demand-supply imbalance in BSS by combining the BSS data and the LODES data from \cite{LEHD2017LODES}. As one of the main challenging problems in BSS operations, bike redistribution problem has been illustrated in the earlier work of \citet{vogel2011understanding} and \citet{o2014mining}. Departing from the existing studies, we have focused on the underlying cause of the redistribution problem, i.e., the demand-supply imbalance of BSS. We transformed the patterns and characteristics to the useful information for building the relation between the imbalance and the spatial distributions of workplace and residence that than reflect the commuting behaviors during peak periods. From the viewpoint of DDDS, although the demand-supply imbalance is output by road users, it is in fact the result of the users' best responses to the inputs from system operators and policymakers. Using the information as decision support, BSS operators can optimize their rebalancing efforts, and policymakers can make strategic policy addressing the rebalancing problems more effectively (for example, one of useful policies is to encourage mixed land use through zoning on workplace and residence). 

In Section \ref{sec:OperationalActivity}, we have analyzed the impacts of some real-world operating activities. Various operating strategies \citep{shaheen2013public,de2016bike} have been implemented in different cities. Most previous research on operating activities focused on route optimization of rebalancing vehicles for reducing cost including emission \citep{shui2017dynamic,elhenawy2018heuristic} and lost demand \citep{ghosh2017dynamic}. Our work have focused on evaluating qualitative and quantitative impacts on specific operating activities, including corral services and accompanying rebalancing efforts, at specific BSS stations. 
From the viewpoint of DDDS, operating activities are the inputs from BSS operators. Using the results as decision support, BSS operators can re-allocate their services to optimize overall operating efficiency and QoS at BSS stations. 

In Section \ref{sec:UseTime}, we have separated use, maintenance (e.g., rebalancing and bike repairing) and idle times based on practical conditions and extracted their distributions in parametric models. In the recent work of \citet{faghih2017empirical}, the separation between user and rebalancing trips was conducted by applying an approximated heuristic approach on aggregate data in five-minute time intervals, which could lead to a significant underestimation of the bike usage by users. 

In Section \ref{sec:UserTripTypes}, we have shown that trip purposes are likely associated with user and trip types in the trip data in DC. Specifically, member users and O-D trips are likely associated with utilitarian trips, whereas casual users and O-O trips are likely associated with recreational trips. Based on the survey data \citep{buck2013bikeshare}, CaBi member and short-term users are likely to cycle for utilitarian and recreational trip purposes, respectively. From the study on GPS trajectory data of 3,596 trips by CaBi users \citep{Wergin2017}, casual trips are longer in trip time and mostly around the National Mall, whereas member trips are faster and mostly in popular mixed-use neighborhoods. Similar geospatial distributions for causal and member trips have been shown in Fig. \ref{fig:BikeState_5Y_CasualUser_TopRoutes}. Our study on empirical use time distributions has shown that respectively 84.84\% and 97.64\% of O-O and O-D trips are below 30 minutes for member users for retaining a low trip cost under the current price scheme of CaBi. Such user behavior might lead to the self-organized formation of polycentricity in Fig. \ref{fig:BikeState_5Y_CasualUser_TopRoutes} from the perspective of urban spatial structure \citep{anas1998urban}.  

Our study has also disclosed some additional information on the broad usage of bikeshare in multimodal transportation environments, based on the analysis on the top O-D pairs with the highest flows by member users in Section \ref{sec:odFlows}. By applying a few simple rules, we have found that most of the top O-D pairs in BSS provide first- and last-mile connections between transportation hubs and residence neighborhoods in DC. Such multimodal connections indicate a complementary relationship \citep{barber2018unraveling} between bikeshare and transit that is potentially beneficial to cyclists and operators \citep{Ravensbergen2018Biking}.
The results in Fig.\ref{fig:BikeState_5Y_DiffOD_Time_MemberUser_PDF} indicate that bike trip time has low variations between O-D pairs, and the parametric models of O-D trip time between BSS stations could be adopted and used for real-time intermodal planning \citep{griffin2016planning} by road users. 

On bike mobility, we have derived the travel speed information using the BSS data and an open API from \citet{GMap2017API}. Traditionally, mobility information could be estimated from survey data \citep{moritz1997survey} or extracted from GPS trajectory data \citep{Wergin2017}, but the two methods may be considerably constrained respectively by the subjective sampling and by some challenges in data collection such as privacy concerns \citep{seidl2016privacy}, data volume \citep{romanillos2016big}, and etc. Compared to existing methods, our method provides a relatively objective and economic way to evaluate bike mobility in BSS, especially for utilitarian trips. The underlying assumption of our method is that road users would follow the shortest bike route recommendation by Google Maps. In practice, most utilitarian users, e.g., commuters, would mostly stick to 2-3 near optimal routes in their trips \citep{Anowar2017}, as travel time has the most importance to these users \citep{bikeshare2016capital,Anowar2017} and near optimal routes would often have similar travel time with the shortest bike route in urban road networks.
The patterns and characteristics also show that travel speed could be significantly improved by using the dedicated bike lane, which provides an important decision support for policymakers to choose the option in bike infrastructure improving mobility in biking networks for the city and fostering the mode shift toward a reduced private car usage. Notice that safety has been a major barrier to the adoption of biking as a new form of mobility for road users \citep{fishman2016bikeshare}. We should keep in mind that bike safety requires significant improvements, according to the broad spatial distribution of bike crashes and the strong correlation between bike crashes and trip demand, as indicated from the results in Section \ref{sec:Safety}.

\subsection{Generalizability to Other Systems}

It would be valuable to know that how much the data analysis and results from one city could be generalized to other systems in different cities. From the perspective of data availability, the datasets used in this analysis are essential and commonly available for most existing systems. The analysis methods and processes generating the outputs of patterns and characteristics are applicable to other different systems. However, in consideration of the potential differences in the inputs of key stakeholders among different systems, we should be very careful to generalize interpretations on different parts of the results regarding MoE and other supports. Let us first take the BSS redistribution as an example. The redistribution problem has been illustrated in different cities from the studies of \citet{vogel2011understanding} and \citet{o2014mining}. In Section \ref{sec:TripDemand}, we have found that the demand-supply imbalance in BSS station clusters is caused by the commuting behavior of road users during AM and PM peak periods and can be reflected from the underlying nonuniform and spatially distributions of workplace and residence clusters, which commonly exists in other cities from the LODES data \citep{schleith2014commuting}. In Section \ref{sec:OperationalActivity}, we have examined the corral service and related bike rebalancing behaviors to analyze the impacts of operating activities. The analysis is generalizable to the other cities only if they have similar valet and corral services \citep{de2016bike}. Next, let us briefly summarize the other findings presented in Section \ref{sec:anaysis_results} on generality. The results are applicable to other cities as data-driven decision supports but at different supporting extents to different stakeholders. The analysis of use and idle times in Section \ref{sec:UseTime} is quite general for different systems, and it would be interesting to examine the differences in the parameters of the model among different systems. In Section \ref{sec:UserTripTypes}, we have focused on understanding the trip purposes for different user and trip types, of which the generality is under the condition of using the 30-minute free-of-charge pricing scheme that has been adopted by most BSS systems \citep{shaheen2013public}. 
Notice that changes in pricing scheme would have substantial impacts on the BSS usage patterns \citep{wu2017explore}. Investigating the other systems using different pricing schemes is definitely beneficial to BSS planning and operations, but it is not in the scope of this paper. In Section \ref{sec:odFlows}, the analysis demonstrates a self-organized formation of urban spatial structure \citep{anas1998urban} in a monocentric or polycentric form depending on the underlying urban landscapes of a city. The analysis on O-D trips discloses the broad usage of bikeshare in multimodal transportation systems to connect with major transportation hubs (Metrorail and railway stations), highlighting the important role of bikesharing on providing first- and last-mile connections between residence places and transportation hubs in DC area. Such multimodal connections can be generalized to the other cities possessing rapid transit systems \citep{barber2018unraveling,adnan2018preferences}. The analysis in Section \ref{sec:Mobility} can be applied to other cities to gauge the attractiveness of biking mode from the perspective of improving mobility, especially to the cities where road users are suffering heavy traffic congestion from vehicle flow. The analysis in Section \ref{sec:Safety} provides general understanding of road safety for bike users.

The implementations of DDDS could be varying for different cities, in consideration of the difference in stakeholders, demographics, urban landscapes, goals, data sources and etc. On the shared aspects among cities, such as pricing schemes and trip purposes, integrating data across cities would be helpful to identify the hidden factors in similar settings or to understand the key impacts of different settings.

\section{Conclusions} \label{sec:conclusions}

Aiming to improve bikeshare and help the transformation of urban transportation system to be more sustainable, we conducted a comprehensive analysis to examine underlying patterns and characteristics of a bikeshare network and 
to acquire implications of the patterns and characteristics for data-driven decision supports (DDDS). As a case study, we used the trip history from the CaBi system in the Washington DC area as our main data source of data analysis, and other data, including Google Maps APIs, the LODES data and the crash data in Open Data DC, as auxiliary data sources to extract related information. With appropriate statistical methods and geographic techniques, we mined travel patterns and characteristics from data on seven important aspects for BSS, which include trip flow and demand, operating activities, use and idle times, trip purpose, O-D flows, mobility, and safety. For each aspect, we explored the results to discuss qualitative and quantitative impacts of the inputs from key stakeholders of BSS on main MoEs such as trip costs, mobility, safety, quality of service, and operational efficiency, where key stakeholders include road users, system operators, and city planners and policymakers. We also disclosed some new patterns and characteristics of BSS to advance the knowledge on travel behaviors.

On trip demand and flow, we showed spatial and temporal patterns in trip demand of bikeshare, and found that trip flow of bikeshare follows a scale-free power-law distribution---a common pattern for modeling human mobility. We computed demand-supply ratio for each station, and discussed the advantages of using the ratio as a complementary method for identifying bikeshare demand-supply imbalance. Moreover, by combining CaBi and LODES data, we found that the clustered regions reflecting bikeshare demand-supply imbalance result from large-scale human mobility behaviors, such as commuting. 
On bikeshare operations, we investigated the effects and implications of operating activities by taking the corral service provided by CaBi as an example. The results on the regular corral service for high-demand seasons provided straightforward MoEs for different stations, which as evidence-based supports can help system operators to make redistribution strategies among different stations for a better operational efficiency. It turned out that more rebalancing efforts would be needed to redistribute extra bikes. The results on the special corral service for high-attendance events demonstrated that corral service is a useful and effective operating activity for encouraging more people to choose biking to participate events. 
On use and idle times, our study indicated that the empirical distributions of the times can be well fitted into parametric models, which can be used for simulation studies. For CaBi, the use time ratio is 3.33\%, meaning that its operational efficiency could be improved enormously. For the improvement, we briefly discussed its concerns and potential solutions. 
On trip purpose, although it is known that biking is usually used for two types of purposes --- utilitarian or recreational, CaBi does not provide any direct corresponding records. We showed that trip purposes can be extracted from data analysis through identifying the differences of patterns in trip flows and in use-time distributions of bikeshare. 
On O-D flows, we found that core clusters and community structure can be identified by analyzing the top O-D pairs in the ranking of the highest O-D flows in a bikeshare network. In addition, we found that although the use-time distributions for the top O-D pairs by member users all follow a typical pattern of utilitarian trips, the O-D trip flows exhibit diverse patterns and can provide us more information of trip purpose and land use.
On biking mobility, we performed statistical analysis on the O-D pairs of primary utilitarian trips to compute the distributions of trip speeds. For the primary utilitarian trips in the bikeshare network, the median speed extracted from the data analysis is 8.31 mph. The biking mobility is competitive as compared with driving a car in over-congested urban areas, but still has a big room to improve. The speed would be enhanced on dedicated bike lanes.
On biking safety, we performed spatial analysis on biking crash data and found a strong spatial correlation between trip demand and the hot spots of bike crashes. We thereupon briefly considered the means to improve biking safety.  

Finally, we discussed and summarized the values of our findings for promoting biking as a frequently used transportation mode and transforming our urban mobility into a better and more sustainable transportation system. We discussed some critical roles and implications of the patterns and characteristics for DDDS from the relations between BSS and key stakeholders. We briefly discussed the importance of adopting a human-centric perspective to the usage of DDDS and of considering negotiations among stakeholders in policy decision making process. We summarized the new patterns and characteristics of BSS disclosed in this study and the added knowledge for bridging the gap between current understanding on patterns and characteristics of BSS and the needs from modeling and applications on turning the patterns and characteristics into evidence-based decision supports in the context of DDDS. We also discussed how much the analysis and results from the current study can be generalized to the BSS systems in other cities.

Several aspects of the current work warrant further study. First, data-driven solutions might be developed to facilitate the operations, maintenance, and expansions of BSS. Second, for urban planners and policy makers, data-driven recommendations might be provided to support or assist improving the bicycle infrastructures and upgrading the multimodal urban transportation systems for a better safety and mobility and expanding the coverage of community structure efficiently. Finally, there is a huge benefit from the data fusion combining bikeshare data with other data sources such as surveys or participatory sensing data, especially for better understanding the driving forces for the shift of transportation towards more sustainable urban mobility.

\section*{References}
\bibliographystyle{apacite}

\begin{thebibliography}{}

\bibitem[\protect\citeauthoryear{Adnan, Altaf, and Bellemans}{Adnan
  et~al.}{2018}]{adnan2018preferences}
Adnan, M., S.~Altaf, and T.~Bellemans (2018).
\newblock Preferences of travelers for using pro-rail bikesharing system for
  their last-mile travel in small/medium sized cities of {Belgium}.
\newblock Paper presented at: {\em Transportation Research Board (TRB) Annual Meeting}, Number
  4616, Washington, DC, 2018. Washington, DC: Transportation Research Board.

\bibitem[\protect\citeauthoryear{Anas, Arnott, and Small}{Anas
  et~al.}{1998}]{anas1998urban}
Anas, A., R.~Arnott, and K.~Small (1998).
\newblock Urban spatial structure.
\newblock {\em Journal of Economic Literature\/}~{\em 36\/}(3), 1426--1464.

\bibitem[\protect\citeauthoryear{Anowar, Eluru, and Hatzopoulou}{Anowar
  et~al.}{2017}]{Anowar2017}
Anowar, S., N.~Eluru, and M.~Hatzopoulou (2017).
\newblock Quantifying the value of a clean ride: How far would you bicycle to
  avoid exposure to traffic-related air pollution?
\newblock Paper presented at: {\em Transportation Research Board (TRB) Annual Meeting}, Number
  3351, Washington, DC, 2017. Washington, DC: Transportation Research Board.

\bibitem[\protect\citeauthoryear{Asensio and Matas}{Asensio and
  Matas}{2008}]{asensio2008commuters}
Asensio, J. and A.~Matas (2008).
\newblock Commuters' valuation of travel time variability.
\newblock {\em Transportation Research Part E\/}~{\em 44\/}(6), 1074--1085.

\bibitem[\protect\citeauthoryear{Attewell, Glase, and McFadden}{Attewell
  et~al.}{2001}]{attewell2001bicycle}
Attewell, R.~G., K.~Glase, and M.~McFadden (2001).
\newblock Bicycle helmet efficacy: A meta-analysis.
\newblock {\em Accident Analysis \& Prevention\/}~{\em 33\/}(3), 345--352.

\bibitem[\protect\citeauthoryear{Barber and Starrett}{Barber and
  Starrett}{2018}]{barber2018unraveling}
Barber, E. and R.~Starrett (2018).
\newblock Unraveling the relationship between bike share and rail transit use:
  A {Chicago} case study.
\newblock Paper presented at: {\em Transportation Research Board (TRB) Annual Meeting}, Number
  5682, Washington, DC, 2018. Washington, DC: Transportation Research Board.

\bibitem[\protect\citeauthoryear{Bhat, Astroza, and Hamdi}{Bhat
  et~al.}{2017}]{bhat2017spatial}
Bhat, C.~R., S.~Astroza, and A.~S. Hamdi (2017).
\newblock A spatial generalized ordered-response model with skew normal kernel
  error terms with an application to bicycling frequency.
\newblock {\em Transportation Research Part B\/}~{\em 95}, 126--148.

\bibitem[\protect\citeauthoryear{Buck, Buehler, Happ, Rawls, Chung, and
  Borecki}{Buck et~al.}{2013}]{buck2013bikeshare}
Buck, D., R.~Buehler, P.~Happ, B.~Rawls, P.~Chung, and N.~Borecki (2013).
\newblock Are bikeshare users different from regular cyclists? {A} first look
  at short-term users, annual members, and area cyclists in the {Washington,
  D.C.,} region.
\newblock {\em Transportation Research Record\/}~{\em 2387}, 112--119.

\bibitem[\protect\citeauthoryear{Buehler and Hamre}{Buehler and
  Hamre}{2015}]{buehler2015business}
Buehler, R. and A.~Hamre (2015).
\newblock Business and bikeshare user perceptions of the economic benefits of
  {Capital Bikeshare}.
\newblock {\em Transportation Research Record\/}~{\em 2520}, 100--111.

\bibitem[\protect\citeauthoryear{{Capital Bikeshare}}{{Capital
  Bikeshare}}{2017a}]{CaBi2017Corrals}
{Capital Bikeshare} (2017a).
\newblock {\em Bike Corrals}.
\newblock Washington, DC: Capital Bikeshare.
\newblock Available online at:
  \url{https://help.capitalbikeshare.com/hc/en-us/articles/115001475712-Bike-Corrals}
  (Accessed: 2017-11-10).

\bibitem[\protect\citeauthoryear{{Capital Bikeshare}}{{Capital
  Bikeshare}}{2017b}]{CaBi2017Data}
{Capital Bikeshare} (2017b).
\newblock {\em System Data}.
\newblock Washington, DC: Capital Bikeshare.
\newblock Available online at:
  \url{http://www.capitalbikeshare.com/system-data} (Accessed: 2017-11-10).

\bibitem[\protect\citeauthoryear{Cesme, Dock, Westrom, Lee, and Barrios}{Cesme
  et~al.}{2017}]{cesme2017data}
Cesme, B., S.~Dock, R.~Westrom, K.~Lee, and J.~A. Barrios (2017).
\newblock Data-driven urban performance measures: Case study application in the
  District of Columbia.
\newblock {\em Transportation Research Record\/}~(2605), 45--53.

\bibitem[\protect\citeauthoryear{Conley and Agrawal}{Conley and
  Agrawal}{2016}]{conley2016view}
Conley, C.~L. and A.~W. Agrawal (2016).
\newblock The view from a bike: How bike-share membership changes perceptions
  of and interactions with the community.
\newblock Paper presented at: {\em Transportation Research Board (TRB) Annual Meeting}, Number
  6673, Washington, DC, 2016. Washington, DC: Transportation Research Board.

\bibitem[\protect\citeauthoryear{{DDOT (District Department of
  Transportation)}}{{DDOT}}{2015a}]{DDOT2014BikeFact}
{DDOT (District Department of Transportation)} (2015a).
\newblock {\em 2014 Bike Program Fact Sheet}.
\newblock Washington, DC: DDOT.
\newblock Available online at:
  \url{https://ddot.dc.gov/publication/2014-bike-program-fact-sheet} (Accessed:
  2017-11-10).

\bibitem[\protect\citeauthoryear{{DDOT (District Department of
  Transportation)}}{{DDOT}}{2015b}]{DDOT2015CabiPlan}
{DDOT (District Department of Transportation)} (2015b).
\newblock {\em District of Columbia Capital Bikeshare Development Plan}.
\newblock Washington, DC: DDOT.
\newblock Available online at: \url{https://ddot.dc.gov/capitalbikeshare}
  (Accessed: 2017-11-10).

\bibitem[\protect\citeauthoryear{{DDOT (District Department of
  Transportation)}}{{DDOT}}{2016}]{DDOT2015SafetyFact}
{DDOT (District Department of Transportation)} (2016).
\newblock {\em Traffic Safety Statistics (2013-2015)}.
\newblock Washington, DC: DDOT.
\newblock Available online at:
  \url{https://ddot.dc.gov/publication/traffic-safety-report-statistics}
  (Accessed: 2017-11-10).

\bibitem[\protect\citeauthoryear{{DDOT (District Department of
  Transportation)}}{{DDOT}}{2017}]{DDOT2017Mobility}
{DDOT (District Department of Transportation)} (2017).
\newblock {\em District Mobility: Multimodal Transportation in the District}.
\newblock Washington, DC: DDOT.
\newblock Available online at: \url{https://districtmobility.org} (Accessed:
  2017-11-10).

\bibitem[\protect\citeauthoryear{de~Chardon, Caruso, and Thomas}{de~Chardon
  et~al.}{2016}]{de2016bike}
de~Chardon, C.~M., G.~Caruso, and I.~Thomas (2016).
\newblock Bike-share rebalancing strategies, patterns, and purpose.
\newblock {\em Journal of Transport Geography\/}~{\em 55}, 22--39.

\bibitem[\protect\citeauthoryear{DeMaio}{DeMaio}{2009}]{demaio2009bike}
DeMaio, P. (2009).
\newblock Bike-sharing: History, impacts, models of provision, and future.
\newblock {\em Journal of Public Transportation\/}~{\em 12\/}(4), 41--56.

\bibitem[\protect\citeauthoryear{El-Assi, Mahmoud, and Habib}{El-Assi
  et~al.}{2017}]{el2017effects}
El-Assi, W., M.~S. Mahmoud, and K.~N. Habib (2017).
\newblock Effects of built environment and weather on bike sharing demand: A
  station level analysis of commercial bike sharing in {Toronto}.
\newblock {\em Transportation\/}~{\em 44}, 589--613.

\bibitem[\protect\citeauthoryear{Elhenawy, Bichiou, and Rakha}{Elhenawy
  et~al.}{2018}]{elhenawy2018heuristic}
Elhenawy, M., Y.~Bichiou, and H.~A. Rakha (2018).
\newblock A heuristic algorithm for rebalancing bike sharing systems.
\newblock Paper presented at: {\em Transportation Research Board (TRB) Annual Meeting}, Number
  0958, Washington, DC, 2018. Washington, DC: Transportation Research Board.

\bibitem[\protect\citeauthoryear{Faghih-Imani and Eluru}{Faghih-Imani and
  Eluru}{2016}]{faghih2016incorporating}
Faghih-Imani, A. and N.~Eluru (2016).
\newblock Incorporating the impact of spatio-temporal interactions on bicycle
  sharing system demand: A case study of {New York CitiBike} system.
\newblock {\em Journal of Transport Geography\/}~{\em 54}, 218--227.

\bibitem[\protect\citeauthoryear{Faghih-Imani, Hampshire, Marla, and
  Eluru}{Faghih-Imani et~al.}{2017}]{faghih2017empirical}
Faghih-Imani, A., R.~Hampshire, L.~Marla, and N.~Eluru (2017).
\newblock An empirical analysis of bike sharing usage and rebalancing: Evidence
  from {Barcelona} and {Seville}.
\newblock {\em Transportation Research Part A\/}~{\em 97}, 177--191.

\bibitem[\protect\citeauthoryear{{{FHWA (Federal Highway
  Administration)}}}{{{FHWA}}}{2016}]{FHWA2017Congestion}
{{FHWA (Federal Highway Administration)}} ({2016}).
\newblock {\em {Urban Congestion Reports}}.
\newblock Washington, DC: FHWA.
\newblock Available online at:
  {\url{https://ops.fhwa.dot.gov/perf_measurement/ucr/} (Accessed:
  2017-11-10)}.

\bibitem[\protect\citeauthoryear{Fishman}{Fishman}{2016}]{fishman2016bikeshare}
Fishman, E. (2016).
\newblock Bikeshare: A review of recent literature.
\newblock {\em Transport Reviews\/}~{\em 36}, 92--113.

\bibitem[\protect\citeauthoryear{Fishman and Schepers}{Fishman and
  Schepers}{2016}]{fishman2016global}
Fishman, E. and P.~Schepers (2016).
\newblock Global bike share: What the data tells us about road safety.
\newblock {\em Journal of Safety Research\/}~{\em 56}, 41--45.

\bibitem[\protect\citeauthoryear{Fishman, Washington, and Haworth}{Fishman
  et~al.}{2014}]{fishman2014bike}
Fishman, E., S.~Washington, and N.~Haworth (2014).
\newblock Bike share's impact on car use: Evidence from the {United States,
  Great Britain, and Australia}.
\newblock {\em Transportation Research Part D\/}~{\em 31}, 13--20.

\bibitem[\protect\citeauthoryear{Fishman, Washington, Haworth, and
  Mazzei}{Fishman et~al.}{2014}]{fishman2014barriers}
Fishman, E., S.~Washington, N.~Haworth, and A.~Mazzei (2014).
\newblock Barriers to bikesharing: An analysis from {Melbourne and Brisbane}.
\newblock {\em Journal of Transport Geography\/}~{\em 41}, 325--337.

\bibitem[\protect\citeauthoryear{Fournier, Christofa, and Knodler}{Fournier
  et~al.}{2017}]{fournier2017sinusoidal}
Fournier, N., E.~Christofa, and M.~A. Knodler (2017).
\newblock A sinusoidal model for seasonal bicycle demand estimation.
\newblock {\em Transportation Research Part D\/}~{\em 50}, 154--169.

\bibitem[\protect\citeauthoryear{Garc{\'\i}a-Palomares, Guti{\'e}rrez, and
  Latorre}{Garc{\'\i}a-Palomares et~al.}{2012}]{garcia2012optimizing}
Garc{\'\i}a-Palomares, J.~C., J.~Guti{\'e}rrez, and M.~Latorre (2012).
\newblock Optimizing the location of stations in bike-sharing programs: a GIS
  approach.
\newblock {\em Applied Geography\/}~{\em 35\/}(1), 235--246.

\bibitem[\protect\citeauthoryear{Gebhart and Noland}{Gebhart and
  Noland}{2014}]{gebhart2014impact}
Gebhart, K. and R.~B. Noland (2014).
\newblock The impact of weather conditions on bikeshare trips in {Washington,
  DC}.
\newblock {\em Transportation\/}~{\em 41\/}(6), 1205--1225.

\bibitem[\protect\citeauthoryear{Ghosh, Varakantham, Adulyasak, and
  Jaillet}{Ghosh et~al.}{2017}]{ghosh2017dynamic}
Ghosh, S., P.~Varakantham, Y.~Adulyasak, and P.~Jaillet (2017).
\newblock Dynamic repositioning to reduce lost demand in bike sharing systems.
\newblock {\em Journal of Artificial Intelligence Research\/}~{\em 58},
  387--430.

\bibitem[\protect\citeauthoryear{Gigerenzer and Todd}{Gigerenzer and
  Todd}{1999}]{gigerenzer1999simple}
Gigerenzer, G. and P.~M. Todd (1999).
\newblock {\em Simple Heuristics that Make Us Smart}.
\newblock Oxford University Press.

\bibitem[\protect\citeauthoryear{Girvan and Newman}{Girvan and
  Newman}{2002}]{girvan2002community}
Girvan, M. and M.~E. Newman (2002).
\newblock Community structure in social and biological networks.
\newblock {\em Proceedings of the National Academy of Sciences\/}~{\em
  99\/}(12), 7821--7826.

\bibitem[\protect\citeauthoryear{Gonzalez, Hidalgo, and Barabasi}{Gonzalez
  et~al.}{2008}]{gonzalez2008understanding}
Gonzalez, M.~C., C.~A. Hidalgo, and A.-L. Barabasi (2008).
\newblock Understanding individual human mobility patterns.
\newblock {\em Nature\/}~{\em 453\/}(7196), 779--782.

\bibitem[\protect\citeauthoryear{{Google}}{{Google}}{2017}]{GMap2017API}
{Google} (2017).
\newblock {\em Google Maps API}.
\newblock Mountain View, CA: Google.
\newblock Available online at: \url{https://developers.google.com/maps} (Accessed: 2017-11-10).

\bibitem[\protect\citeauthoryear{Griffin and Sener}{Griffin and
  Sener}{2016}]{griffin2016planning}
Griffin, G.~P. and I.~N. Sener (2016).
\newblock Planning for bike share connectivity to rail transit.
\newblock {\em Journal of Public Transportation\/}~{\em 19\/}(2), 1.

\bibitem[\protect\citeauthoryear{Hamilton and Wichman}{Hamilton and
  Wichman}{2018}]{hamilton2017bicycle}
Hamilton, T.~L. and C.~J. Wichman (2018).
\newblock Bicycle infrastructure and traffic congestion: Evidence from {DC's
  Capital Bikeshare}.
\newblock {\em Journal of Environmental Economics and Management\/}~{\em 87\/}, 72--93.

\bibitem[\protect\citeauthoryear{{Holiday Weather Website}}{{Holiday Weather
  Website}}{2017}]{DC2017Weather}
{Holiday Weather Website} (2017).
\newblock {\em Washington DC: Annual Weather Averages}.
\newblock Newport, UK: Target Services Ltd.
\newblock Available online at:
  \url{http://www.holiday-weather.com/washington_dc/averages/} (Accessed:
  2017-11-10).

\bibitem[\protect\citeauthoryear{Jestico, Nelson, and Winters}{Jestico
  et~al.}{2016}]{jestico2016mapping}
Jestico, B., T.~Nelson, and M.~Winters (2016).
\newblock Mapping ridership using crowdsourced cycling data.
\newblock {\em Journal of Transport Geography\/}~{\em 52}, 90--97.

\bibitem[\protect\citeauthoryear{{LDA Consulting}}{{LDA
  Consulting}}{2016}]{bikeshare2016capital}
{LDA Consulting} (2016).
\newblock {\em 2016 Capital Bikeshare Member Survey Report}.
\newblock Washington, DC: Capital Bikeshare.
\newblock Available online at:
  \url{https://www.capitalbikeshare.com/system-data} (Accessed: 2017-11-10).

\bibitem[\protect\citeauthoryear{Lee, Sener, and Jones}{Lee
  et~al.}{2017}]{lee2017understanding}
Lee, R.~J., I.~N. Sener, and S.~N. Jones (2017).
\newblock Understanding the role of equity in active transportation planning in
  the {United States}.
\newblock {\em Transport Reviews\/}~{\em 37\/}(2), 211--226.

\bibitem[\protect\citeauthoryear{Lepri, Staiano, Sangokoya, Letouz{\'e}, and
  Oliver}{Lepri et~al.}{2017}]{lepri2017tyranny}
Lepri, B., J.~Staiano, D.~Sangokoya, E.~Letouz{\'e}, and N.~Oliver (2017).
\newblock The tyranny of data? the bright and dark sides of data-driven
  decision-making for social good.
\newblock In Cerquitelli T., Quercia D., and Pasquale F. (Eds.), {\em Transparent Data Mining for Big and Small Data}, pp.\  3--24.
  Cham, Switzerland: Springer.

\bibitem[\protect\citeauthoryear{Liu and Shi}{Liu and
  Shi}{2017}]{Liu2017Impact}
Liu, J.~H. and W.~Shi (2017).
\newblock Impact of bike facilities on residential property prices.
\newblock {\em Transportation Research Record\/}~{\em 2662}, 50--58.

\bibitem[\protect\citeauthoryear{Louf and Barthelemy}{Louf and
  Barthelemy}{2014}]{louf2014congestion}
Louf, R. and M.~Barthelemy (2014).
\newblock How congestion shapes cities: From mobility patterns to scaling.
\newblock {\em Scientific Reports\/}~{\em 4}, 5561.

\bibitem[\protect\citeauthoryear{Lowry, Furth, and Hadden-Loh}{Lowry
  et~al.}{2016}]{lowry2016prioritizing}
Lowry, M.~B., P.~Furth, and T.~Hadden-Loh (2016).
\newblock Prioritizing new bicycle facilities to improve low-stress network
  connectivity.
\newblock {\em Transportation Research Part A\/}~{\em 86}, 124--140.

\bibitem[\protect\citeauthoryear{Ma, Liu, and Erdo{\u{g}}an}{Ma
  et~al.}{2015}]{ma2015bicycle}
Ma, T., C.~Liu, and S.~Erdo{\u{g}}an (2015).
\newblock Bicycle sharing and public transit: Does {Capital Bikeshare} affect
  {Metrorail} ridership in {Washington, DC}?
\newblock {\em Transportation Research Record\/}~{\em 2534}, 1--9.

\bibitem[\protect\citeauthoryear{Martin, Cohen, Botha, and Shaheen}{Martin
  et~al.}{2016}]{martin2016bikesharing}
Martin, E., A.~Cohen, J.~L. Botha, and S.~Shaheen (2016).
\newblock {\em Bikesharing and Bicycle Safety}.
\newblock Technical Report CA-MTI-15-1204, San Jose,
  CA: San Jose State University.
\newblock Available online at:
  \url{http://transweb.sjsu.edu/research/bikesharing-and-bicycle-safety}
  (Accessed: 2017-11-10).

\bibitem[\protect\citeauthoryear{Martin and Shaheen}{Martin and
  Shaheen}{2014}]{martin2014evaluating}
Martin, E.~W. and S.~A. Shaheen (2014).
\newblock Evaluating public transit modal shift dynamics in response to
  bikesharing: a tale of two us cities.
\newblock {\em Journal of Transport Geography\/}~{\em 41}, 315--324.

\bibitem[\protect\citeauthoryear{McKenzie}{McKenzie}{2014}]{mckenzie2014modes}
McKenzie, B. (2014).
\newblock {\em Modes Less Traveled -- Bicycling and Walking to Work in the United
  States: 2008--2012}.
\newblock Technical Report ACS-25, Washington, DC: United States Census Bureau.
\newblock Available online at:
  \url{https://www.census.gov/prod/2014pubs/acs-25.pdf}
  (Accessed: 2017-11-10).

\bibitem[\protect\citeauthoryear{McLaughlin}{McLaughlin}{1987}]{mclaughlin1987learning}
McLaughlin, M.~W. (1987).
\newblock Learning from experience: Lessons from policy implementation.
\newblock {\em Educational Evaluation and Policy Analysis\/}~{\em 9\/}(2),
  171--178.

\bibitem[\protect\citeauthoryear{Meddin and DeMaio}{Meddin and DeMaio}{2017}]{Meddin2017bike}
Meddin, R. and P.~DeMaio (2017).
\newblock {\em The Bike-Sharing World Map}.
\newblock Available online at: \url{http://www.bikesharingworld.com} (Accessed: 2017-11-10).

\bibitem[\protect\citeauthoryear{Miranda-Moreno, Nosal, Schneider, and
  Proulx}{Miranda-Moreno et~al.}{2013}]{miranda2013classification}
Miranda-Moreno, L., T.~Nosal, R.~Schneider, and F.~Proulx (2013).
\newblock Classification of bicycle traffic patterns in five {North American}
  cities.
\newblock {\em Transportation Research Record\/}~{\em 2339}, 68--79.

\bibitem[\protect\citeauthoryear{Moritz}{Moritz}{1997}]{moritz1997survey}
Moritz, W. (1997).
\newblock Survey of {North American} bicycle commuters: Design and aggregate
  results.
\newblock {\em Transportation Research Record\/}~{\em 1578}, 91--101.

\bibitem[\protect\citeauthoryear{Mueller, Rojas-Rueda, Cole-Hunter, and
  de~Nazelle}{Mueller et~al.}{2015}]{mueller2015health}
Mueller, N., D.~Rojas-Rueda, T.~Cole-Hunter, and A.~de~Nazelle (2015).
\newblock Health impact assessment of active transportation: A systematic
  review.
\newblock {\em Preventive Medicine\/}~{\em 76}, 103--114.

\bibitem[\protect\citeauthoryear{{NHTSA (National Highway Traffic Safety
  Administration)}}{{NHTSA}}{2017}]{NHTSA2015TSF}
{NHTSA (National Highway Traffic Safety Administration)} (2017).
\newblock {\em Traffic Safety Facts 2015}.
\newblock Technical Report DOT-HS-812-384, Washington, DC: NTSHA.
\newblock Available online at:
  \url{https://crashstats.nhtsa.dot.gov/Api/Public/Publication/812384}
  (Accessed: 2017-11-10).

\bibitem[\protect\citeauthoryear{Noland, Smart, and Guo}{Noland
  et~al.}{2017}]{Noland2017}
Noland, R., M.~Smart, and Z.~Guo (2017).
\newblock What do people use bikesharing for? {Analysis} of origin and
  destination pairs in {New York City}.
\newblock Paper presented at: {\em Transportation Research Board (TRB) Annual Meeting}, Number 716,
  Washington, DC, 2015. Washington, DC: Transportation Research Board.

\bibitem[\protect\citeauthoryear{O'Brien, Cheshire, and Batty}{O'Brien
  et~al.}{2014}]{o2014mining}
O'Brien, O., J.~Cheshire, and M.~Batty (2014).
\newblock Mining bicycle sharing data for generating insights into sustainable
  transport systems.
\newblock {\em Journal of Transport Geography\/}~{\em 34}, 262--273.

\bibitem[\protect\citeauthoryear{{OCTO (Office of the Chief Technology Officer,
  DC)}}{{OCTO, DC}}{2017}]{DC2017Data}
{OCTO (Office of the Chief Technology Officer), DC} (2017).
\newblock {\em Open Data DC}.
\newblock Washington, DC: OCTO.
\newblock Available online at: \url{http://opendata.dc.gov} (Accessed:
  2017-11-10).

\bibitem[\protect\citeauthoryear{Perez, Buck, and Ma}{Perez
  et~al.}{2017}]{Perez2017}
Perez, B.~O., D.~Buck, and Y.~Ma (2017).
\newblock Mind the gap: Assessing the impacts of bicycle accessibility and
  mobility on mode share in {Washington, DC}.
\newblock Paper presented at: {\em Transportation Research Board (TRB) Annual Meeting}, Number
  1440, Washington, DC, 2017. Washington, DC: Transportation Research Board.

\bibitem[\protect\citeauthoryear{Poirier}{Poirier}{2018}]{Poirier2018Bicycle}
Poirier, J. (2018).
\newblock Bicycle lanes and business success: A {San Francisco} examination.
\newblock Paper presented at: {\em Transportation Research Board (TRB) Annual Meeting}, Number
  3763, Washington, DC, 2018. Washington, DC: Transportation Research Board.

\bibitem[\protect\citeauthoryear{Portilla, Valencia, Espinosa, N{\'u}{\~n}ez,
  and De~Schutter}{Portilla et~al.}{2016}]{portilla2016model}
Portilla, C., F.~Valencia, J.~Espinosa, A.~N{\'u}{\~n}ez, and B.~De~Schutter
  (2016).
\newblock Model-based predictive control for bicycling in urban intersections.
\newblock {\em Transportation Research Part C\/}~{\em 70}, 27--41.

\bibitem[\protect\citeauthoryear{Power}{Power}{2008}]{power2008understanding}
Power, D.~J. (2008).
\newblock Understanding data-driven decision support systems.
\newblock {\em Information Systems Management\/}~{\em 25\/}(2), 149--154.

\bibitem[\protect\citeauthoryear{Pucher and Buehler}{Pucher and
  Buehler}{2017}]{pucher2017cycling}
Pucher, J. and R.~Buehler (2017).
\newblock Cycling towards a more sustainable transport future.
\newblock {\em Transport Reviews\/}~{\em 37\/}(6), 689--694.

\bibitem[\protect\citeauthoryear{Ravensbergen}{Ravensbergen}{2018}]{Ravensbergen2018Biking}
Ravensbergen, L. (2018).
\newblock Biking to ride: Investigating the challenges and barriers of
  integrating cycling with regional rail transit.
\newblock Paper presented at: {\em Transportation Research Board (TRB) Annual Meeting}, Number
  2701, Washington, DC, 2018. Washington, DC: Transportation Research Board.

\bibitem[\protect\citeauthoryear{Rixey and Prabhakar}{Rixey and
  Prabhakar}{2017}]{Rixey2017}
Rixey, R. and N.~Prabhakar (2017).
\newblock Impacts of the level of traffic stress on bikeshare ridership in the
  case of {Capital Bikeshare} in {Montgomery County, Maryland}.
\newblock Paper presented at: {\em Transportation Research Board (TRB) Annual Meeting}, Number
  5454, Washington, DC, 2017. Washington, DC: Transportation Research Board.

\bibitem[\protect\citeauthoryear{Romanillos, Zaltz~Austwick, Ettema, and
  De~Kruijf}{Romanillos et~al.}{2016}]{romanillos2016big}
Romanillos, G., M.~Zaltz~Austwick, D.~Ettema, and J.~De~Kruijf (2016).
\newblock Big data and cycling.
\newblock {\em Transport Reviews\/}~{\em 36\/}(1), 114--133.

\bibitem[\protect\citeauthoryear{Schepers, Fishman, Den~Hertog, Wolt, and
  Schwab}{Schepers et~al.}{2014}]{schepers2014safety}
Schepers, J., E.~Fishman, P.~Den~Hertog, K.~K. Wolt, and A.~Schwab (2014).
\newblock The safety of electrically assisted bicycles compared to classic
  bicycles.
\newblock {\em Accident Analysis \& Prevention\/}~{\em 73}, 174--180.

\bibitem[\protect\citeauthoryear{Schleith and Horner}{Schleith and
  Horner}{2014}]{schleith2014commuting}
Schleith, D. and M.~Horner (2014).
\newblock Commuting, job clusters, and travel burdens: Analysis of spatially
  and socioeconomically disaggregated longitudinal employer-household dynamics
  data.
\newblock {\em Transportation Research Record\/}~(2452), 19--27.

\bibitem[\protect\citeauthoryear{Schrank, Eisele, Lomax, and Bak}{Schrank
  et~al.}{2015}]{schrank20152015}
Schrank, D., B.~Eisele, T.~Lomax, and J.~Bak (2015).
\newblock {\em 2015 Urban Mobility Scorecard}.
\newblock College Station, TX: Texas A\&M Transportation Institute.
\newblock Available online at:
  \url{tti.tamu.edu/documents/mobility-scorecard-2015.pdf}
  (Accessed: 2017-11-10).

\bibitem[\protect\citeauthoryear{Seidl, Jankowski, and Tsou}{Seidl
  et~al.}{2016}]{seidl2016privacy}
Seidl, D.~E., P.~Jankowski, and M.-H. Tsou (2016).
\newblock Privacy and spatial pattern preservation in masked {GPS} trajectory
  data.
\newblock {\em International Journal of Geographical Information
  Science\/}~{\em 30\/}(4), 785--800.

\bibitem[\protect\citeauthoryear{Shaheen, Cohen, and Martin}{Shaheen
  et~al.}{2013}]{shaheen2013public}
Shaheen, S., A.~Cohen, and E.~Martin (2013).
\newblock Public bikesharing in {North America}: Early operator understanding
  and emerging trends.
\newblock {\em Transportation Research Record\/}~{\em 2387}, 83--92.

\bibitem[\protect\citeauthoryear{Shui and Szeto}{Shui and
  Szeto}{2018}]{shui2017dynamic}
Shui, C. and W.~Szeto (2018).
\newblock Dynamic green bike repositioning problem -- a hybrid rolling horizon
  artificial bee colony algorithm approach.
\newblock {\em Transportation Research Part D\/}~{\em 60}, 119--136.

\bibitem[\protect\citeauthoryear{Stinson and Bhat}{Stinson and
  Bhat}{2004}]{stinson2004frequency}
Stinson, M. and C.~Bhat (2004).
\newblock Frequency of bicycle commuting: Internet-based survey analysis.
\newblock {\em Transportation Research Record\/}~{\em 1878}, 122--130.

\bibitem[\protect\citeauthoryear{{USCB (United States Census Bureau)}}{{USCB}}{2017}]{LEHD2017LODES}
{USCB (United States Census Bureau)} (2017).
\newblock {\em Longitudinal Employer-Household Dynamics (LEHD)}.
\newblock Washington, DC: USCB.
\newblock Available online at: \url{https://lehd.ces.census.gov} (Accessed:
  2017-11-10).

\bibitem[\protect\citeauthoryear{Vogel, Greiser, and Mattfeld}{Vogel
  et~al.}{2011}]{vogel2011understanding}
Vogel, P., T.~Greiser, and D.~C. Mattfeld (2011).
\newblock Understanding bike-sharing systems using data mining: Exploring
  activity patterns.
\newblock {\em Procedia-Social and Behavioral Sciences\/}~{\em 20}, 514--523.

\bibitem[\protect\citeauthoryear{Wang and Zhou}{Wang and
  Zhou}{2017}]{wang2017bike}
Wang, M. and X.~Zhou (2017).
\newblock Bike-sharing systems and congestion: Evidence from {US} cities.
\newblock {\em Journal of Transport Geography\/}~{\em 65}, 147--154.

\bibitem[\protect\citeauthoryear{Wergin and Buehler}{Wergin and
  Buehler}{2017}]{Wergin2017}
Wergin, J. and R.~Buehler (2017).
\newblock Where do bikeshare bikes actually go? {Analysis} of {Capital
  Bikeshare} trips using {GPS} data.
\newblock Paper presented at: {\em Transportation Research Board (TRB) Annual Meeting}, Number
  1167, Washington, DC, 2017. Washington, DC: Transportation Research Board.

\bibitem[\protect\citeauthoryear{Wu, Kang, and Wang}{Wu
  et~al.}{2017}]{wu2017explore}
Wu, Y.-H., L.~Kang, and P.-C. Wang (2017).
\newblock Exploratory multivariate analysis of bikesharing system use: Trip
  characteristics and effect of pricing scheme change.
\newblock Paper presented at: {\em Transportation Research Board (TRB) Annual Meeting}, Number
  6683, Washington, DC, 2017. Washington, DC: Transportation Research Board.

\bibitem[\protect\citeauthoryear{Xie, Liu, and Wang}{Xie
  et~al.}{2014}]{xie2014cooperative}
Xie, X.-F., J.~Liu, and Z.-J. Wang (2014).
\newblock A cooperative group optimization system.
\newblock {\em Soft Computing\/}~{\em 18\/}(3), 469--495.

\bibitem[\protect\citeauthoryear{Xie and Wang}{Xie and Wang}{2015}]{Xie2015}
Xie, X.-F. and Z.-J. Wang (2015).
\newblock An empirical study of combining participatory and physical sensing to
  better understand and improve urban mobility networks.
\newblock Paper presented at: {\em {Transportation Research Board (TRB) Annual Meeting}}, Number
  3238, Washington, DC, 2015. Washington, DC: Transportation Research Board.

\bibitem[\protect\citeauthoryear{Yi and Shirk}{Yi and Shirk}{2018}]{yi2018data}
Yi, Z. and M.~Shirk (2018).
\newblock Data-driven optimal charging decision making for connected and
  automated electric vehicles: A personal usage scenario.
\newblock {\em Transportation Research Part C\/}~{\em 86}, 37--58.

\bibitem[\protect\citeauthoryear{Zhao, Wang, and Deng}{Zhao
  et~al.}{2015}]{zhao2015exploring}
Zhao, J., J.~Wang, and W.~Deng (2015).
\newblock Exploring bikesharing travel time and trip chain by gender and day of
  the week.
\newblock {\em Transportation Research Part C\/}~{\em 58}, 251--264.

\bibitem[\protect\citeauthoryear{Zhou, Jia, Juan, Fu, and Xiao}{Zhou
  et~al.}{2017}]{zhou2017data}
Zhou, C., H.~Jia, Z.~Juan, X.~Fu, and G.~Xiao (2017).
\newblock A data-driven method for trip ends identification using large-scale
  smartphone-based {GPS} tracking data.
\newblock {\em IEEE Transactions on Intelligent Transportation Systems\/}~{\em
  18\/}(8), 2096--2110.

\end{thebibliography}

\end{document}